\documentclass[journal, 12pt, onecolumn, draftclsnofoot]{IEEEtran}
\usepackage{authblk}
\usepackage[utf8]{inputenc}
\usepackage{mathtools}
\usepackage{graphicx}
\usepackage{xcolor}
\usepackage{amsthm}
\usepackage{algorithm}
\usepackage{algpseudocode}
\usepackage{amssymb}
\usepackage{cite}
\usepackage{multirow}
\usepackage{bm}
\usepackage{tablefootnote}
\usepackage{threeparttable}

\usepackage{hyperref}

\DeclareMathOperator{\pr}{Pr}

\newtheorem{lemma}{Lemma}
\newtheorem{theorem}{Theorem}

\newtheorem{remark}{Remark}

\theoremstyle{definition}
\newtheorem{example}{Example}

\begin{document}
\title{\Large On Single Server Private Information Retrieval with  Private Coded Side Information}
\author{\normalsize Yuxiang Lu and Syed A. Jafar}
\affil{\small Center for Pervasive Communications and Computing (CPCC), UC Irvine\\
Email: \{yuxiang.lu, syed\}@uci.edu}
\date{}

\allowdisplaybreaks

\maketitle
\begin{abstract}
    Motivated by an open problem and a conjecture, this work studies the problem of single server private information retrieval with private coded side information (PIR-PCSI) that was recently introduced by Heidarzadeh et al. The goal of PIR-PCSI is to allow a user  to efficiently  retrieve a desired message $\bm{W}_{\bm{\theta}}$, which is one of $K$ independent messages  that are stored at a server, while utilizing private side information of a linear combination of a uniformly chosen size-$M$ subset ($\bm{\mathcal{S}}\subset[K]$) of messages. The settings PIR-PCSI-I and PIR-PCSI-II correspond to the constraints that $\bm{\theta}$ is generated  uniformly from $[K]\setminus\bm{\mathcal{S}}$, and  $\bm{\mathcal{S}}$, respectively. In each case, $(\bm{\theta},\bm{\mathcal{S}})$ must be kept private from the server. The capacity is defined as the supremum over message and field sizes, of achievable rates (number of bits of desired message retrieved per bit of download) and is characterized by Heidarzadeh et al. for PIR-PCSI-I in general, and for PIR-PCSI-II for $M>(K+1)/2$ as $(K-M+1)^{-1}$. For $2\leq M\leq (K+1)/2$ the capacity of PIR-PCSI-II remains open, and  it is conjectured that even in this case the capacity is $(K-M+1)^{-1}$. We show the capacity of PIR-PCSI-II is equal to $2/K$ for $2 \leq M \leq \frac{K+1}{2}$, which is strictly larger than the conjectured value, and  does not depend on $M$ within this parameter regime. Remarkably, half the side-information is found to be redundant. We also characterize the infimum capacity (infimum over fields instead of supremum), and the capacity with private coefficients. The results are generalized to PIR-PCSI-I ($\theta\in[K]\setminus\mathcal{S}$) and PIR-PCSI ($\theta\in[K]$) settings.
\end{abstract}

\section{Introduction}
As cloud services and distributed data storage become increasingly prevalent, growing concerns about users' privacy have sparked much recent interest  in the problem of Private Information Retrieval (PIR). Originally introduced in  \cite{PIRfirst,PIRfirstjournal}, the goal of PIR is to allow a user to efficiently retrieve a desired message from a server or a set of servers where multiple messages are stored, without revealing any information about which message is desired. In the information theoretic framework, which requires perfect privacy and assumes  long messages, the capacity of PIR is the maximum number of bits of desired information that can be retrieved per bit of download from the server(s) \cite{Sun_Jafar_PIR}. Capacity characterizations have recently been obtained for various forms of PIR, especially for the multi-server setting  \cite{Shah_Rashmi_Kannan, Sun_Jafar_PIR, Tajeddine_Gnilke_Karpuk_Hollanti, Sun_Jafar_TPIR, Sun_Jafar_SPIR, 
Banawan_Ulukus_BPIR, 
Wang_Skoglund_PIRSPIRAd, 
Wang_Skoglund_TSPIR,  
FREIJ_HOLLANTI, Sun_Jafar_MDSTPIR, Wang_Skoglund_MDS, Jia_Sun_Jafar_XSTPIR, Jia_Jafar_MDSXSTPIR, Yang_Shin_Lee,
Jia_Jafar_GXSTPIR, Tandon_CachePIR,  Wei_Banawan_Ulukus,  Tajeddine_Gnilke_Karpuk, Yao_Liu_Kang_Collusion_Pattern, PIR_survey}. 

PIR in the basic single server setting would be most valuable if it could be made efficient. However, it was already shown in the earliest works on PIR \cite{PIRfirst, PIRfirstjournal} that in the single server case there is no better alternative to the trivial solution of downloading everything, which is prohibitively expensive. Since the optimal solution turns out to be trivial, single server PIR generally received less attention from the information theoretic perspective, until recently. 
Interest in the capacity of single-server PIR was revived  by the seminal contribution of Kadhe et al. in \cite{Kadhe_Garcia_Heidarzadeh_ElRouayheb_Sprintson_PIR_SI} which showed that the presence of \emph{side information} at the user can significantly improve the efficiency of PIR, and that capacity characterizations under side information  are far from trivial. This crucial observation inspired much work on understanding the role of side-information in PIR \cite{Chen_Wang_Jafar_Side, heidarzadeh2018oncapacity, li2018single, li2020single, kazemi2019single, heidarzadeh2019single, heidarzadeh2018capacity, heidarzadeh2019capacity, Wei_Banawan_Ulukus_Side, PIR_PCSI}, which remains an active topic of research. Among the  recent advances in this area is the study of single-server PIR with private coded side information (PIR-PCSI) that was initiated by Heidarzadeh, Kazemi and Sprintson in \cite{PIR_PCSI}.  Heidarzadeh et al. obtain sharp capacity characterizations for PIR-PCSI in many cases, and also note an open problem, along with an intriguing conjecture that motivates our work in this paper.

In the PIR-PCSI problem, a single server stores $K$ independent messages $\bm{W}_1, \cdots, \bm{W}_K$, each represented by $L$ i.i.d. uniform symbols from a finite field $\mathbb{F}_q$. A user wishes to efficiently retrieve a desired message $\bm{W}_{\bm{\theta}}$, while utilizing private side information $(\bm{\mathcal{S}}, \bm{\Lambda}, \bm{Y}^{[\bm{\mathcal{S}},\bm{\Lambda}]})$ that is unknown to the server, comprised of a linear combination $ \bm{Y}^{[\bm{\mathcal{S}},\bm{\Lambda}]}=\sum_{m=1}^M\bm{\lambda}_m\bm{W}_{i_m}$ of a uniformly chosen size-$M$ subset of messages, $\bm{\mathcal{S}}=\{\bm{i}_1,\bm{i}_2,\cdots,\bm{i}_M\}\subset[K], \bm{i}_1<\bm{i}_2<\cdots<\bm{i}_M$, with the coefficient vector $\bm{\Lambda}=(\bm{\lambda}_1, \cdots, \bm{\lambda}_M)$ whose elements are chosen i.i.d. uniform from $\mathbb{F}_q^\times$, i.e., the multiplicative subgroup of $\mathbb{F}_{q}$.  Depending on whether $\bm{\theta}$ is drawn uniformly from $[K]\setminus\bm{\mathcal{S}}$ or uniformly from $\bm{\mathcal{S}}$, there are two settings,  known as PIR-PCSI-I and PIR-PCSI-II, respectively. In each case, $(\bm{\theta}, \bm{\mathcal{S}})$ must be kept private. Capacity of PIR is typically defined as the maximum number of bits of desired message that can be retrieved per bit of download from the server(s), and includes a supremum over message size $L$. Since the side-information formulation    specifies a finite field $\mathbb{F}_q$, the capacity of PIR-PCSI can potentially depend on the field. A field-independent notion of capacity  is introduced in  \cite{PIR_PCSI} by allowing a supremum over all finite fields. For PIR-PCSI-I, where $\bm{\theta}\notin \bm{\mathcal{S}}$, Heidarzadeh et al. fully characterize the  capacity as $(K-M)^{-1}$ for $1 \leq M \leq K-1$. For PIR-PCSI-II, the   capacity is characterized as $(K-M+1)^{-1}$ for $\frac{K+1}{2} < M \leq K$. Capacity characterization for the remaining case of $2 \leq M \leq \frac{K+1}{2}$ is noted as an open problem in \cite{PIR_PCSI}, and it is conjectured that the capacity in this case is also $(K-M+1)^{-1}$. 

The main motivation of our work is to settle this conjecture and obtain the capacity characterization for PIR-PCSI-II when $2 \leq M \leq \frac{K+1}{2}$. Given the importance of better understanding the role of side information for single-server PIR, additional motivation comes from the following questions:  What is the infimum capacity (infimum over all finite fields instead of supremum)? What if the coefficient vector $\bm{\Lambda}$ (whose privacy is not required in \cite{PIR_PCSI}) is also required to be private? Can the  side-information be reduced, e.g., to save storage, without reducing capacity?

\begin{table*}[!t]
    \caption{Capacity results for PIR-PCSI-I, PIR-PCSI-II and PIR-PCSI}
    \label{tab:capacity}
    \centering
    \scalebox{0.76}{
    \begin{tabular}{|c|c|c|}
    \hline
    PIR-PCSI-I ($1 \leq M \leq K-1$) & PIR-PCSI-II ($2 \leq M \leq K$) & PIR-PCSI ($1 \leq M \leq K$)\\ \hline
    $C_{\mbox{\tiny PCSI-I}}^{\sup} = \frac{1}{K-M}$, \cite{PIR_PCSI} 
    & $C_{\mbox{\tiny PCSI-II}}^{\sup} =
        \begin{cases}
            \frac{2}{K}, & 2 \leq M \leq \frac{K+1}{2}, \text{Thm. \ref{thm:cap_PCSI2_sup}}\\	
            \frac{1}{K-M+1}, & \frac{K+1}{2} < M \leq K,\text{\cite{PIR_PCSI}}
        \end{cases}$ 
    & $C_{\mbox{\tiny PCSI}}^{\sup} = 
    \begin{cases}
        \frac{1}{K-1}, & M=1,\\
        \frac{1}{K-M+1}, & 2 \leq M \leq K,
    \end{cases}$, Thm. \ref{thm:cap_PCSI_sup}\\ \hline
    $C_{\mbox{\tiny PCSI-I}}^{\inf} =
        \begin{cases}
            \frac{1}{K-1}, & 1 \leq M \leq \frac{K}{2},\\
            \big(K - \frac{M}{K-M}\big)^{-1}, & \frac{K}{2} < M \leq K-1,
        \end{cases}$, Thm. \ref{thm:cap_PCSI1_inf}
    & $C_{\mbox{\tiny PCSI-II}}^{\inf} = \frac{M}{(M-1)K}$, Thm. \ref{thm:cap_PCSI2_inf} & $C_{\mbox{\tiny PCSI}}^{\inf} = \frac{1}{K-1}$, Thm. \ref{thm:cap_PCSI_inf} \\ \hline
    
    \begin{tabular}{c}$C_{\mbox{\tiny PCSI-I}}^{\inf} \leq C_{\mbox{\tiny PCSI-I}}(q) \leq C_{\mbox{\tiny PCSI-I}}^{\sup}$\\
    \\
    \hline
    \\
    When $M = K-1$, $C_{\mbox{\tiny PCSI-I}}(q) = 1$, Rmk. \ref{rmk:pcsi1_arb_field}
    \end{tabular} & 
    \begin{tabular}{c}$C_{\mbox{\tiny PCSI-II}}^{\inf} \leq C_{\mbox{\tiny PCSI-II}}(q) \leq C_{\mbox{\tiny PCSI-II}}^{\sup}$\\
    \hline
    When $M=K$,\\
    $C_{\mbox{\tiny PCSI-II}}(q)=\begin{cases}
    1/(K-1), &q=2,\\
    1, &q\neq 2,
    \end{cases}$, Thm. \ref{thm:MK}\\
    \hline
    When $M = 3, K = 4$,\\
    $C_{\mbox{\tiny PCSI-II}}(q) =\begin{cases}
    3/8, &q=2,\\
    1/2,&q\neq 2,
    \end{cases}$, Thm. \ref{thm:M3K4}
    \end{tabular} & 
    \begin{tabular}{c}$C_{\mbox{\tiny PCSI}}^{\inf} \leq C_{\mbox{\tiny PCSI}}(q) \leq C_{\mbox{\tiny PCSI}}^{\sup}$
    \end{tabular} \\ \hline
    
    \begin{tabular}{c}$ C_{\mbox{\tiny PCSI-I}}^{\mbox{\tiny pri}, \sup} = C_{\mbox{\tiny PCSI-I}}^{\inf}$\\
    $\frac{1}{K-1} \leq C_{\mbox{\tiny PCSI-I}}^{\mbox{\tiny pri}, \inf} \leq \min\bigg(C_{\mbox{\tiny PCSI-I}}^{\inf}, \frac{1}{K-2}\bigg)$\end{tabular}, Thm. \ref{thm:pcsi1_pub_pri} & $ C_{\mbox{\tiny PCSI-II}}^{\mbox{\tiny pri}}(q) = C_{\mbox{\tiny PCSI-II}}^{\inf}$, Thm. \ref{thm:pcsi2_pub_pri} & $ C_{\mbox{\tiny PCSI}}^{\mbox{\tiny pri}}(q) = C_{\mbox{\tiny PCSI}}^{\inf}$, Thm. \ref{thm:pcsi_pub_pri} \\ \hline

    \end{tabular}}
    \begin{tablenotes}
        \item {\it Notation summary:} $C$ stands for capacity, $q$ in the parentheses denotes the problem lies in $\mathbb{F}_{q}$, the subscript denotes the type of  problem (PIR-PCSI-I, PIR-PCSI-II, or PIR-PCSI). In the superscript, `$\inf$' (resp. `$\sup$') denotes that the infimum (resp. supremum) of the capacity over all valid $q$ is considered. The term `pri' as a superscript indicates that it is the capacity when coefficients must also be kept private. Thm. (resp. Rmk.) points out the  theorem (resp. remark), where the result appears.
    \end{tablenotes}
\end{table*}

The contributions of this work are summarized in Table \ref{tab:capacity}, along with prior results from \cite{PIR_PCSI}. As our main contribution we show that the  capacity of PIR-PCSI-II for $2 \leq M \leq \frac{K+1}{2}$ is equal to $2/K$, which is strictly higher than the conjectured value in this parameter regime. The result reveals two surprising aspects  of this parameter regime. First, whereas previously known capacity characterizations of PIR-PCSI-II (and PIR-PCSI-I) in \cite{PIR_PCSI} are all strictly increasing with $M$ (the size of the support set of side information), here the capacity does not depend on $M$. Second, in this parameter regime (and also when $M=\left\lfloor (K+1)/2\right\rfloor +1$), half of the side information turns out to be redundant, i.e., the supremum capacity remains the same even if the user discards half of the side information. We also show that if more than half of the side information is discarded, then the supremum capacity is strictly smaller. By contrast, in other regimes no redundancy exists in the side information, i.e., any reduction in side information would lead to a  loss in supremum capacity. The results regarding the redundancy in the side information when the supremum capacity is achieved are summarized in Table \ref{tab:redundancy} in Section \ref{sec:main} as the definition of redundancy will be clear then.

The optimal rate $2/K$ is shown to be achievable for any finite field $\mathbb{F}_q$ where $q$ is an \emph{even} power of a prime. The achievable scheme requires downloads that are ostensibly non-linear in $\mathbb{F}_q$, but in its essence the scheme is linear, as can be seen by interpreting $\mathbb{F}_q$ as a $2$ dimensional vector space over the base field $\mathbb{F}_{\sqrt{q}}$, over which the downloads are indeed linear. Intuitively, the scheme may be understood as follows. A rate of $2/K$ means a download of $K/2$, which is achieved by downloading \emph{half} of every message (one of the two dimensions in the $2$ dimensional vector space over $\mathbb{F}_{\sqrt{q}}$). The key idea is \emph{interference alignment} -- for the undesired messages that appear in the side information, the halves that are downloaded are perfectly \emph{aligned} with each other, whereas for the desired message, the half that is downloaded is not aligned with the downloaded halves of the undesired messages. For messages that are not included in the side information, any random half can be downloaded to preserve privacy. 

With a bit of oversimplification for the sake of intuition, suppose  there are $K=4$ messages, that can be represented as $2$-dimensional vectors $\bm{A}=[\bm{a}_1~~ \bm{a}_2], \bm{B}=[\bm{b}_1~~ \bm{b}_2], \bm{C}=[\bm{c}_2~~ \bm{c}_2], \bm{D}=[\bm{d}_1~~ \bm{d}_2]$, the side information is comprised of  $M=3$ messages, say at first $\bm{A}+\bm{B}+\bm{C}=[\bm{a}_1+\bm{b}_1+\bm{c}_1~~ \bm{a}_2+\bm{b}_2+\bm{c}_2]$, and the desired message is $\bm{A}$. Then the user could recover $\bm{A}$ by downloading $\bm{a}_1, \bm{b}_2, \bm{c}_2$ and either $\bm{d}_1$ or $\bm{d}_2$, i.e., half of each message for a total download of $K/2=2$ (normalized by message size). We may also note that half of the side information is redundant, i.e., the user only needs $\bm{a}_2+\bm{b}_2+\bm{c}_2$, and can discard the rest. But there is a problem with this oversimplification -- this toy example  seemingly loses privacy because the matching indices reveal that $\bm{b}_2$ aligns with $\bm{c}_2$ but not $\bm{a}_1$. This issue is resolved by noting that the side information is in fact $\bm{\lambda}_1\bm{A}+\bm{\lambda}_2\bm{B}+\bm{\lambda}_3\bm{C}=\bm{A}'+\bm{B}'+\bm{C}'$. Suppose $\bm{\lambda}_1, \bm{\lambda}_2, \bm{\lambda}_3$ are random (unknown to the server) independent linear transformations (\emph{matrices}) that  independently `\emph{rotate}' $\bm{A}, \bm{B}, \bm{C}$ vectors into $\bm{A}',\bm{B}',\bm{C}'$ vectors, respectively, such that the projections (combining coefficients) of each along any particular dimension become independent of each other. In other words, $\bm{a}_i', \bm{b}_i', \bm{c}_i'$ are independent projections of $\bm{A}, \bm{B}, \bm{C}$, and downloading, say $(\bm{a}_1', \bm{b}_2', \bm{c}_2', \bm{d}_2')$ reveals to the server no information about their relative alignments in the side information. From the server's perspective, each downloaded symbol is simply an independent random linear combination of the two components of each message. Intuitively, since the random rotation is needed to maintain privacy, it is important that $\bm{\lambda}_i$ are matrices, not scalars (because scalars only scale, they do not rotate vectors). This is not directly  the case in $\mathbb{F}_q$ because $\bm{\lambda}_i$ are scalars in $\mathbb{F}_q$. However, viewed as a $2$ dimensional vector space over $\mathbb{F}_{\sqrt{q}}$, the $\bm{\lambda}_i$ indeed act as invertible $2\times 2$ matrices that act on the vectors $\bm{A}, \bm{B}, \bm{C}, \bm{D}$, rotating each vector randomly and independently, thus ensuring privacy. 

In order for $\mathbb{F}_{\sqrt{q}}$ to be a valid finite field we need $q$ to be an \emph{even} power of a prime. This suffices to characterize the  capacity because the capacity definition in \cite{PIR_PCSI} allows a supremum over all fields. However, the question remains about whether the rate $2/K$ is  achievable over every finite field. To understand this better, we explore an alternative definition of capacity (called infimum capacity in this work) which considers the infimum (instead of supremum) over all $\mathbb{F}_q$. We find that the infimum capacity of PIR-PCSI-II is always equal to $M/((M-1)K)$. Evidently, for $M=2$ the capacity is  field independent because the infimum and supremum over fields produce the same capacity result. In general however, the infimum capacity can be strictly smaller, thus confirming field-dependence. The worst case corresponds to the binary field $\mathbb{F}_2$.  Intuitively, the reason that the infimum capacity corresponds to the binary field is that over $\mathbb{F}_2$ the non-zero coefficients $\bm{\lambda}_m$ must all be equal to one, and thus the coefficients are essentially known to the server.  On the other hand, we also present an example with $q=3$ (and $M=3, K=4$) where $2/K$ is achievable (and optimal), to show that the achievability of $2/K$ for $M>2$ is not limited to just field sizes that are even powers of a prime number. We also show that for PIR-PCSI-II, the the infimum capacity with  private $(\bm{\theta},\bm{\mathcal{S}})$ is the same as the (supremum or infimum) capacity with private $(\bm{\theta},\bm{\mathcal{S}},\bm{\Lambda})$, i.e., when the coefficients $\bm{\Lambda}$  must also be kept private from the server. 

Next we consider PIR-PCSI-I where $\bm{\theta}$ is drawn from $[K]\setminus\bm{\mathcal{S}}$. The supremum capacity of PIR-PCSI-I is fully characterized in \cite{PIR_PCSI}. In this case, we show that there is no redundancy in the CSI. As in PIR-PCSI-II, we find that the infimum capacity of PIR-PCSI-I is strictly smaller than the supremum capacity in general, and the binary field $\mathbb{F}_2$ yields the worst case. Unlike PIR-PCSI-II, however, the infimum capacity of PIR-PCSI-I with private $(\bm{\theta},\bm{\mathcal{S}})$ does not always match the infimum capacity with  private $(\bm{\theta},\bm{\mathcal{S}},\bm{\Lambda})$. For example, if $M=K-1$, then both the supremum and infimum capacities of PIR-PCSI-I are equal to $1$ for private $(\bm{\theta},\bm{\mathcal{S}})$, but if the  coefficient vector $\bm{\Lambda}$ must also be kept private then the infimum capacity is no more than $1/(K-2)$. Thus, the loss in capacity from requiring privacy of coefficients can be quite significant.

To complete the picture, we finally consider the capacity of PIR-PCSI where $\bm\theta$ is drawn uniformly from $[K]$.  In PIR-PCSI the server is not allowed to learn anything about whether or not $\bm{\theta}\in\bm{\mathcal{S}}$.  The supremum capacity of PIR-PCSI is found to be $(K-M+1)^{-1}$ for $2 \leq M \leq K$. Remarkably, this is not just the smaller  of the two capacities of PIR-PCSI-I and PIR-PCSI-II, so there is an additional cost to be paid for hiding from the server whether $\bm{\theta} \in \bm{\mathcal{S}}$ or $\bm{\theta} \notin \bm{\mathcal{S}}$. Depending on the relative values of $M$ and $K$, in this case we find that the redundancy in CSI can be as high as $1/2$ or as low as $0$.  The infimum capacity of PIR-PCSI is smaller than the supremum capacity, the binary field $\mathbb{F}_2$ yields the worst case, and as in PIR-PCSI-II, the infimum capacity with private $(\bm{\theta},\bm{\mathcal{S}})$ is the same as the (supremum or infimum) capacity with private $(\bm{\theta},\bm{\mathcal{S}},\bm{\Lambda})$.

This paper is organized as follows: Section \ref{sec:state} states PIR-PCSI, PIR-PCSI-I, PIR-PCSI-II problems in \cite{PIR_PCSI}. Section \ref{sec:main} states our capacity and redundancy (in the CSI) results for PIR-PCSI-II, PIR-PCSI-I, PIR-PCSI with fourteen theorems which are proved in Section \ref{sec:cap_PCSI2_sup} to Section \ref{proof:pcsi_pub_pri}. Section \ref{sec:con} concludes this paper and gives possible future directions.

\emph{Notation}: For a positive integer $a$, let $[a]$ denote the set $\{1,2,\cdots,a\}$. For two integers $a, b$ where $a < b$, $[a:b]$ denotes the set $\{a, a+1, \cdots, b\}$.  For a set $\mathcal{S} = \{i_1, i_2, \cdots, i_n\}$, $|\mathcal{S}|$ denotes the cardinality of $\mathcal{S}$. $\mathbf{I}_{M}$ denotes the $M \times M$ identity matrix, and $\mathbf{0}_{M}$ denotes the $M \times M$ all-zero matrix. For a matrix $\mathbf{A}$, let $\mathbf{A}(i,:)$ be the $i^{th}$ row of $\mathbf{A}$. For a set $\mathcal{A}$ whose elements are integers, let $\mathcal{A}(i)$ denote the $i^{th}$ element of $\mathcal{A}$ in ascending order. Let $\mathbb{F}_{q}$ denote the finite field of order $q$ and $\mathbb{F}_{q}^{\times}$ contain all the non-zero elements of $\mathbb{F}_{q}$. The notation $\mathbb{F}_q^{a\times b}$ represents the set of all $a\times b$ matrices with elements in $\mathbb{F}_q$. The notation $\mathbb{F}_q^{a\times 1}$ may be shortened to $\mathbb{F}_q^a$. Let $\mathfrak{S}$ be the set of all the subsets with cardinality $M$ of $[K]$, i.e., $|\mathfrak{S}| = \tbinom{K}{M}$, and let $\mathfrak{C}$ be the set of all length $M$ sequences with elements in $\mathbb{F}_{q}^{\times}$, i.e., $|\mathfrak{C}| = (q-1)^M$. For an index set $S\subset[K]$, define the subscript notation $X_S=\{X_s\mid s\in S\}$. All entropies are in $q$-ary units. For a random variable $\bm{A}$, ${\mathbb{E}}[\bm{A}]$ is the expectation of $\bm{A}$, $\pr(\bm{A} = A)$ denotes the probability of $\bm{A}$ being $A$.

\section{Problem Statement}\label{sec:state}
\subsection{Capacity of PIR-PCSI-I, PIR-PCSI-II, PIR-PCSI}
A single server stores $K$ independent messages $\bm{W}_1, \bm{W}_2, \cdots, \bm{W}_{K}\in\mathbb{F}_q^L$, each comprised of $L$ i.i.d. uniform symbols  from $\mathbb{F}_{q}$, i.e., $\bm{W}_k=(\bm{W}_k(1),\bm{W}_k(2),\cdots,\bm{W}_k(L))^T$, and each $\bm{W}_k(\ell)$, which denotes the $\ell^{th}$ instance of the $k^{th}$ message, is drawn i.i.d. uniform from $\mathbb{F}_q$. The number of instances $L$ may be chosen freely by the coding scheme. We refer to $\mathbb{F}_{q}$ as the \emph{base field}. In terms of entropies,
\begin{align}
    &H(\bm{W}_{1}) = H(\bm{W}_{2}) = \cdots = H(\bm{W}_{K}) = L,\\
    &H(\bm{W}_{[K]}) = \sum_{k \in [K]}H(\bm{W}_{k}) = KL.
\end{align}

 A user  wishes to retrieve a message $\bm{W}_{\bm{\theta}}$ for a privately generated index $\bm{\theta}$. The user has a linear combination of $M$ messages available as coded side information (CSI). $M$ is globally known. The CSI is comprised of $(\bm{\mathcal{S}}, \bm{\Lambda},  \bm{Y}^{[\bm{\mathcal{S}},\bm{\Lambda}]})$,  defined as follows. The \emph{support index set} $\bm{\mathcal{S}}$, drawn uniformly from $\mathfrak{S}$, is a subset of $[K]$, of cardinality $M$. The vector of coefficients $\bm{\Lambda}=(\bm{\lambda}_1,\bm{\lambda}_2,\cdots,\bm{\lambda}_M)$ is drawn uniformly from $\mathfrak{C}$, and  applied across all $L$ instances, i.e., the same linear combining coefficients appear in each of the $L$ instances of the CSI.
The linear combination available to the user is
\begin{align}
    \bm{Y}^{[\bm{\mathcal{S}},\bm{\Lambda}]}\triangleq \bm{\lambda}_1\bm{W}_{\bm{\mathcal{S}}(1)} + \bm{\lambda}_2\bm{W}_{\bm{\mathcal{S}}(2)} + \cdots + \bm{\lambda}_M\bm{W}_{\bm{\mathcal{S}}(M)},\label{eq:sideinfo_CSI}
\end{align}
where we recall the notation that $\bm{\mathcal{S}}(m)$ denotes the $m^{th}$ element of $\bm{\mathcal{S}}$, in ascending order, i.e., $\bm{\mathcal{S}}(1)<\bm{\mathcal{S}}(2)<\cdots<\bm{\mathcal{S}}(M)$. 
We assume that $(\bm{\theta}, \bm{\mathcal{S}})$, $\bm{\Lambda}$, $\bm{W}_{[K]}$ are  independent.
\begin{align}
    H(\bm{\theta}, \bm{\mathcal{S}}, \bm{\Lambda}, \bm{W}_{[K]}) = H(\bm{\theta}, \bm{\mathcal{S}}) + H(\bm{\Lambda}) + H(\bm{W}_{[K]}).
\end{align}

There are three formulations of the problem depending on how $\bm{\theta}$ is chosen by the user.
\begin{enumerate}
\item{\bf PIR-PCSI-I}: $\bm{\theta}$ is chosen uniformly from $[K]\setminus\bm{\mathcal{S}}$.
\item{\bf PIR-PCSI-II}: $\bm{\theta}$ is chosen uniformly from $\bm{\mathcal{S}}$.
\item{\bf PIR-PCSI}: $\bm{\theta}$ is chosen uniformly from $[K]$.
\end{enumerate}
When referring to all three formulations, we will  refer to the problem as {\bf PIR-PCSI*} for brevity. In such statements, PCSI* can be replaced with PCSI-I, PCSI-II, or PCSI  to obtain corresponding statements for each of the three formulations.

The server knows the distributions but not the realizations of $\bm\theta, \bm{\mathcal{S}}, \bm{\Lambda},  \bm{Y}^{[\bm{\mathcal{S}},\bm{\Lambda}]}$.
It is required that $(\bm{\theta},\bm{\mathcal{S}})$ be kept jointly private from the server. Note that the privacy of $\bm{Y}^{[\bm{\mathcal{S}},\bm{\Lambda}]}$ or the coefficient vector $\bm{\Lambda}$ is not required. While the server initially knows nothing about the realization of $\bm{\Lambda}$, a PIR-PCSI* scheme may reveal some information about the coefficients, especially if it allows for efficient retrieval without leaking any information about $(\bm{\theta},\bm{\mathcal{S}})$. Leaking information about $\bm{\Lambda}$ has  implications for reusability of side-information, an issue that is explored recently in  \cite{Anoosheh_reusable}.

In order to retrieve $\bm{W_\theta}$, we assume as in \cite{PIR_PCSI} that the user generates a random query $\bm{Q}$ that is independent of the messages. Specifically,
\begin{align}
    I(\bm{W}_{[K]}; \bm{Q}, \bm{\theta}, \bm{\mathcal{S}}, \bm{\Lambda}) = 0.\label{eq:indQ}
\end{align}
Let $\mathcal{Q}$ denote the alphabet of $\bm{Q}$.

Because the messages are i.i.d. uniform, and the coefficients are non-zero, according to the construction of $\bm{Y}^{[\bm{\mathcal{S}}, \bm{\Lambda}]}$, it follows that
\begin{align}
L&=H(\bm{Y}^{[\bm{\mathcal{S}}, \bm{\Lambda}]}),\\
& = H(\bm{Y}^{[\bm{\mathcal{S}}, \bm{\Lambda}]} \mid \bm{Q}, \bm{\mathcal{S}}, \bm{\Lambda},  \bm{W}_{{[K]}\setminus\{\bm{\mathcal{S}}(m)\}}), \forall m \in [M].\label{eq:indY} 
\end{align}

The user uploads $\bm{Q}$ to the server. Mathematically, the privacy constraint is expressed as,
\begin{align}
    &\text{[$(\bm{\theta}, \bm{\mathcal{S}})$ Privacy]} &&I\left(\bm{\theta}, \bm{\mathcal{S}}; \bm{Q}, \bm{W}_{[K]}\right) = 0.\label{eq:tsprivacy}
\end{align}
The server returns an answer $\bm{\Delta}$ as a function of $\bm{Q}$ and the messages, i.e.,
\begin{align}
    H\left(\bm{\Delta} \mid \bm{Q}, \bm{W}_{[K]}\right) = 0.
\end{align}
The answer $\bm{\Delta}$ takes values in an alphabet set $\mathcal{A}_{\bm{Q}}$ that depends on the query $\bm{Q}$. The download cost, measured in $q$-ary symbols is $\log_q|\mathcal{A}_{\bm{Q}}|$. Since $\mathcal{A}_{\bm{Q}}$ is a function of ${\bm{Q}}$, note that different queries may result in different download costs.

Upon receiving the answer, the user must be able to decode the desired message $\bm{W}_{\bm\theta}$. 
\begin{align}
    &\text{[Correctness]} &&H(\bm{W}_{\bm{\theta}} \mid \bm{\Delta}, \bm{Q}, \bm{Y}^{[\bm{\mathcal{S}}, \bm{\Lambda}]},  \bm{\mathcal{S}}, \bm{\Lambda},\bm{\theta}) = 0.
\end{align}
We are interested in the \emph{average} download cost, $D$, across all queries, which is defined and bounded as follows.
\begin{align}
    D &\triangleq {\mathbb{E}}_{\bm{Q}}\big[\log_q|\mathcal{A}_{\bm{Q}}|\big]\\
    &= \sum_{Q \in \mathcal{Q}}\pr(\bm{Q} = Q)\log_q|\mathcal{A}_{{Q}}|\\
    &\geq \sum_{Q \in \mathcal{Q}}\pr(\bm{Q} = Q)H(\bm{\Delta} \mid \bm{Q} = Q)\label{eq:DgeqH}\\
    &= H(\bm{\Delta} \mid \bm{Q}).
\end{align}
In \eqref{eq:DgeqH} we used the fact that the entropy of a random variable is no more than the logarithm of the cardinality of its alphabet, corresponding to the fact that the uniform distribution maximizes entropy. This bound will be useful for converse proofs.

The rate achieved by a PIR scheme is defined as,
\begin{align}
    R \triangleq \frac{L}{D}
\end{align}
The capacity is the supremum of achievable rates over all message sizes $L$,
\begin{align}
    C_{\mbox{\tiny PCSI*}}(q) = \sup_{L, \mbox{\tiny achievable $R$}}R.
\end{align}
The capacity can depend on the field $\mathbb{F}_q$ which affects the nature of side information. Field-independent measures of capacity may be obtained by taking a supremum (as in \cite{PIR_PCSI}) or infimum over all finite fields. These are called supremum and infimum capacity, respectively.
\begin{align}
    C_{\mbox{\tiny PCSI*}}^{\sup} &= \sup_{q}C_{\mbox{\tiny PCSI*}}(q),\\
    C_{\mbox{\tiny PCSI*}}^{\inf} &= \inf_{q}C_{\mbox{\tiny PCSI*}}(q).
\end{align}

\begin{remark}
Throughout this paper, we will use the notation $\mathbb{F}_q$ (and accordingly the symbol $q$) only to represent the field in which the message symbols, and in particular the linear combinations that constitute the CSI lie. The encoding operations may occasionally take place in a different field, typically a sub-field (e.g., $\mathbb{F}_{\sqrt{q}}$ if it exists) or an extension field (e.g., $\mathbb{F}_{q^l}$) of $\mathbb{F}_q$, which will be identified as such.
\end{remark}

\subsection{Capacity of PIR-PCSI* with Private Coefficients}
Recall that in the formulation of PIR-PCSI* as presented above, while $(\bm{\theta},\bm{\mathcal{S}})$ must be kept private, the privacy of the coefficient vector $\bm{\Lambda}$ is not required. As an important benchmark, we  consider the setting where the privacy of coefficients must also be preserved. In this setting, the privacy constraint is modified so that instead of \eqref{eq:tsprivacy} we require the following.
\begin{align}
    &\text{[$(\bm{\theta}, \bm{\mathcal{S}}, \bm{\Lambda})$ Privacy]} &&I\left(\bm{\theta}, \bm{\mathcal{S}}, \bm{\Lambda}; \bm{Q}, \bm{W}_{[K]}\right) = 0.\label{eq:tscprivacy}
\end{align}
The capacity under this privacy constraint is referred to as the capacity with private coefficients and is denoted as $C_{\mbox{\tiny PCSI*}}^{\mbox{\tiny pri}}(q)$, which is potentially a function of the field size $q$. The supremum and infimum (over  $q$) of $C_{\mbox{\tiny PCSI*}}^{\mbox{\tiny pri}}(q)$ are denoted as $C_{\mbox{\tiny PCSI*}}^{\mbox{\tiny pri},\sup}, C_{\mbox{\tiny PCSI*}}^{\mbox{\tiny pri},\inf}$, respectively.

\subsection{Redundancy of CSI}
In addition to the capacity of PIR-PCSI*, we also wish to determine how much (if any) of the side information  is redundant, i.e., can be discarded without any loss in the \emph{supremum capacity}. 

For all $\mathcal{S}\in\mathfrak{S}, \Lambda\in\mathfrak{C}$, let $f_{\mathcal{S},\Lambda}: \mathbb{F}_{q}^{L} \rightarrow \overline{\mathcal{Y}}$ be  arbitrary functions that take the CSI $\bm{Y}^{[{\mathcal{S}}, {\Lambda}]}$ as input and output some   $\overline{\bm{Y}}^{[{\mathcal{S}}, {\Lambda}]}\in\overline{\mathcal{Y}}$. These functions could be used to discard some parts of the side-information, and retain other parts, e.g., to reduce storage cost.
\begin{align}
\overline{\bm{Y}}^{[{\mathcal{S}}, {\Lambda}]} = f_{\mathcal{S}, \Lambda}(\bm{Y}^{[{\mathcal{S}}, {\Lambda}]}).
\end{align}
Let us refer to all these functions collectively as $\mathcal{F}=(f_{\mathcal{S}, \Lambda})_{\mathcal{S}\in\mathfrak{S}, \Lambda\in\mathfrak{C}}$. 
Define, $\overline{C}_{\mbox{\tiny PCSI*}}(q,\mathcal{F})$ as the capacity (supremum of achievable rates) if the decoding must be based on $\overline{\bm{Y}}^{[{\mathcal{S}}, {\Lambda}]}$ instead of $\bm{Y}^{[{\mathcal{S}}, {\Lambda}]}$, i.e., the correctness condition is modified to
\begin{align}
H(\bm{W}_{\bm{\theta}} \mid \bm{\Delta}, \bm{Q}, \overline{\bm{Y}}^{[\bm{\mathcal{S}}, \bm{\Lambda}]}, \bm{\mathcal{S}}, \bm{\Lambda},\bm{\theta}) = 0.
\end{align}
We say that $\mathcal{F}$ uses $\alpha$-CSI, where
\begin{align}
\alpha=\max_{\mathcal{S}\in\mathfrak{S}, \Lambda\in\mathfrak{C}} H(\overline{\bm{Y}}^{[{\mathcal{S}}, {\Lambda}]})/L
\end{align}
Whereas storing $\bm{Y}^{[{\mathcal{S}}, {\Lambda}]}$ requires $L$ $q$-ary symbols, note that storing $\overline{\bm{Y}}^{[{\mathcal{S}}, {\Lambda}]}$ requires only $\alpha L$ storage, i.e., storage is reduced by a factor $\alpha$. Define the $\alpha$-CSI constrained capacity as
\begin{align}
\overline{C}_{\mbox{\tiny PCSI*}}(q,\alpha)&=\sup_{
\mathcal{F}: ~\mbox{\footnotesize uses no more than $\alpha$-CSI}} \overline{C}_{\mbox{\tiny PCSI*}}(q,\mathcal{F})
\end{align}
In other words, $\overline{C}_{\mbox{\tiny PCSI*}}(q,\alpha)$ is the capacity when the user is allowed to retain no more than a fraction $\alpha$ of the CSI $\bm{Y}^{[{\mathcal{S}}, {\Lambda}]}$.
The notion of $\alpha$-CSI constrained capacity is of broader interest on its own. However, in this work we will explore only the redundancy of CSI with regard to the supremum capacity. We say that `$\alpha$-CSI is sufficient' if 
\begin{align}
\sup_q\overline{C}_{\mbox{\tiny PCSI*}}(q,\alpha)&={C}_{\mbox{\tiny PCSI*}}^{\sup}
\end{align}
Define $\alpha^*$ as the smallest value of $\alpha$ such that $\alpha$-CSI is sufficient.
The redundancy of PCSI is defined as $\rho_{\mbox{\tiny PCSI*}}=1-\alpha^*$.
Note that the opposite extremes of $\rho_{\mbox{\tiny PCSI*}}=1$ and $\rho_{\mbox{\tiny PCSI*}}=0$ correspond to situations where all of the side information is redundant, and where none of the side information is redundant, respectively.

For later use, it is worthwhile to note that for any scheme that uses no more than $\alpha$-CSI,  because $\overline{\bm{Y}}^{[{\mathcal{S}}, {\Lambda}]}$ is a function of ${\bm{Y}}^{[{\mathcal{S}}, {\Lambda}]}$, it follows from \eqref{eq:indY} that for all\footnote{We say $(Q,\mathcal{S},\Lambda)$ is feasible if $\Pr((\bm{Q}, \bm{\mathcal{S}}, \bm{\Lambda}) = (Q,\mathcal{S},\Lambda))>0$.} feasible $(Q,\mathcal{S},\Lambda)$,
{\small
\begin{align}
H\bigg(\overline{\bm{Y}}^{[{\mathcal{S}}, {\Lambda}]} \mid (\bm{Q},\bm{\mathcal{S}},\bm{\Lambda})=(Q,\mathcal{S},\Lambda)\bigg)=H(\overline{\bm{Y}}^{[{\mathcal{S}}, {\Lambda}]} )\leq\alpha L.\label{eq:invaYR}
\end{align}
}
This is because of the property that if $A$ is independent of $B$, then any function of $A$ is also independent of $B$. In this case,  \eqref{eq:indY} tells us that ${\bm{Y}}^{[{\mathcal{S}}, {\Lambda}]}$ is independent of ${\bf Q}$, therefore so is $\overline{{\bm{Y}}}^{[{\mathcal{S}}, {\Lambda}]}$.

\section{Main Results}\label{sec:main}
\begin{table*}[!t]
    \caption{Redundancy results for PIR-PCSI-I, PIR-PCSI-II and PIR-PCSI}
    \label{tab:redundancy}
    \centering
    \scalebox{0.93}{
    \begin{tabular}{|c|c|c|}
    \hline
    PIR-PCSI-I ($1 \leq M \leq K-1$) & PIR-PCSI-II ($2 \leq M \leq K$) & PIR-PCSI ($1 \leq M \leq K$)\\ \hline
    $ \rho_{\mbox{\tiny PCSI-I}}= 0,$ Thm. \ref{thm:redundancy1}
    &$\rho_{\mbox{\tiny PCSI-II}} =
        \begin{cases}
            1/2, & 1< M\leq (K+2)/2,\\
            0, & (K+2)/2<M\leq K,
        \end{cases}$, Thm. \ref{thm:red} 
    & \begin{tabular}{c}$\rho_{\mbox{\tiny PCSI}} = \frac{1}{2},  M = 2,$\\
        $\rho_{\mbox{\tiny PCSI}} \leq \frac{1}{M}, 3 \leq M \leq \frac{K+2}{2},$\\
        $\rho_{\mbox{\tiny PCSI}} = 0, \text{otherwise},$\end{tabular}, Thm. \ref{thm:redundancy}\\\hline
    \end{tabular}}
\end{table*}
Our main results are presented as theorems in this section, and  summarized in Table \ref{tab:capacity} and Table \ref{tab:redundancy} for quick reference. We start with  PIR-PCSI-II (where $\bm{\theta}\in\bm{\mathcal{S}}$), which is the main motivation for this work. Note that the case $M=1$ is trivial, because in that case the user already has the desired message. Therefore, for PIR-PCSI-II we will always assume that $M>1$.
\subsection{PIR-PCSI-II (where $\bm{\theta}$ is drawn uniformly from $\bm{\mathcal{S}}$)}
\begin{theorem}\label{thm:cap_PCSI2_sup}
    The supremum capacity of PIR-PCSI-II  is
    \begin{align}
        C_{\mbox{\tiny PCSI-II}}^{\sup} &=\max\left(\frac{2}{K},\frac{1}{K-M+1}\right)\\
        &=
        \begin{cases}
            \frac{2}{K}, & 1 < M \leq \frac{K+1}{2},\\	
            \frac{1}{K-M+1}, & \frac{K+1}{2} < M \leq K,\text{\cite{PIR_PCSI}}
        \end{cases}
    \end{align}
\end{theorem}

The case $(K+1)/2<M\leq K$ was already  settled by Heidarzadeh et al. in \cite{PIR_PCSI}, and is included in Theorem \ref{thm:cap_PCSI2_sup} primarily for the sake of completeness. Our contribution to Theorem \ref{thm:cap_PCSI2_sup} is for the case $1< M\leq (K+1)/2$ which was noted as an open problem in \cite{PIR_PCSI} along with a conjecture that the supremum capacity for this case may also be equal to $1/(K-M+1)$. Theorem \ref{thm:cap_PCSI2_sup} settles this open problem and resolves the conjecture by establishing that the supremum capacity in this case is $2/K$. The proof of Theorem \ref{thm:cap_PCSI2_sup} for the case $2\leq M\leq (K+1)/2$ appears in Section \ref{sec:cap_PCSI2_sup}.

Note that when $2\leq M\leq (K+1)/2$, the supremum capacity value $2/K$  is strictly higher than the conjectured value $1/(K-M+1)$, and does not depend on the support size $M$ of the coded side information. Achievability of $2/K$ is shown in Section \ref{sec:cap_PCSI2_sup} for any field $\mathbb{F}_q$ where $q$ is an even power of a prime, by viewing $\mathbb{F}_q$ as a $2$ dimensional vector space over $\mathbb{F}_{\sqrt{q}}$. Note that $q$ needs to be an even power of a prime, in order for $\mathbb{F}_{\sqrt{q}}$ to be a valid finite field. Specifically, we choose $L=1$, so each message is comprised of $1$ symbol from $\mathbb{F}_q$, equivalently $2$ symbols from $\mathbb{F}_{\sqrt{q}}$, which can be represented as  a $2\times 1$ vector over $\mathbb{F}_{\sqrt{q}}$, while the  coefficients $\bm{\lambda}_m\in\mathbb{F}_q, m \in [M]$  take the role of $2\times 2$ matrices in $\mathbb{F}_{\sqrt{q}}$ that \emph{rotate} the vectors corresponding to the messages $\bm{W}_{\bm{\mathcal{S}}_m}$ involved in the CSI, thus randomizing their relative alignments.\footnote{As an alternative, suppose instead we consider each message as comprised of $L=2$ symbols from $\mathbb{F}_q$, which also allows us to work with a $2$ dimensional vector space (over $\mathbb{F}_q$). However, since the coefficients are scalars in $\mathbb{F}_q$ and constant across $\ell\in[L]$, in this $2$ dimensional vector space the coefficients translate to only scaled versions of $2\times 2$ identity matrices, which does not yield the  \emph{rotations} that are essential for privacy.} Half of the desired message $\bm{W}_{\bm{\theta}}$ is recovered by downloading  the corresponding halves of undesired messages that align (interfere) with that half of $\bm{W}_{\bm{\theta}}$ (so that the interference can be subtracted),  while the other half of $\bm{W}_{\bm{\theta}}$ is downloaded directly. The private rotations due to $\bm{\Lambda}$ in the CSI hide the alignments from the server. For the messages that are not involved in the CSI any random half can be downloaded. Since a random half of every message is downloaded, no information is leaked about $(\bm{\theta},\bm{\mathcal{S}})$ and the scheme is private.

Intuitively, since one half of the desired message is directly downloaded, it stands to reason that the corresponding half of the CSI may be redundant and could be discarded by the user, thus saving storage cost. Indeed, this intuition turns out to be correct, as encapsulated in the next theorem which characterizes precisely how much of the side information in each parameter regime is redundant, i.e., can be discarded without any loss in the supremum capacity specified in Theorem \ref{thm:cap_PCSI2_sup}.

\begin{theorem}\label{thm:red}
For the supremum capacity of PIR-PCSI-II, the redundancy in coded side information is characterized as,
\begin{align}
\rho_{\mbox{\tiny PCSI-II}}&=
\begin{cases}
          1/2, & 1< M\leq (K+2)/2,\\
           0, & (K+2)/2<M\leq K.
        \end{cases}
\end{align}
\end{theorem}
In particular, $\rho_{\mbox{\tiny PCSI-II}}=1/2$ implies that exactly half of the side information is redundant, and $\rho_{\mbox{\tiny PCSI-II}}=0$ implies that there is no redundancy in the side information. The proof of Theorem \ref{thm:red} appears in Section \ref{proof:red}. Thus, for all $(M,K)$ parameters where the supremum capacity is equal to $2/K$, half of the coded side information is redundant. Note that in the boundary case where $M=(K+2)/2$, we have $2/K=1/(K-M+1)$, i.e., this boundary case could be included in either of the two cases in Theorem \ref{thm:cap_PCSI2_sup}. Remarkably, these are the only cases where we have any redundancy in coded side information. According to Theorem \ref{thm:red}, there is no redundancy when $(K+2)/2<M\leq K$. 

As our next result for PIR-PCSI-II, we characterize the infimum capacity $C_{\mbox{\tiny PCSI-II}}^{\inf}$ in the following theorem.

\begin{theorem}\label{thm:cap_PCSI2_inf}
    The infimum capacity of PIR-PCSI-II,
    \begin{align}
        C_{\mbox{\tiny PCSI-II}}^{\inf} = C_{\mbox{\tiny PCSI-II}}(q=2) = \frac{M}{(M-1)K}.
    \end{align}
\end{theorem}
The proof of Theorem \ref{thm:cap_PCSI2_inf} appears in Section \ref{sec:cap_PCSI2_inf}. Evidently, the infimum capacity of PIR-PCSI-II matches its capacity over the binary field. Intuitively, one might expect that the binary field would represent the worst case because over $\mathbb{F}_2$, the coefficients $\bm{\lambda}_m$, which must be non-zero, can only take the value $1$. Thus, the coefficients are known to the server. It is also worth noting that constant $\bm{\Lambda}$ trivially satisfy $(\bm{\theta}, \bm{\mathcal{S}}, \bm{\Lambda})$ privacy whenever $(\bm{\theta}, \bm{\mathcal{S}})$ privacy is satisfied.

Note that for $M=2$, the infimum capacity of PIR-PCSI-II matches the supremum capacity, therefore for any field $\mathbb{F}_q$, we have the exact capacity characterization, $C_{\mbox{\tiny PCSI-II}}(q)=C_{\mbox{\tiny PCSI-II}}^{\inf}=C_{\mbox{\tiny PCSI-II}}^{\sup}$. However, in general the infimum capacity is strictly smaller. The gap can be significant, for example when $M=K$  the supremum capacity is $1$ while the infimum capacity is $1/(K-1)$. In general the capacity for arbitrary fields, arbitrary support size $M$ and arbitrary number of messages $K$ remains open. Intuitively, we expect that the capacity for most fields should be either equal to or close to the supremum capacity, whereas fields where the capacity is closer to the infimum capacity should be relatively rare. For certain $M,K$ values, however, we are able to characterize the capacity of PIR-PCSI-II for arbitrary fields. These results are presented in the next two theorems. Notably, for these specific $M,K$, while the binary field $\mathbb{F}_2$ yields the infimum capacity, for all other fields ($\mathbb{F}_q$, $q>2$), the capacity  matches the supremum capacity, i.e., $C_{\mbox{\tiny PCSI-II}}(q)=C_{\mbox{\tiny PCSI-II}}^{\sup}$.

\begin{theorem}\label{thm:MK}
For PIR-PCSI-II with $M=K$,
\begin{align}
C_{\mbox{\tiny PCSI-II}}(q)&=\begin{cases}
1/(K-1)=C_{\mbox{\tiny PCSI-II}}^{\inf}, &q=2,\\
1=C_{\mbox{\tiny PCSI-II}}^{\sup},&q\neq 2.
\end{cases}
\end{align}
\end{theorem}
The proof of Theorem \ref{thm:MK} appears in Section \ref{proof:MK}.

\begin{theorem}\label{thm:M3K4}
For PIR-PCSI-II with $M=3, K=4$,
\begin{align}
C_{\mbox{\tiny PCSI-II}}(q)&=\begin{cases}
3/8=C_{\mbox{\tiny PCSI-II}}^{\inf}, &q=2,\\
1/2=C_{\mbox{\tiny PCSI-II}}^{\sup},&q\neq 2.
\end{cases}
\end{align}
\end{theorem}
Note that $M=3,K=4$ is a boundary case for which $1/(K-M+1)=2/K$, therefore the supremum capacity is achievable by both the Modified Specialized GRS Codes scheme presented in \cite{PIR_PCSI} and by the interference alignment scheme that appears in the proof of Theorem \ref{thm:cap_PCSI2_sup}. However, the former requires field size $q\geq K=4$, and the latter requires $q$ to be an even power of a prime. Aside from $q=2$ which corresponds to the infimum capacity, this leaves only $q=3$, which is neither greater than or equal to $4$ nor an even power of a prime, as the only new result in Theorem \ref{thm:M3K4}. 
The proof for $q=3$ appears in Section \ref{proof:M3K4}. 

Building on the observation that the infimum capacity corresponds to the binary field where the coefficients are essentially constants such that the $(\bm{\theta}, \bm{\mathcal{S}}, \bm{\Lambda})$ privacy is automatically satisfied, we next explore the capacity of PIR-PCSI-II for the case of private coefficients. The result appears as the next theorem.
\begin{theorem}\label{thm:pcsi2_pub_pri}
    The capacity of PIR-PCSI-II, for the setting with private coefficients, is given by
    \begin{align}
        C_{\mbox{\tiny PCSI-II}}^{\mbox{\tiny pri}}(q) = C_{\mbox{\tiny PCSI-II}}^{\mbox{\tiny pri}, \inf} = C_{\mbox{\tiny PCSI-II}}^{\mbox{\tiny pri}, \sup} = C_{\mbox{\tiny PCSI-II}}^{\inf}.
    \end{align}
\end{theorem}
The proof of Theorem \ref{thm:pcsi2_pub_pri} appears in Section \ref{proof:pcsi2_pub_pri}. Note that the capacity with private coefficients does not depend on the field (infimum and supremum are the same). Compared with the case where the coefficients are not required to be kept private, i.e., the case where only $(\bm{\theta}, \bm{\mathcal{S}})$ privacy is required, there is a loss of the supremum capacity, which represents the cost of also keeping the coefficients private.


\subsection{PIR-PCSI-I (where $\bm{\theta}$ is drawn uniformly from $[K]\setminus \bm{\mathcal{S}}$)}
In this section we consider the setting of PIR-PCSI-I (where $\bm{\theta} \in [K]\setminus \bm{\mathcal{S}}$). Note that the case $M=K$ is not valid, because in that case the desired message is also contained in the support set. Therefore, for PIR-PCSI-I we will always restrict $1 \leq M \leq K-1$.

The supremum capacity of PIR-PCSI-I is already found in \cite{PIR_PCSI} as $C_{\mbox{\tiny PCSI}}^{\sup} = (K-M)^{-1}$ and is achievable by Specialized GRS Codes. We start by characterizing the redundancy in the side information  in the following theorem, whose proof appears  in Section \ref{proof:redundancy1}.

\begin{theorem}\label{thm:redundancy1}
    For the supremum capacity of PIR-PCSI-I, there is no redundancy in coded side information i.e., $ \rho_{\mbox{\tiny PCSI-I}}= 0$.
\end{theorem}

Next we characterize the infimum capacity of PIR-PCSI-I.
\begin{theorem}\label{thm:cap_PCSI1_inf}
    The infimum capacity of PIR-PCSI-I,
    \begin{align}
        C_{\mbox{\tiny PCSI-I}}^{\inf} &= C_{\mbox{\tiny PCSI-I}}(q=2) \notag\\
        &= \max\left(\frac{1}{K-1}, \left(K-\frac{M}{K-M}\right)^{-1}\right)\notag\\
        &= 
        \begin{cases}
            \frac{1}{K-1}, & 1 \leq M \leq \frac{K}{2},\\
            \left(K - \frac{M}{K-M}\right)^{-1}, & \frac{K}{2} < M \leq K-1.
        \end{cases}
    \end{align}
\end{theorem}
The proof of Theorem \ref{thm:cap_PCSI1_inf} appears in Section \ref{sec:cap_PCSI1_inf}. The infimum capacity of PIR-PCSI-I also matches its capacity over binary field. The intuition why $\mathbb{F}_{2}$ represents the worst case, is similar to the PIR-PCSI-II setting.

\begin{remark}\label{rmk:pcsi1_arb_field}
    Note that for $M=K-1$, the infimum capacity of PIR-PCSI-I matches the supremum capacity, therefore for any field $\mathbb{F}_{q}$, we have the exact capacity characterization, $C_{\mbox{\tiny PCSI-I}}(q) = C_{\mbox{\tiny PCSI-I}}^{\inf} = C_{\mbox{\tiny PCSI-I}}^{\sup} = 1$. However, in general the infimum capacity is strictly smaller and the gap can be significant. For example, when $M=K/2$ the supremum capacity is $2/K$ while the infimum capacity is $1/(K-1)$, i.e., for large $K$, the infimum capacity is nearly half of the supremum capacity.
\end{remark}

We next explore the capacity of PIR-PCSI-I for the case of private coefficients. 
\begin{theorem}\label{thm:pcsi1_pub_pri}
    The supremum capacity of PIR-PCSI-I, for the setting with private coefficients, is given by
    \begin{align}
        C_{\mbox{\tiny PCSI-I}}^{\mbox{\tiny pri}, \sup} = C_{\mbox{\tiny PCSI-I}}^{\inf},
    \end{align}
    while the infimum capacity of the private coefficients setting can be bounded as
    \begin{align}
        \frac{1}{K-1} \leq C_{\mbox{\tiny PCSI-I}}^{\mbox{\tiny pri}, \inf} \leq \min\bigg(C_{\mbox{\tiny PCSI-I}}^{\inf}, \frac{1}{K-2}\bigg).
    \end{align}
\end{theorem}
The proof of Theorem \ref{thm:pcsi1_pub_pri} appears in Section \ref{proof:pcsi1_pub_pri}. Unlike PIR-PCSI-II, for PIR-PCSI-I the capacity with private coefficients may depend on the field, and may be strictly smaller than the infimum capacity. For example, if $M=K-1$, then the infimum capacity is $1$, but the infimum capacity with private coefficients is no more than $1/(K-2)$. Remarkably, infimum capacity with private coefficients does not correspond to the binary field $\mathbb{F}_2$, i.e., there exist other fields that yield strictly lower capacities than $\mathbb{F}_2$ for PIR-PCSI-I when the coefficients are fully private.

\subsection{PIR-PCSI (where $\bm{\theta}$ is drawn uniformly  from $[K]$)}
To complete the picture, in this section we characterize the capacity of PIR-PCSI which was not studied in \cite{PIR_PCSI}. Since $\bm{\theta} \in [K]$, any $1\leq M \leq K$ is valid. We start with the supremum capacity.

\begin{theorem}\label{thm:cap_PCSI_sup}
    The supremum capacity of PIR-PCSI is
    \begin{align}
        C_{\mbox{\tiny PCSI}}^{\sup} &= \max\bigg(\frac{1}{K-1}, \frac{1}{K-M+1}\bigg)\notag\\
        &=\begin{cases}
            \frac{1}{K-1}, & M=1,\\
            \frac{1}{K-M+1}, & 2 \leq M \leq K,
        \end{cases}
    \end{align}
\end{theorem}
The proof of Theorem \ref{thm:cap_PCSI_sup} appears in Section \ref{sec:cap_PCSI_sup}.  For $M=1$, this problem is dominated by the PIR-PCSI-I setting, and the capacity is $(K-1)^{-1}$.

The redundancy of  CSI to achieve the supremum capacity of PIR-PCSI is bounded in the following theorem.
\begin{theorem}\label{thm:redundancy}
    The redundancy of the CSI to achieve the supremum capacity of PIR-PCSI is bounded as
    \begin{align}
        \rho_{\mbox{\tiny PCSI}} &= \frac{1}{2}, && M = 2\notag\\
        \rho_{\mbox{\tiny PCSI}} &\leq \frac{1}{M}, && 3 \leq M \leq \frac{K+2}{2},\notag\\
        \rho_{\mbox{\tiny PCSI}} &= 0, &&\text{otherwise}.
    \end{align}
\end{theorem}
The proof of Theorem \ref{thm:redundancy} appears in Section \ref{proof:redundancy}. Evidently, for different values of $M$ the redundancy can be as high as $1/2$ and as low as $0$.

The infimum capacity of PIR-PCSI is found next.
\begin{theorem}\label{thm:cap_PCSI_inf}
    The infimum capacity of PIR-PCSI corresponds to $q = 2$, and,
    \begin{align}
        C_{\mbox{\tiny PCSI}}^{\inf} = C_{\mbox{\tiny PCSI}}(q=2) = \frac{1}{K-1}.
    \end{align}
\end{theorem}
The proof of Theorem \ref{thm:cap_PCSI_inf} appears in Section \ref{sec:cap_PCSI_inf}. 

Note that for $M=1$, the infimum capacity of PIR-PCSI matches the supremum capacity, therefore for any field $\mathbb{F}_{q}$, we have the exact capacity characterization, $C_{\mbox{\tiny PCSI}}(q) = C_{\mbox{\tiny PCSI}}^{\inf} = C_{\mbox{\tiny PCSI}}^{\sup}$. However, in general the infimum capacity is strictly smaller and the gap can be significant. For example, when $M=K$ the supremum capacity is $1$ while the infimum capacity is $1/(K-1)$.

Finally, the capacity of PIR-PCSI for the case of private coefficients is characterized.
\begin{theorem}\label{thm:pcsi_pub_pri}
    The capacity of PIR-PCSI, for the setting with private coefficients, is given by
    \begin{align}
        C_{\mbox{\tiny PCSI}}^{\mbox{\tiny pri}}(q) = C_{\mbox{\tiny PCSI}}^{\inf}.
    \end{align}
\end{theorem}
The proof of Theorem \ref{thm:pcsi_pub_pri} appears in Section \ref{proof:pcsi_pub_pri}. Similar to PIR-PCSI-II, and unlike PIR-PCSI-I,  for PIR-PCSI the capacity  with private coefficients does not depend on the field, and is always equal to the infimum capacity. 

Let us conclude this section with Table \ref{tab:open_prob} which summarizes the solved and open cases of various PIR-PCSI* problems considered in this work.
\begin{table*}[!t]
    \caption{A summary of  solved and open problems for PIR-PCSI*}
    $~$\\[-0.5cm]
    \label{tab:open_prob}
    \centering
    \scalebox{0.80}{
    \begin{tabular}{|c|c|c|c|}
    \hline
     & PIR-PCSI-I & PIR-PCSI-II & PIR-PCSI\\ \hline
    Supremum Capacity & Solved & Solved & Solved\\\hline
    Infimum Capacity & Solved & Solved &Solved\\\hline
    Field-dependent Capacity & Only solved for $M = K-1$ & Only solved for $M = K$ and $M = 3, K =4$ & Open\\ \hline
    Capacity with private coeff.  & Open for infimum capacity and arbitrary $q$ & Solved & Solved\\ \hline
    Redundancy in CSI & Solved & Solved & Open for $3 \leq M \leq \frac{K+2}{2}$\\\hline
    \end{tabular}}
\end{table*}

\section{Proof of Theorem \ref{thm:cap_PCSI2_sup}}\label{sec:cap_PCSI2_sup}

\subsection{Converse}
The following lemma, which is essentially Lemma 1 of \cite{PIR_PCSI}, states that for PIR-PCSI*, for every feasible $(Q, \mathcal{S}, \theta)$, there must exist at least one coefficient vector that allows successful decoding.

\begin{lemma}\label{lem:privacy}
\begin{align}
&\mbox{PIR-PCSI: } \forall (Q,\mathcal{S},\theta)\in\mathcal{Q}\times \mathfrak{S}\times[K], ~~\exists \Lambda\in\mathfrak{C}, \mbox{ s. t. }\notag\\
&\hspace{0.2cm} H(\bm{W}_{\theta} \mid \bm{\Delta}, \bm{Y}^{[\mathcal{S},\Lambda]}, \bm{Q}=Q) = 0.\label{eq:lemma1pcsi}\\
&\mbox{PIR-PCSI-I: }\forall (Q,\mathcal{S},\theta)\in\mathcal{Q}\times \mathfrak{S}\times([K]\setminus\mathcal{S}),~~\exists \Lambda\in\mathfrak{C}, \mbox{ s. t. }\notag\\
&\hspace{0.2cm} H(\bm{W}_{\theta} \mid \bm{\Delta}, \bm{Y}^{[\mathcal{S},\Lambda]}, \bm{Q}=Q) = 0.\label{eq:lemma1pcsi1}\\
&\mbox{PIR-PCSI-II: }\forall (Q,\mathcal{S},\theta)\in\mathcal{Q}\times \mathfrak{S}\times\mathcal{S},~~\exists \Lambda\in\mathfrak{C}, \mbox{ s. t. }\notag\\
&\hspace{0.2cm} H(\bm{W}_{\theta} \mid \bm{\Delta}, \bm{Y}^{[\mathcal{S},\Lambda]}, \bm{Q}=Q) = 0.\label{eq:lemma1pcsi2}
\end{align}
\end{lemma}

\proof Since the server  knows $\bm{\Delta}, \bm{Q}$ and can test all possible realizations of $\bm{\theta}, \bm{\mathcal{S}}, \bm{\Lambda}$ for decodability, if no  coefficient vector exists for a particular $(\bm{\theta} = \theta,\bm{\mathcal{S}} = \mathcal{S})$ then that $(\theta,\mathcal{S})$ can be ruled out by the server. This contradicts the privacy constraint.$\hfill\square$

Let us prove the converse for $2 \leq M \leq \frac{K+1}{2}$.

Consider any particular realization $Q \in \mathcal{Q}$ of $\bm{Q}$. For all $i\in[M-1]$, consider $\mathcal{S}_i=[i:i+M-1]$ and $\theta=i$, and let $\Lambda_i$ be a coefficient vector that satisfies \eqref{eq:lemma1pcsi2} according to Lemma \ref{lem:privacy}, so that 
\begin{align}
    H(\bm{W}_i \mid \bm{\Delta}, \bm{Y}^{[\mathcal{S}_{i},\Lambda_i]}, \bm{Q}=Q) = 0.\label{eq:decodability1}
\end{align}
Writing $\bm{Y}^{[\mathcal{S}_{i},\Lambda_i]}$ as $\bm{Y}_i$ for compact notation, we have 
\begin{align}
    &H(\bm{W}_{[M-1]}, \bm{Y}_{[M-1]} \mid \bm{\Delta}, \bm{Q}=Q)\\
    &\leq \sum_{i \in [M-1]}H(\bm{W}_i, \bm{Y}_{i} \mid \bm{\Delta}, \bm{Q}=Q)\label{eq:decodability_cr_cr}\\
    &= \sum_{i \in [M-1]}H(\bm{Y}_i \mid \bm{\Delta}, \bm{Q}=Q)\notag\\
    &\quad + \sum_{i \in [M-1]}H(\bm{W}_i \mid \bm{\Delta}, \bm{Y}_i, \bm{Q}=Q)\label{eq:decodability_cr}\\
    &= \sum_{i \in [M-1]}H(\bm{Y}_i \mid \bm{\Delta}, \bm{Q}=Q)\label{eq:decodability2}\\
    &\leq (M-1)L 
    \label{eq:bound_Y}
\end{align}
where \eqref{eq:decodability_cr_cr} results from chain rule and the property that conditioning reduces entropy. Step \eqref{eq:decodability_cr} is simply the chain rule of entropy. \eqref{eq:decodability2} is implied by \eqref{eq:decodability1}, and \eqref{eq:bound_Y} is true since $\forall i \in [M-1], \bm{Y}_i\in\mathbb{F}_{q}$.

Next we note\footnote{$2M-2 \leq K$ since we consider the case where $2 \leq M \leq \frac{K+1}{2}$.} that $\bm{W}_{[2M-2]}$  can be obtained from ($\bm{W}_{[M-1]}, \bm{Y}_{[M-1]})$, as follows:  $\bm{W}_{M}$ is obtained by subtracting $\bm{W}_{[M-1]}$ terms from $\bm{Y}_1$ which is a linear combination of $\bm{W}_{[M]}$; $\bm{W}_{M+1}$ by subtracting $\bm{W}_{[2:M]}$ terms from $\bm{Y}_2$ which is a linear combination of $\bm{W}_{[2:M+1]}$; $\cdots$; and finally $\bm{W}_{2M-2}$ by subtracting $\bm{W}_{[M-1:2M-3]}$ terms from $\bm{Y}_{M-1}$ which is a linear combination of $\bm{W}_{[M-1:2M-2]}$. Thus, 
\begin{align}
    &H(\bm{W}_{[2M-2]} \mid \bm{\Delta}, \bm{Q}=Q) \\
    &\leq H(\bm{W}_{[M-1]}, \bm{Y}_{[M-1]}\mid \bm{\Delta}, \bm{Q}=Q)\\
    &\overset{\eqref{eq:bound_Y}}{\leq}  (M-1)L,  \hspace{1cm}\forall Q \in \mathcal{Q}.
\end{align}
Averaging over $\bm{Q}$, we have
\begin{align}
    H(\bm{W}_{[2M-2]} \mid \bm{\Delta}, \bm{Q}) \leq (M-1)L.\label{eq:decodability3}
\end{align}
We can follow the same argument for any $2M-2$ out of the $K$ messages, thus \eqref{eq:decodability3} must be true for any $2M-2$ of $K$ messages. Thus, by submodularity,
\begin{align}
    H(\bm{W}_{[K]} \mid \bm{\Delta}, \bm{Q}) &\leq \frac{K(M-1)}{2M-2}L \notag\\
    &= \frac{K}{2}L.\label{eq:resinfo_bound}
\end{align}
Next we have,
\begin{align}
    &H(\bm{W}_{[K]} \mid \bm{\Delta}, \bm{Q}) \notag\\
    &= H(\bm{W}_{[K]}, \bm{\Delta} \mid \bm{Q}) - H(\bm{\Delta} \mid \bm{Q})\\
    &= H(\bm{W}_{[K]} \mid \bm{Q}) + H(\bm{\Delta} \mid \bm{W}_{[K]}, \bm{Q}) - H(\bm{\Delta} \mid \bm{Q})\\
    &= H(\bm{W}_{[K]} \mid \bm{Q}) - H(\bm{\Delta} \mid \bm{Q})\label{eq:download_func_WQ}\\
    &= H(\bm{W}_{[K]}) - H(\bm{\Delta} \mid \bm{Q})\label{eq:independence_WQ}\\
    &= KL - H(\bm{\Delta} \mid \bm{Q}),\label{eq:entropy_W}
\end{align}
 where the first two steps apply the chain rule of entropy, \eqref{eq:download_func_WQ} results from the fact that $\bm{\Delta}$ is a function of the messages and query, and \eqref{eq:independence_WQ} follows from the independence of messages and queries as specified in \eqref{eq:indQ}.
Thus, we have
\begin{align}
    D&\geq H(\bm{\Delta} \mid \bm{Q})\notag\\
    &\overset{\eqref{eq:entropy_W}}{=} KL - H(\bm{W}_{[K]} \mid \bm{\Delta}, \bm{Q})\\
    &\overset{\eqref{eq:resinfo_bound}}{\geq} \frac{K}{2}L.\label{eq:delta_bound}
\end{align}
Thus, the rate achieved must be bounded as $R=L/D\leq 2/K$. Since this is true for every achievable scheme, $C_{\mbox{\tiny PCSI-II}}(q) \leq 2/K$ for $2 \leq M \leq \frac{K+1}{2}$. The remaining case, $C_{\mbox{\tiny PCSI-II}}(q) \leq (K-M+1)^{-1}$ for $\frac{K+1}{2} < M \leq K$ is already shown in \cite{PIR_PCSI}.

\subsection{Achievability}
We present an interference alignment based scheme that works for arbitrary $2 \leq M \leq K$ and is capacity achieving for $2 \leq M \leq \frac{K+1}{2}$. The capacity-achieving scheme for the remaining case is already shown in \cite{PIR_PCSI}. The scheme requires that $q$ should be an even power of a prime number, so that $\mathbb{F}_{\sqrt{q}}$ is also a finite field. Recall that according to polynomial representations of finite fields, $\mathbb{F}_q=\mathbb{F}_{\sqrt{q}}[x]/g(x)$ for some degree $2$ irreducible polynomial $g(x)=x^2+a_1x+a_0\in\mathbb{F}_{\sqrt{q}}[x]$, and  $\mathbb{F}_q$ can be repesented as $\mathbb{F}_q=\{\mu x+\gamma \mid \forall\mu,\gamma\in\mathbb{F}_{\sqrt{q}}\}$. Alternatively, $\mathbb{F}_q$ can be seen as a $2$ dimensional vector space over $\mathbb{F}_{\sqrt{q}}$. Any element ${c}= {\mu_c} x + {\gamma_c} \in \mathbb{F}_{q}$, where ${\mu_c}, {\gamma_c} \in \mathbb{F}_{\sqrt{q}}$, has a corresponding $2\times 1$ vector representation ${V_{c}}\in\mathbb{F}_{\sqrt{q}}^{2\times 1}$  and a $2\times 2$ matrix  representation ${M_{c}}\in\mathbb{F}_{\sqrt{q}}^{2\times 2}$ as follows (see p. 65 of  \cite{Finite_Fields}).
\begin{align}
{V}_c&=\begin{bmatrix}\gamma_c\\ \mu_c\end{bmatrix},&&{M}_c=\begin{bmatrix}
        {\gamma}_{{c}} & -{\mu}_{{c}} a_{0}\\
        {\mu}_{\bm{c}} & {\gamma}_{{c}} - {\mu}_{{c}} a_{1}
    \end{bmatrix}
\end{align}
such that for any $a,b,c\in\mathbb{F}_q$ we have $a=bc$, if and only if
\begin{align}
V_a&=M_bV_c.
\end{align}
Let us start with the following lemma.
\begin{lemma}\label{lem:uniform12}
If $\bm{c}$ is chosen uniformly randomly over $\mathbb{F}_q^\times$, then each row of $M_{\bm{c}}$ is uniformly distributed over $\mathbb{F}_{\sqrt{q}}^{1\times 2}\setminus\{[0,0]\}$.
\end{lemma}
\proof Since $a_0,a_1$ are given constants, the second row, $M_{\bm{c}}(2,:)=[\mu_{\bm{c}}, \gamma_{\bm{c}}-\mu_{\bm{c}}a_1]$ is an invertible function of $V_{\bm{c}}$. Next, note that $a_0\neq 0$ because otherwise $g(x)=x(x+a_1)$ would not be irreducible. Therefore, the first row, $M_{\bm{c}}(1,:)=[\gamma_{\bm{c}}, -\mu_{\bm{c}}a_0]$ is also an invertible function of $V_{\bm{c}}$. Finally, since $\bm{c}$ is uniform over $\mathbb{F}_q^\times$, it follows that $V_{\bm{c}}$ is uniform over $\mathbb{F}_{\sqrt{q}}^{1\times 2}\setminus\{[0,0]\}$, and as an invertible function of $V_{\bm{c}}$ that maps  non-zero vectors to  non-zero vectors, so is each row of $M_{\bm{c}}$. 
 $\hfill\square$

The scheme proposed in this section needs only $L=1$, so let us say $L=1$. 
Recall that the coded side information (CSI) $\bm{Y}^{[\bm{\mathcal{S}}, \bm{\Lambda}]} \triangleq \bm{\lambda}_{1}\bm{W}_{\bm{i}_1} + \cdots + \bm{\lambda}_{M}\bm{W}_{\bm{i}_M}$ where $\bm{\mathcal{S}} = \{\bm{i}_1, \cdots, \bm{i}_M\}$ and $\bm{i}_1<\bm{i}_2<\cdots<\bm{i}_M$.

Since $L=1$, each message is a symbol in $\mathbb{F}_{q}$. Thus each message $\bm{W}_{k}, k \in [K]$ has  vector representation ${V}_{\bm{W}_k} \in \mathbb{F}_{\sqrt{q}}^{2\times 1}$. The first and second entry of ${V}_{\bm{W}_k}$, namely ${V}_{\bm{W}_k}(1)$ and ${V}_{\bm{W}_k}(2)$ respectively, are both elements in $\mathbb{F}_{\sqrt{q}}$ and $\bm{W}_k = {V}_{\bm{W}_k}(2)x + {V}_{\bm{W}_k}(1)$. 

Each coefficient $\bm{\lambda}_m, m \in [M]$ is drawn from $\mathbb{F}_{q}^{\times}$, and can be represented as  ${M}_{\bm{\lambda}_m} \in \mathbb{F}_{\sqrt{q}}^{2\times 2}$ such that ${M}_{\bm{\lambda}_m}{V}_{\bm{W}_{\bm{i}_m}} \in \mathbb{F}_{\sqrt{q}}^{2\times 1}$ is the vector representation of $\bm{\lambda}_{m}\bm{W}_{\bm{i}_m} \in \mathbb{F}_{q}$.

Thus,
\begin{align}
    {V}_{\bm{Y}} = {M}_{\bm{\lambda}_{1}}{V}_{\bm{W}_{\bm{i}_1}} + \cdots + {M}_{\bm{\lambda}_{M}}{V}_{\bm{W}_{\bm{i}_M}} \in \mathbb{F}_{\sqrt{q}}^{2\times 1},\label{eq:vec_rep_CSI}
\end{align}
is the vector representation of  $\bm{Y}^{[\bm{\mathcal{S}}, \bm{\Lambda}]} \in \mathbb{F}_{q}$.

Let $M_{\bm{\lambda}_{m}}(1,:), M_{\bm{\lambda}_{m}}(2,:)$ denote the first and second row of $M_{\bm{\lambda}_{m}}$ respectively, and $M_{\bm{\lambda}_{m}}(r,:)V_{\bm{W}_{\bm{i}_m}}$  the dot product of the $r^{th}$ row of $M_{\bm{\lambda}_{m}}$ with $V_{\bm{W}_{\bm{i}_m}}$ . Then the first and second entry of $V_{\bm{Y}}$ are
\begin{align}
    V_{\bm{Y}}(1) = M_{\bm{\lambda}_{1}}(1,:)V_{\bm{W}_{\bm{i}_1}} + \cdots + M_{\bm{\lambda}_{M}}(1,:)V_{\bm{W}_{\bm{i}_M}},\label{eq:vec_rep_CSI_1}\\
    V_{\bm{Y}}(2) = M_{\bm{\lambda}_{1}}(2,:)V_{\bm{W}_{\bm{i}_1}} + \cdots + M_{\bm{\lambda}_{M}}(2,:)V_{\bm{W}_{\bm{i}_M}},
\end{align}
respectively.

To privately  retrieve $\bm{W}_{\bm{\theta}}$ for some $\bm{\theta} \in \bm{\mathcal{S}}$, the user's download $\bm{\Delta}$ is 
\begin{align}
    \bm{\Delta} = (\mathbf{L}_{k}V_{\bm{W}_{k}})_{k \in [K]},
\end{align}
where $\mathbf{L}_{\bm{i}_m}=M_{\bm{\lambda}_{m}}(1,:)$ for $\bm{i}_m \in \bm{\mathcal{S}} \setminus \{\bm{\theta}\}$,  and $\mathbf{L}_{\bm{i}_m}=M_{\bm{\lambda}_{m}}(2,:)$ for $\bm{i}_m=\bm{\theta}$. For $k \in [K] \setminus \bm{\mathcal{S}}$, $\mathbf{L}_{k}$ is uniformly drawn from $\mathbb{F}_{\sqrt{q}}^{1\times 2}\setminus\{[0~~0]\}$.

Upon receiving $\bm{\Delta}$, by subtracting the $\{M_{\bm{\lambda}_{m}}(1,:)V_{\bm{W}_{\bm{i}_m}}\}_{\bm{i}_m \in \bm{\mathcal{S}} \setminus \{\bm{\theta}\}}$ terms from $V_{\bm{Y}}(1)$, the user is able to obtain $M_{\bm{\lambda}_{\bm{t}}}(1,:)V_{\bm{W}_{\bm{\theta}}}$, where $\bm{i}_t=\bm{\theta}$. Together with $M_{\bm{\lambda}_{\bm{t}}}(2,:)V_{\bm{W}_{\bm{\theta}}}$, which is directly downloaded, the user is able to recover $M_{\bm{\lambda}_{\bm{t}}}V_{\bm{W}_{\bm{\theta}}}$, i.e., $\bm{\lambda}_{\bm{t}}\bm{W}_{\bm{\theta}}$, and since $\bm{\lambda}_{\bm{t}}$ is a non-zero value in $\mathbb{F}_q$ that is known to the user, the user is able to retrieve the desired message $\bm{W}_{\bm{\theta}}$.

Since $\bm{\lambda}_[M]$ are  i.i.d. uniform over $\mathbb{F}_q^\times$, it follows from Lemma \ref{lem:uniform12} that all $\mathbf{L}_k, k\in[K]$ are i.i.d. uniform over $\mathbb{F}_{\sqrt{q}}^{1\times 2}\setminus\{[0~~0]\}$. Thus, the queries are independent of $(\bm{\theta},\bm{\mathcal{S}})$, and the privacy constraint is satisfied.

\begin{remark}\label{rmk:PCSI2_margin}
    The scheme is also capacity achieving for the boundary case $\frac{K+1}{2} < M \leq \frac{K+2}{2}$ (i.e., $2M = K+2$) because in this case, $2/K = (K-M+1)^{-1}$.
\end{remark}

\begin{remark}\label{rmk:half_CSI}
    The scheme only uses $V_{\bm{Y}}(1)$ specified in \eqref{eq:vec_rep_CSI}, i.e., $V_{\bm{Y}}(2)$ is never used so it can be discarded by the user. Thus, at least half of the side-information is redundant.
\end{remark}

Let us consider an example for illustration.
\begin{example}\label{exp:F4}
Suppose $q = 4$, $L=1$. There are $K=3$ messages $\bm{A},\bm{B},\bm{C}\in\mathbb{F}_4$.   We have $M=2$. Say the CSI is the linear combination  $\bm{Y}=\bm{\lambda}_1\bm{A}+\bm{\lambda}_2\bm{B}$, with $\bm{\lambda}_1,\bm{\lambda}_2$ i.i.d. uniform in $\mathbb{F}_4^\times$, and the desired message is $\bm{A}$.

We note that $\mathbb{F}_4=\mathbb{F}_2[x]/(x^2+x+1)$ has the $4$ elements: $0, 1, x, 1+x$, which have matrix representations:
{\small 
\begin{align*}
M_0= \begin{bmatrix}
        0&0\\
        0&0\\
    \end{bmatrix}, M_1= \begin{bmatrix}
        1&0\\
        0&1\\
    \end{bmatrix}, M_{x}= \begin{bmatrix}
        0&1\\
        1&1\\
        \end{bmatrix},M_{1+x}= \begin{bmatrix}
        1&1\\
        1&0\\
        \end{bmatrix}.
\end{align*}
}
Note that if ${\bf c}$ is uniform over $\mathbb{F}_4^\times=\{1,x,1+x\}$ then the first row of $M_c$, i.e., $M_{\bm{c}}(1,:)$ is uniform over $\{[1~~0],[0~~1],[1~~1]\}=\mathbb{F}_{2}^{1\times 2}\setminus\{[0~~0]\}$, and so is the second row, $M_{\bm{c}}(2,:)$, as claimed by Lemma \ref{lem:uniform12}.
Define $\bm{A}=\bm{A}_1+\bm{A}_2x$, where $\bm{A}_1,\bm{A}_2\in\mathbb{F}_2$, so that $V_{\bm{A}}=[\bm{A}_1~~\bm{A}_2]^T$, and use similar definitions for $\bm{B},\bm{C}$ as well.

Let $\bm{\lambda}_1\bm{A} = \bm{A}^{\prime} = \bm{A}_{1}^{\prime}+\bm{A}_2^{\prime}x$. The vector representation of it can thus be written as $V_{\bm{A}^{\prime}} = M_{\bm{\lambda}_1}V_{\bm{A}} = [\bm{A}_{1}^{\prime}~~\bm{A}_2^{\prime}]^{\mathrm{T}}$. Note that  $\bm{A}_{1}^{\prime} = M_{\bm{\lambda}_1}(1,:)[\bm{A}_1 ~~ \bm{A}_2]^{\mathrm{T}}$ and $M_{\bm{\lambda}_1}(1,:)$ is uniform over $\{[1~~0],[0~~1],[1~~1]\}$, thus $\bm{A}_{1}^{\prime}$ is uniform over $\{\bm{A}_1,\bm{A}_2,\bm{A}_1+\bm{A}_2\}$. $\bm{A}_{2}^{\prime}$ is uniform over the same set because $M_{\bm{\lambda}_1}(1,:)$ and $M_{\bm{\lambda}_1}(2,:)$ have the same distribution. Similarly, let $\bm{\lambda}_2\bm{B} = \bm{B}_1^{\prime} + \bm{B}_2^{\prime}x$, and note that $\bm{B}_1^{\prime}, \bm{B}_2^{\prime}$ are individually uniform over $\{\bm{B}_1,\bm{B}_2,\bm{B}_1+\bm{B}_2\}$. Then the side information can be denoted as $\bm{Y} = (\bm{A}_{1}^{\prime} + \bm{B}_{1}^{\prime}) + (\bm{A}_{2}^{\prime} + \bm{B}_{2}^{\prime})x$. According to our scheme, $\bm{B}_{1}^{\prime} = M_{\bm{\lambda}_2}(1,:)V_{\bm{B}}$ is downloaded which enables the user to retrieve $\bm{A}_{1}^{\prime}$ by subtracting it from the first dimension of $\bm{Y}$. The $\bm{A}_{2}^{\prime} = M_{\bm{\lambda}_1}(2,:)V_{\bm{A}}$ is also downloaded. $\bm{A}_{1}^{\prime}, \bm{A}_{2}^{\prime}$ together enable the user to get $\bm{A}^{\prime}$ and thus $\bm{A}$. Note that in our scheme, a non-zero random linear combination of $\bm{C}_1, \bm{C}_2$ is also downloaded. Thus, the download, made up of $\bm{A}_2^{\prime}, \bm{B}_{1}^{\prime}$ and a linear combination of $\bm{C}_1, \bm{C}_2$ is uniform over $\{\bm{A}_1,\bm{A}_2,\bm{A}_1+\bm{A}_2\}\times\{\bm{B}_1,\bm{B}_2,\bm{B}_1+\bm{B}_2\}\times \{\bm{C}_1,\bm{C}_2,\bm{C}_1+\bm{C}_2\}$. For any other realization of $(\bm{\theta},\bm{\mathcal{S}})$, a similar argument applies. Thus, the download is always uniform over the same set, regardless of the realization of $(\bm{\theta},\bm{\mathcal{S}})$, which guarantees privacy.

For example, let us say $\bm{\lambda}_1=1+x, \bm{\lambda}_2=x$, then $V_{\bm{Y}}(1)= \bm{A}_1+\bm{A}_2+\bm{B}_2$. The user downloads, say $\bm{\Delta}=(\bm{A}_1,\bm{B}_2,\bm{C}_1+\bm{C}_2)$ which allows $\bm{A}$ to be retrieved with the help of the side information $V_{\bm{Y}}(1)$. However, from the server's perspective, the following possibilities are equally likely, as the download $\bm{\Delta}=(\bm{A}_1,\bm{B}_2,\bm{C}_1+\bm{C}_2)$ enables the user to decode the desired message under all conditions.
\begin{align}
    \begin{array}{c|c|c}
        \hline
        \mbox{Support Set} & \mbox{CSI} &\mbox{Desired}\\
        \hline
        \{\bm{A},\bm{B}\} & (1+x)\bm{A} + x\bm{B}  & \bm{A}\\
        \hline
        \{\bm{A},\bm{B}\} &  \bm{A} + \bm{B} & \bm{B}\\
        \hline
        \{\bm{B},\bm{C}\} & \bm{B} + (1+x)\bm{C} & \bm{B}\\
        \hline
        \{\bm{B},\bm{C}\} & x\bm{B} + x\bm{C}& \bm{C}\\
        \hline
        \{\bm{A},\bm{C}\} & (1+x)\bm{A} + (1+x)\bm{C} & \bm{A}\\
        \hline
        \{\bm{A},\bm{C}\} &  \bm{A} + x\bm{C}& \bm{C}\\
        \hline
    \end{array}
\end{align}
\end{example}

\section{Proof of Theorem \ref{thm:red}}\label{proof:red}
We need Lemma \ref{lem:alpha_min_1} and \ref{lem:alpha_min_2} to bound the redundancy $\rho_{\mbox{\tiny PCSI-II}}$ from above, (equivalently, lower-bound $\alpha^*$) for $2 \leq M \leq \frac{K+2}{2}$ and $\frac{K+2}{2} < M \leq K$, respectively. 

\begin{lemma}\label{lem:alpha_min_1}
    For $2 \leq M \leq \frac{K+2}{2}$, the redundancy $\rho_{\mbox{\tiny PCSI-II}}\leq 1/2$.
\end{lemma}

Intuitively, the entropy of the download is $H(\bm{\Delta}) = \frac{K}{2}L$. On average, at most $L/2$ symbols of each message are contained in the download. In order to fully recover the desired message, the user must have at least another $L/2$ $q$-ary symbols as the side information.

\proof Recall that the capacity for this case is $2/K$, i.e., the optimal average download cost is $D/L=K/2$. Since this is the infimum across all achievable schemes, there must exist achievable schemes that achieve $D/L\leq K/2+\epsilon$ for any $\epsilon>0$. So consider an achievable scheme such that $\alpha$-CSI is sufficient and the average download cost  $D/L\leq K/2+\epsilon$ for some $L$. Since $D/L\leq K/2+\epsilon$, we have
\begin{align}
&LK/2+\epsilon L\notag\\
&\geq D\\
    &\geq H(\bm{\Delta} \mid \bm{Q}) \\
    &\geq I(\bm{\Delta}; \bm{W}_{[K]} \mid \bm{Q}) \label{eq:non_neg_en}\\
    &= \sum_{k \in [K]}I(\bm{\Delta}; \bm{W}_k \mid \bm{Q}, \bm{W}_{[k-1]})\label{eq:red2_mi_cr}\\
    &= \sum_{k \in [K]}\bigg(H(\bm{W}_k \mid \bm{Q}, \bm{W}_{[k-1]})-H(\bm{W}_k \mid \bm{\Delta}, \bm{Q}, \bm{W}_{[k-1]})\bigg)\\
    &= \sum_{k \in [K]}\bigg(H(\bm{W}_k)-H(\bm{W}_k \mid \bm{\Delta}, \bm{Q}, \bm{W}_{[k-1]})\bigg)\label{eq:independent}\\
    &\geq \sum_{k \in [K]}\bigg(H(\bm{W}_k)-H(\bm{W}_k \mid \bm{\Delta}, \bm{Q})\bigg)\label{eq:red2_en_cr}\\
    &= \sum_{k \in [K]}I(\bm{W}_k; \bm{\Delta}, \bm{Q}),\\
    &\geq K I(\bm{W}_{k^*}; \bm{\Delta}, \bm{Q})
    \label{eq:alpha_half}
\end{align}
where \eqref{eq:non_neg_en} follows from the non-negativity of entropy, \eqref{eq:red2_mi_cr} follows from the chain rule of mutual information, \eqref{eq:independent} holds since all the messages and the query are mutually independent, \eqref{eq:red2_en_cr} results from conditioning reduces entropy and \eqref{eq:alpha_half} is true by setting
\begin{align}
k^*=\arg\min_{k\in[K]}I(\bm{W}_k; \bm{\Delta}, \bm{Q})
\end{align}
From \eqref{eq:alpha_half} we have,
\begin{align}
    &H(\bm{W}_{k^*} \mid \bm{\Delta}, \bm{Q})\\
    &= H(\bm{W}_{k^*}) - I(\bm{W}_{k^*}; \bm{\Delta}, \bm{Q})\\
    &\geq L - (L/2 + \epsilon L/K)\\
    &= L/2 - \epsilon L/K.
\end{align}

Thus, there must exist a feasible query $Q$ such that 
\begin{align}
   H(\bm{W}_{k^*} \mid \bm{\Delta}, \bm{Q}=Q) \geq L/2-\epsilon L/K. \label{eq:smallest}
\end{align}
Let $\mathcal{S} =\{i_1, \cdots, i_{M-1}, k^{*}\}\subset[K]$, such that $|\mathcal{S}|=M$. Then according to Lemma \ref{lem:privacy} and \eqref{eq:invaYR}, there must exist $\Lambda \in \mathfrak{C}$ such that 
\begin{align}
    &H(\bm{W}_{k^*} \mid \bm{\Delta}, \overline{\bm{Y}}^{[\mathcal{S},\Lambda]}, \bm{Q}=Q) = 0,\label{eq:dec_smallest}\\
    &H(\overline{\bm{Y}}^{[\mathcal{S},\Lambda]} \mid \bm{Q} = Q) = H(\overline{\bm{Y}}^{[\bm{\mathcal{S}},\bm{\Lambda}]}) \leq \alpha L.
\end{align}
Combining \eqref{eq:smallest} and \eqref{eq:dec_smallest}, we have
\begin{align}
    I(\overline{\bm{Y}}^{[\mathcal{S},\Lambda]}; \bm{W}_{k^*} \mid \bm{\Delta}, \bm{Q}=Q) \geq L/2-\epsilon L/K.
\end{align}
Thus 
\begin{align}
    \alpha L &\geq H(\overline{\bm{Y}}^{[\mathcal{S},\Lambda]} \mid \bm{Q} = Q)\notag\label{eq:indYQR}\\
    &\geq I(\overline{\bm{Y}}^{[\mathcal{S},\Lambda]}; \bm{W}_{k^*} \mid \bm{\Delta}, \bm{Q}=Q) \geq \frac{L}{2}-\epsilon L/K
\end{align}
which implies that $\alpha \geq 1/2 - \epsilon/K$. In order to approach capacity, we must have $\epsilon\rightarrow 0$, therefore we need $\alpha\geq 1/2$. Since this is true for any $\alpha$ such that $\alpha$-CSI is sufficient, it is also true for $\alpha^*$, and therefore the redundancy is $\rho_{\mbox{\tiny PCSI-II}}\leq 1/2$.

\begin{lemma}\label{lem:alpha_min_2}
    For $\frac{K+2}{2} < M \leq K$, the redundancy $\rho_{\mbox{\tiny PCSI-II}}\leq 0$.
    \end{lemma}

\proof Recall that the capacity for this case is $(K-M+1)^{-1}$, i.e., the optimal average download cost is $D/L=K-M+1$. Consider an achievable scheme such that $\alpha$-CSI is sufficient and the average download cost $D/L\leq K-M+1+\epsilon$ for some $L$. Since $D/L\leq K-M+1+\epsilon$, we have $L(K-M+1)+\epsilon L\geq D\geq H(\bm{\Delta} \mid \bm{Q})$. Thus, there exists a feasible $Q$ such that 
\begin{align}
   H(\bm{\Delta} \mid \bm{Q} = Q) \leq (K-M+1)L+\epsilon L.\label{eq:red2_2_delta_bound}
\end{align}
For all $i \in [K-M+1]$, let $\mathcal{S}_{i} = [i:i+M-1]$. Also, let $\mathcal{S}_{K-M+2} = \{1\} \cup [K-M+2:K]$. For all $i \in [K-M+2]$, let $\Lambda_i \in \mathfrak{C}$ satisfy
\begin{align}
    H(\bm{W}_i \mid \bm{\Delta}, \overline{\bm{Y}}^{[\mathcal{S}_i,\Lambda_i]}, \bm{Q} = Q) = 0.
\end{align}
Such $\Lambda_i$'s must exist according to Lemma \ref{lem:privacy}. 

Writing $\overline{\bm{Y}}^{[\mathcal{S}_i,\Lambda_i]}$ as $\overline{\bm{Y}}_{i}$ for compact notation, we have 
\begin{align}
    H(\bm{W}_{[K-M+2]} \mid \bm{\Delta}, \overline{\bm{Y}}_{[K-M+2]}, \bm{Q}=Q) = 0. \label{eq:redundancy2_1}
\end{align}
According to \eqref{eq:invaYR}, 
\begin{align}
    H(\overline{\bm{Y}}_i \mid \bm{Q} = Q)  \leq \alpha L.\label{eq:red2_2_invaYR}
\end{align}
so we have
\begin{align}
    &(K-M+1)L+\epsilon L + H(\overline{\bm{Y}}_{[K-M+1]} \mid \bm{Q} = Q)+\alpha L\notag\\
    &\geq H(\bm{\Delta} \mid \bm{Q} = Q) + H(\overline{\bm{Y}}_{[K-M+1]} \mid \bm{Q} = Q)\notag\\
    &\quad\quad+ H(\overline{\bm{Y}}_{K-M+2} \mid \bm{Q} = Q)\label{eq:red2_2_combine}\\
    &\geq H(\bm{\Delta}, \overline{\bm{Y}}_{[K-M+2]} \mid \bm{Q} = Q)\\
    &\geq I(\bm{\Delta}, \overline{\bm{Y}}_{[K-M+2]}; \bm{W}_{[K-M+2]},\overline{\bm{Y}}_{[K-M+2]}\mid \bm{Q}=Q)\label{eq:red2_2_non_neg_en}\\
    &= H(\bm{W}_{[K-M+2]},\overline{\bm{Y}}_{[K-M+2]} \mid \bm{Q}=Q)\label{eq:redundancy2_2}\\
    &\geq H(\bm{W}_{[K-M+2]},\overline{\bm{Y}}_{[K-M+1]} \mid \bm{Q}=Q)\\
    &= H(\bm{W}_{[K-M+2]} \mid \bm{Q}=Q)\notag\\
    &\quad\quad + H(\overline{\bm{Y}}_{[K-M+1]} \mid \bm{W}_{[K-M+2]}, \bm{Q}=Q)\label{eq:red2_2_en_cr_1}\\
    &\geq (K-M+2)L\notag\\
    &\quad\quad + H(\overline{\bm{Y}}_{[K-M+1]} \mid \bm{W}_{[M-1]}, \bm{Q}=Q),\label{eq:region}
\end{align}
where \eqref{eq:red2_2_combine} follows from \eqref{eq:red2_2_delta_bound} and \eqref{eq:red2_2_invaYR}. \eqref{eq:red2_2_non_neg_en} results from the non-negativity of entropy. \eqref{eq:redundancy2_2} follows from \eqref{eq:redundancy2_1}. \eqref{eq:red2_2_en_cr_1} is true according to the chain rule. Step \eqref{eq:region} uses the independence of messages and queries according to \eqref{eq:indQ}, and the fact that  $M-1 \geq K-M+2$, because we require $M>(K+2)/2$. We further bound
\begin{align}
    &H(\overline{\bm{Y}}_{[K-M+1]} \mid \bm{W}_{[M-1]}, \bm{Q}=Q)\notag\\
    &=H(\overline{\bm{Y}}_{1} \mid\bm{W}_{[M-1]}, \bm{Q}=Q) + \cdots \notag\\
    &\quad\quad + H(\overline{\bm{Y}}_{K-M+1} \mid \bm{W}_{[M-1]}, \overline{\bm{Y}}_{[K-M]}, \bm{Q}=Q)\label{eq:res2_2_en_cr_2}\\
    &\geq \sum_{i=1}^{K-M+1}H(\overline{\bm{Y}}_i\mid \bm{W}_{[i+M-2]}, \bm{Q}=Q)\label{eq:linearfunc}\\
        &= \sum_{i=1}^{K-M+1}H(\overline{\bm{Y}}_i\mid  \bm{Q}=Q)\label{eq:pcsi2_red_indYO}\\
        &\geq H(\overline{\bm{Y}}_{[K-M+1]}\mid  \bm{Q}=Q)\label{eq:plugin}
\end{align}
\eqref{eq:res2_2_en_cr_2} follows from the chain rule. \eqref{eq:linearfunc} holds because $\overline{\bm{Y}}_{[i-1]}$ is a function of $\bm{W}_{[i+M-2]}$ for all $i \in [2:K-M+1]$. Step \eqref{eq:pcsi2_red_indYO} follows from \eqref{eq:invaYR}. Substituting from \eqref{eq:plugin} into \eqref{eq:region}, and subtracting $H(\overline{\bm{Y}}_{[K-M+1]}\mid  \bm{Q}=Q)$ from both sides, we have 
\begin{align}
    &(K-M+1)L + \epsilon L+ \alpha L \geq (K-M+2)L ,
\end{align}
which gives $\alpha \geq 1-\epsilon$. In order to approach capacity, we must have $\epsilon\rightarrow 0$, so we need $\alpha\geq 1$, and since this is true for any $\alpha$ such that $\alpha$-CSI is sufficient, it is also true for $\alpha^*$. Thus, the redundancy is bounded as $\rho_{\mbox{\tiny PCSI-II}}\leq 0$. $\hfill\square$

According to Remark \ref{rmk:PCSI2_margin} and \ref{rmk:half_CSI}, $\alpha = 1/2$ is sufficient for $2 \leq M \leq \frac{K+2}{2}$ and by the construction of CSI (a linear combination of messages), $\alpha \leq 1$. Theorem \ref{thm:red} is thus proved.

\section{Proof of Theorem \ref{thm:cap_PCSI2_inf}}\label{sec:cap_PCSI2_inf}
We prove Theorem \ref{thm:cap_PCSI2_inf} by first showing that  $C_{\mbox{\tiny PCSI-II}}(q=2)\leq M/((M-1)K)$ and then presenting a PIR-PCSI-II scheme with rate $M/((M-1)K)$ that works for any $\mathbb{F}_{q}$.

\subsection{Converse for $C_{\mbox{\tiny PCSI-II}}(q=2)$}
Note that Lemma \ref{lem:privacy} is true for arbitrary $\mathbb{F}_q$. In $\mathbb{F}_{2}$, we can only have $\bm{\Lambda}=(1,1,\cdots,1)=1_M$, i.e., the length $M$ vector whose elements are all equal to $1$. As a direct result of Lemma \ref{lem:privacy}, for PIR-PCSI-II in $\mathbb{F}_2$,
\begin{align}
    H(\bm{W}_{\mathcal{S}} \mid \bm{\Delta}, \bm{Y}^{[\mathcal{S},1_{M}]}, \bm{Q}=Q) = 0, ~\forall(Q,\mathcal{S}) \in \mathcal{Q}\times\mathfrak{S}.\label{eq:pcsi2_inf_dec}
\end{align}
Thus, $\forall (Q,\mathcal{S}) \in \mathcal{Q}\times\mathfrak{S}$,
\begin{align}
    &H(\bm{W}_{\mathcal{S}} \mid \bm{\Delta}, \bm{Q}=Q)\notag\\
    &= H(\bm{W}_{\mathcal{S}}, \bm{Y}^{[\mathcal{S},1_M]} \mid \bm{\Delta}, \bm{Q}=Q)\label{eq:F2CSI}\\
    &= H(\bm{Y}^{[\mathcal{S},1_M]} \mid \bm{\Delta}, \bm{Q}=Q)\notag\\
    &\quad + H(\bm{W}_{\mathcal{S}} \mid \bm{\Delta}, \bm{Y}^{[\mathcal{S},1_M]}, \bm{Q}=Q)\\
    &\leq L.\label{eq:inf2_resinfo_bound}
\end{align}
\eqref{eq:F2CSI} holds because $\bm{Y}^{[\mathcal{S},1_M]}$ is simply the summation of $\bm{W}_{\mathcal{S}}$. \eqref{eq:inf2_resinfo_bound} follows from \eqref{eq:pcsi2_inf_dec}. Averaging over $\bm{Q}$, we have $H(\bm{W}_{\mathcal{S}} \mid \bm{\Delta}, \bm{Q}) \leq L, \forall \mathcal{S} \in \mathfrak{S}$. By submodularity,
\begin{align}
    H(\bm{W}_{[K]} \mid \bm{\Delta}, \bm{Q}) \leq KL/M.
\end{align}
The download cost can now be lower bounded as,
\begin{align}
 D\geq   H(\bm{\Delta} \mid \bm{Q}) \geq KL - H(\bm{W}_{[K]} \mid \bm{\Delta}, \bm{Q})) \geq \frac{(M-1)KL}{M}.
\end{align}
Thus, we have shown that $C_{\mbox{\tiny PCSI-II}}(q=2) \leq \frac{M}{(M-1)K}$.

\subsection{A PIR-PCSI-II Scheme  for Arbitrary $q$}\label{sec:PCSI2_inf_ach}
In this section, we prove $C_{\mbox{\tiny PCSI-II}}(q) \geq \frac{M}{(M-1)K}$ for all $q$ by proposing a scheme, namely \emph{Generic Linear Combination Based Scheme}, that can achieve the rate $\frac{M}{(M-1)K}$ for any $\mathbb{F}_{q}$.

Let us choose $L = Ml$ where $M$ is the size of the support index set and $l$ is a positive integer which can be arbitrarily large. Thus, any message $\bm{W}_k, k \in [K]$ can be represented as a length-$M$ column vector $V_{\bm{W}_k} \in \mathbb{F}_{q^l}^{M\times 1}$. Let 
\begin{align}
    V_{\bm{W}_{\bm{\mathcal{S}}}} = 
    \begin{bmatrix}
        V_{\bm{W}_{\bm{i}_1}}^{\mathrm{T}} &  \cdots & V_{\bm{W}_{\bm{i}_M}}^{\mathrm{T}}
    \end{bmatrix}^{\mathrm{T}} \in \mathbb{F}_{q^l}^{M^2\times 1}
\end{align}
where $\bm{\mathcal{S}} = \{\bm{i}_1, \cdots, \bm{i}_M\}$ is the support index set. The CSI $\bm{Y}$ can be represented as  $V_{\bm{Y}} \in \mathbb{F}_{q^l}^{M\times 1}$ such that,
\begin{align}
    V_{\bm{Y}} = \underbrace{
    \begin{bmatrix}
        \bm{\lambda}_{1}\mathbf{I}_{M} & \bm{\lambda}_{2}\mathbf{I}_{M} & \cdots & \bm{\lambda}_{M}\mathbf{I}_{M}
    \end{bmatrix}}_{M}V_{\bm{W}_{\bm{\mathcal{S}}}},
\end{align}
where $\mathbf{I}_{M} \in \mathbb{F}_{q^l}^{M\times M}$ is the $M \times M$ identity matrix. 

The download is specified as,
\begin{align}
    \bm{\Delta} = \{&\mathbf{L}_1^{(1)}V_{\bm{W}_{1}}, \cdots, \mathbf{L}_1^{(M-1)}V_{\bm{W}_{1}}, \notag\\ 
    &\cdots, \mathbf{L}_K^{(1)}V_{\bm{W}_{K}}, \cdots, \mathbf{L}_K^{(M-1)}V_{\bm{W}_{K}}\},
\end{align}
where $\forall k \in [K], m \in [M-1], \mathbf{L}_k^{(m)} \in \mathbb{F}_{q^l}^{1\times M}$ is a length-$M$ row vector, i.e., for any message vector $V_{\bm{W}_{k}} \in \mathbb{F}_{q^l}^{M\times 1}$, $\bm{\Delta}$ contains $M-1$ linear combinations of that message vector.

Suppose the vectors $\mathbf{L}_k^{(m)}$ are chosen such that  $\forall \mathcal{S}=\{j_1, \cdots, j_{M}\} \in \mathfrak{S}$ the following $M^2\times M^2$ square matrix has full rank.
\begin{align}
    \mathbf{G}_{\mathcal{S}} = 
    \begin{bmatrix}
        \lambda_{1}\mathbf{I}_{M} & \cdots & \lambda_{M}\mathbf{I}_{M}\\
         & \mathbf{e}_{1}\otimes\mathbf{L}_{j_1}^{(1)} &\\
         & \cdots &\\
         & \mathbf{e}_{1}\otimes\mathbf{L}_{j_1}^{(M-1)} &\\
         & \cdots &\\
         & \mathbf{e}_{M}\otimes\mathbf{L}_{j_M}^{(1)} &\\
         & \cdots &\\
         & \mathbf{e}_{M}\otimes\mathbf{L}_{j_M}^{(M-1)} &
    \end{bmatrix},\label{eq:F2_inv}
\end{align}
$\mathcal{S} = \{j_1, \cdots, j_{M}\} \in \mathfrak{S}$. Note that $(\lambda_1, \cdots, \lambda_M) \in \mathfrak{C}$ is the realization of $\bm{\Lambda}$, $\mathbf{e}_{m}, m\in [M]$ is the $m^{th}$ row of the $M\times M$ identity matrix and ``$\otimes$'' is the Kronecker product.

The correctness constraint is satisfied because the side-information and the downloads allow the user to obtain  $\mathbf{G}_{\bm{\mathcal{S}}}V_{\bm{W}_{\bm{\mathcal{S}}}}$, which can then be multiplied by the inverse of $\mathbf{G}_{\bm{\mathcal{S}}}$ to obtain $V_{\bm{W}_{\bm{\mathcal{S}}}}$, i.e., $\bm{W}_{\bm{\mathcal{S}}}$ which contains $\bm{W}_{\bm{\theta}}$. Specifically the side-information corresponds to the first $M$ rows of $\mathbf{G}_{\bm{\mathcal{S}}}V_{\bm{W}_{\bm{\mathcal{S}}}}$, the downloads $\mathbf{L}_{\bm{j}_1}^{(1)}V_{\bm{W}_{\bm{j}_1}},\cdots,\mathbf{L}_{\bm{j}_1}^{(M-1)}V_{\bm{W}_{\bm{j}_1}}$ correspond to the next $M-1$ rows of $\mathbf{G}_{\bm{\mathcal{S}}}V_{\bm{W}_{\bm{\mathcal{S}}}}$, and so on.

On the other hand, the privacy constraint is satisfied because the construction is such that for every feasible $\mathcal{S}$, the user is able to decode all $M$ messages $\bm{W}_{\mathcal{S}}$. 

Finally let us evaluate the rate achieved by this scheme. Since the user downloads $\frac{M-1}{M}$ portion of every message, the download cost is $D=LK(M-1)/M$, and the rate achieved is $M/((M-1)K))$. Since this rate is achieved for any $\mathbb{F}_q$, we have the lower bound $C_{\mbox{\tiny PCSI-II}}(q) \geq M/((M-1)K))$.

It remains to show the existence of such $\mathbf{L}_k^{(m)}$, for which we need the following lemma.
\begin{lemma}\label{lem:existence}
    There exist $\{\mathbf{L}_{k}^{(m)}\}_{k \in [K], m \in [M-1]}$ such that  for every $\mathcal{S} = \{j_1, \cdots, j_{M}\} \in \mathfrak{S}$, the matrix $ \mathbf{G}_{\mathcal{S}} $  in \eqref{eq:F2_inv} has full rank, provided 
    \begin{align}
        q^l > \tbinom{K}{M}M(M-1).
    \end{align}
\end{lemma}
\proof The proof is in Appendix \ref{app:existence}. $\hfill\square$

With the help of Lemma  \ref{lem:existence}, Theorem \ref{thm:cap_PCSI2_inf} is  proved. Let us illustrate the scheme with an example.

\begin{example}
Consider $M=2, K=4, L=2l, q = 2$. The $4$ messages are $\bm{A},\bm{B},\bm{C},\bm{D}$ each of which has $L = 2l$ symbols in $\mathbb{F}_{2}$. Let $l\geq 3$.

$\bm{A}$ can be represented as a $2\times 1$ vector $V_{\bm{A}} = [V_{\bm{A}}(1) \quad V_{\bm{A}}(2)]^{\mathrm{T}}$ where $V_{\bm{A}}(1), V_{\bm{A}}(2) \in \mathbb{F}_{2^l}$. Similarly, $\bm{B}, \bm{C}, \bm{D}$ can be represented as $V_{\bm{B}}$, $V_{\bm{C}}$, $V_{\bm{D}}$, respectively. Choose $\alpha_1, \cdots, \alpha_4$ as any elements of $\mathbb{F}_{2^l}$ 
such that $\alpha_1, \alpha_2, \alpha_3, \alpha_4,1,0$ are all distinct. This is feasible if $l\geq 3$ because $\mathbb{F}_{2^l}$ has $2^l\geq 8$ distinct elements that include $1,0$ (the elements of $\mathbb{F}_2$). For all possible realizations of $(\bm{\mathcal{S}}, \bm{\theta})$, the download $\bm{\Delta}$ remains the same as follows. 
\begin{align}
\bm{\Delta} = 
\begin{bmatrix}
    \bm{\Delta}_{A}\\
    \bm{\Delta}_{B}\\
    \bm{\Delta}_{C}\\
    \bm{\Delta}_{D}\\
\end{bmatrix}
=
\begin{bmatrix}
    V_{\bm{A}}(1)+\alpha_1 V_{\bm{A}}(2)\\
    V_{\bm{B}}(1)+\alpha_2 V_{\bm{B}}(2)\\
    V_{\bm{C}}(1)+\alpha_3 V_{\bm{C}}(2)\\
    V_{\bm{D}}(1)+\alpha_4 V_{\bm{D}}(2)
\end{bmatrix},\label{eq:delta_binary}
\end{align}
As the download is the same regardless of the realizations of $\bm{\mathcal{S}}$ and $\bm{\theta}$, the query is actually a constant which is trivially independent of $\bm{\mathcal{S}}, \bm{\theta}$ and thus the privacy is guaranteed.

What remains to be proved is the correctness of this scheme, i.e., this specific download enables the user to decode the desired message under all realizations of $\bm{\mathcal{S}}, \bm{\theta}$. 

Let us consider the case where the support set is $\{\bm{A}, \bm{B}\}$, i.e., the side information is $\bm{A} + \bm{B}$, and the desired message is $\bm{A}$. The side information can be represented as $V_{\bm{Y}} \in \mathbb{F}_{2^{l}}^{2\times 1}$ where
\begin{align}
    V_{\bm{Y}} = 
    \begin{bmatrix}
        1&0&1&0\\
        0&1&0&1
    \end{bmatrix}
    \begin{bmatrix}
    V_{\bm{A}}(1)\\
    V_{\bm{A}}(2)\\
    V_{\bm{B}}(1)\\
    V_{\bm{B}}(2)
    \end{bmatrix} =
    \begin{bmatrix}
        V_{\bm{A}}(1)+V_{\bm{B}}(1)\\
        V_{\bm{A}}(2)+V_{\bm{B}}(2)
    \end{bmatrix}.
\end{align}

$V_{\bm{Y}}$, together with $\bm{\Delta}_{A}, \bm{\Delta}_{B}$, can be written as follows
\begin{align}
\begin{bmatrix}
    V_{\bm{Y}}\\
    \bm{\Delta}_{A}\\
    \bm{\Delta}_{B}
\end{bmatrix}
=
\begin{bmatrix}
    1&0&1&0\\
    0&1&0&1\\
    1&\alpha_1&0&0\\
    0&0&1&\alpha_2
\end{bmatrix}
\begin{bmatrix}
    V_{\bm{A}}(1)\\
    V_{\bm{A}}(2)\\
    V_{\bm{B}}(1)\\
    V_{\bm{B}}(2)
\end{bmatrix},
\end{align}
where the invertibility of the matrix is guaranteed by the condition that $\alpha_1, \alpha_2, \alpha_3, \alpha_4,1,0$ are distinct. The user is thus able to recover both $\bm{A}, \bm{B}$ by inverting the matrix. Evidently, the scheme is also correct even if the support set is $\{\bm{A}, \bm{B}\}$ and the desired message is $\bm{B}$.

Suppose the user has $\bm{A} + \bm{C}$ as the side information. Let the vector representation of the side information in this case be $V_{\bm{Y}^{\prime}} \in \mathbb{F}_{2^l}^{2\times 1}$. With the same download as specified in \eqref{eq:delta_binary}, the user has 
\begin{align}
    \begin{bmatrix}
    V_{\bm{Y}^{\prime}}\\
    \bm{\Delta}_{A}\\
    \bm{\Delta}_{C}
\end{bmatrix}
=
\begin{bmatrix}
    1&0&1&0\\
    0&1&0&1\\
    1&\alpha_1&0&0\\
    0&0&1&\alpha_3
\end{bmatrix}
\begin{bmatrix}
    V_{\bm{A}}(1)\\
    V_{\bm{A}}(2)\\
    V_{\bm{C}}(1)\\
    V_{\bm{C}}(2)
\end{bmatrix},
\end{align}
which again guarantees the decodability of both $\bm{A}, \bm{C}$ as the matrix is invertible. Thus the scheme is also correct when $\bm{\mathcal{S}} = \{\bm{A}, \bm{C}\}$ and the desired message is $\bm{A}$ or $\bm{C}$.

Similarly, for all other $4$ possible realizations of the support set ($\{\bm{A}, \bm{D}\}$, $\{\bm{B}, \bm{C}\}$, $\{\bm{B}, \bm{D}\}$, $\{\bm{C}, \bm{D}\}$) and any valid realization of $\bm{\theta} \in \mathcal{S}$, the same $\bm{\Delta}$ enables the user to decode both messages in the support set. Thus, the scheme is also correct.

\end{example}

\section{Proof of Theorem \ref{thm:MK}}\label{proof:MK}
For the case $q=2$, it suffices to download any $K-1$ messages out of the $K$ messages to achieve the capacity $\frac{1}{K-1}$, since the desired message is either directly downloaded or can be recovered by subtracting the $K-1$ downloaded messages from the CSI.

For $q \neq 2$, to achieve the capacity $1$, it suffices to download a linear combination of all $K$ messages with non-zero coefficients. Specifically, 
\begin{align}
    \bm{\Delta} = \bm{Y} + \bm{\lambda}^{\prime}\bm{W}_{\bm{\theta}},
\end{align}
where $\bm{Y}$ is the CSI and $\bm{\lambda}^{\prime} \in \mathbb{F}_{q}^{\times}$ is a non-zero element in $\mathbb{F}_{q}$ such that $\bm{\lambda}_{\bm{t}} + \bm{\lambda}^{\prime} \neq 0$ (let $\bm{\lambda}_{\bm{t}}$ denote the coefficient in front of $\bm{W}_{\bm{\theta}}$ in the CSI $\bm{Y}$). Such $\bm{\lambda}^{\prime}$ always exists for $q \neq 2$. From the server's perspective, the user is downloading a random linear combination of $K$ messages so the privacy constraint is satisfied. The user is able to decode $\bm{W}_{\bm{\theta}}$ by subtracting $\bm{Y}$ from $\bm{\Delta}$ so the correctness constraint is satisfied. 

\section{Proof of Theorem \ref{thm:M3K4}}\label{proof:M3K4}
Let us denote the $K=4$ messages as $\bm{W}_1=\bm{A},\bm{W}_2=\bm{B},\bm{W}_3=\bm{C},\bm{W}_4=\bm{D}$ for simpler notation. We have $M = 3$, the base field is $\mathbb{F}_{3}$ and the length of each message is $L=1$. Our goal is to prove the achievability of rate $1/2$, i.e., download cost $D=2$ for $L=1$. The user downloads, 
\begin{align}
    \bm{\Delta} = \{&\bm{\Delta}_1 = \bm{A} + \bm{\eta}_{b}\bm{B} + \bm{\eta}_{c}\bm{C}, \notag\\
    &\bm{\Delta}_2 = 2\bm{\eta}_{b}\bm{B} + \bm{\eta}_{c}\bm{C} + \bm{\eta}_{d}\bm{D}\}.\label{eq:queryfixed}
\end{align}
From $\bm{\Delta}$, the user is able to also compute 
\begin{align}
    \bm{L}_1 = \bm{\Delta}_1 + \bm{\Delta}_2 &= \bm{A} + 2\bm{\eta}_{c}\bm{C} + \bm{\eta}_{d}\bm{D},\\
    \bm{L}_2 = \bm{\Delta}_1 + 2\bm{\Delta}_2 &= \bm{A} + 2\bm{\eta}_{b}\bm{B} + 2\bm{\eta}_{d}\bm{D}.
\end{align}
Let $\bm{W}_{\bm{\theta}}$ denote the desired message. Let us normalize $\bm{\lambda_1}=1$ without loss of generality.
The $\bm{\eta}_b, \bm{\eta}_c, \bm{\eta}_d$ values are specified as follows. 
\begin{enumerate}
    \item When $\bm{\mathcal{S}}=\{1,2,3\}$ and $\bm{Y} = \bm{A} + \bm{\lambda}_2\bm{B} + \bm{\lambda}_3\bm{C}$, then $\bm{\eta}_d$ is randomly chosen from $\mathbb{F}_{3}^{\times}=\{1,2\}$ and $\bm{\eta}_b,\bm{\eta}_c$ are chosen so that the desired message $\bm{W}_{\bm\theta}$ can be recovered from $\bm{Y}$ and $\bm{\Delta}_1$ as follows.
     {\small
        \begin{align}
           \bm{W}_{\bm{\theta}} = \bm{A}:& ~(\bm{\eta}_b,\bm{\eta}_c) =( 2 \bm{\lambda}_2, 2 \bm{\lambda}_3),  2\bm{A}=\bm{Y}+\bm{\Delta}_1  \notag\\
         \bm{W}_{\bm{\theta}} = \bm{B}:&~(\bm{\eta}_b,\bm{\eta}_c) =( 2 \bm{\lambda}_2,  \bm{\lambda}_3), \bm{\lambda}_2\bm{B}=2\bm{Y}+\bm{\Delta}_1 \notag\\
  \bm{W}_{\bm{\theta}} = \bm{C}:&~ (\bm{\eta}_b,\bm{\eta}_c) =( \bm{\lambda}_2,  2\bm{\lambda}_3),\bm{\lambda}_3\bm{C}=2\bm{Y}+\bm{\Delta}_1 \notag
         \end{align} 
        }
    
    \item When $\bm{\mathcal{S}}=\{2,3,4\}$ and $\bm{Y} = \bm{B} + \bm{\lambda}_2\bm{C} + \bm{\lambda}_3\bm{D}$, then $\bm{\eta}_b$ is randomly chosen from $\mathbb{F}_{q}^{\times}=\{1,2\}$ and  $\bm{\eta}_c,\bm{\eta}_d$ are chosen so that the desired message $\bm{W}_{\bm\theta}$ can be recovered from $\bm{Y}$ and $\bm{\Delta}_2$ as follows.
         {\small
        \begin{align}
           \bm{W}_{\bm{\theta}} = \bm{B}:& ~(\bm{\eta}_c,\bm{\eta}_d) =( \bm{\eta}_b \bm{\lambda}_2, \bm{\eta}_b \bm{\lambda}_3),  \bm{B}=2\bm{Y}+\bm{\Delta}_2/\bm{\eta}_b \notag \\
         \bm{W}_{\bm{\theta}} = \bm{C}:&~(\bm{\eta}_c,\bm{\eta}_d) =( \bm{\eta}_b \bm{\lambda}_2, 2\bm{\eta}_b \bm{\lambda}_3), 2\bm{\lambda}_2 \bm{C}=\bm{Y}+\bm{\Delta}_2/\bm{\eta}_b \notag \\
                  \bm{W}_{\bm{\theta}} = \bm{D}:&~(\bm{\eta}_c,\bm{\eta}_d) =(2 \bm{\eta}_b \bm{\lambda}_2, \bm{\eta}_b \bm{\lambda}_3), 2\bm{\lambda}_3 \bm{D}=\bm{Y}+\bm{\Delta}_2/\bm{\eta}_b \notag 
         \end{align} 
        }
    \item When $\bm{\mathcal{S}}=\{1,3,4\}$ and $\bm{Y} = \bm{A} + \bm{\lambda}_2\bm{C} + \bm{\lambda}_3\bm{D}$, then $\bm{\eta}_b$ is randomly chosen from $\mathbb{F}_{q}^{\times}$ and   $\bm{\eta}_c,\bm{\eta}_d$ are chosen so that the desired message $\bm{W}_{\bm\theta}$ can be recovered from $\bm{Y}$ and $\bm{L}_1$ as follows.
             {\small
        \begin{align}
           \bm{W}_{\bm{\theta}} = \bm{A}:& ~(\bm{\eta}_c,\bm{\eta}_d) =(  \bm{\lambda}_2,  2\bm{\lambda}_3),  2\bm{A}=\bm{Y}+\bm{L}_1 \notag \\
         \bm{W}_{\bm{\theta}} = \bm{C}:&~(\bm{\eta}_c,\bm{\eta}_d) =(  \bm{\lambda}_2,  \bm{\lambda}_3),  \bm{\lambda}_2\bm{C}=2\bm{Y}+\bm{L}_1 \notag \\
                  \bm{W}_{\bm{\theta}} = \bm{D}:&~(\bm{\eta}_c,\bm{\eta}_d) =( 2\bm{\lambda}_2,  2\bm{\lambda}_3), \bm{\lambda}_3 \bm{D}=2\bm{Y}+\bm{L}_1 \notag 
         \end{align} 
        }

    \item When $\bm{\mathcal{S}}=\{1,2,4\}$ and $\bm{Y} = \bm{A} + \bm{\lambda}_2\bm{B} + \bm{\lambda}_3\bm{D}$, then $\bm{\eta}_c$ is randomly chosen from $\mathbb{F}_{q}^{\times}$ and   $\bm{\eta}_b,\bm{\eta}_d$ are chosen so that the desired message $\bm{W}_{\bm\theta}$ can be recovered from $\bm{Y}$ and $\bm{L}_2$ as follows. 
             {\small
        \begin{align}
           \bm{W}_{\bm{\theta}} = \bm{A}:& ~(\bm{\eta}_b,\bm{\eta}_d) =(  \bm{\lambda}_2,  \bm{\lambda}_3),  2\bm{A}=\bm{Y}+\bm{L}_2 \notag \\
         \bm{W}_{\bm{\theta}} = \bm{B}:&~(\bm{\eta}_b,\bm{\eta}_d) =(  \bm{\lambda}_2,  2\bm{\lambda}_3),  \bm{\lambda}_2\bm{B}=2\bm{Y}+\bm{L}_2 \notag \\
                  \bm{W}_{\bm{\theta}} = \bm{D}:&~(\bm{\eta}_b,\bm{\eta}_d) =( 2\bm{\lambda}_2,  \bm{\lambda}_3), \bm{\lambda}_3 \bm{D}=2\bm{Y}+\bm{L}_2 \notag 
         \end{align} 
        }
\end{enumerate}
Correctness is already shown. For privacy, note that the form of the query is fixed as in \eqref{eq:queryfixed} so the user only needs to specify $\bm{\eta}_b,\bm{\eta}_c,\bm{\eta}_d$, and  those are i.i.d. uniform over $\mathbb{F}_3^{\times}=\{1,2\}$, regardless of $(\bm{\mathcal{S}},\bm{\theta})$. Thus, the scheme is private, and the rate achieved is $1/2$, which completes the proof of Theorem \ref{thm:M3K4}.

\section{Proof of Theorem \ref{thm:pcsi2_pub_pri}}\label{proof:pcsi2_pub_pri}
\subsection{Converse}
Here we prove that 
\begin{align}
    C_{\mbox{\tiny PCSI-II}}^{\mbox{\tiny pri}}(q) \leq C_{\mbox{\tiny PCSI-II}}(q=2) = C_{\mbox{\tiny PCSI-II}}^{\inf}.
\end{align}

The following lemma states that for PIR-PCSI*, for every feasible $Q$ and $(\theta, \mathcal{S})$ value, all  possible coefficient vectors must allow successful decoding.
\begin{lemma}\label{lem:fullypri} Under the constraint of $(\bm{\theta}, \bm{\mathcal{S}, \bm{\Lambda}})$ privacy, 
    \begin{align}
    &\mbox{PIR-PCSI: } \forall (Q,\mathcal{S},\theta,\Lambda)\in\mathcal{Q}\times \mathfrak{S}\times[K]\times\mathfrak{C},\notag\\
    &\hspace{0.2cm} H(\bm{W}_{\theta} \mid \bm{\Delta}, \bm{Y}^{[\mathcal{S},\Lambda]}, \bm{Q}=Q) = 0.\label{eq:pcsi_pri}\\
    &\mbox{PIR-PCSI-I: }\forall (Q,\mathcal{S},\theta,\Lambda)\in\mathcal{Q}\times \mathfrak{S}\times([K]\setminus\mathcal{S})\times\mathfrak{C},\notag\\
    &\hspace{0.2cm} H(\bm{W}_{\theta} \mid \bm{\Delta}, \bm{Y}^{[\mathcal{S},\Lambda]}, \bm{Q}=Q) = 0.\label{eq:pcsi1_pri}\\
    &\mbox{PIR-PCSI-II: }\forall (Q,\mathcal{S},\theta,\Lambda)\in\mathcal{Q}\times \mathfrak{S}\times\mathcal{S}\times\mathfrak{C},\notag\\
    &\hspace{0.2cm} H(\bm{W}_{\theta} \mid \bm{\Delta}, \bm{Y}^{[\mathcal{S},\Lambda]}, \bm{Q}=Q) = 0.\label{eq:pcsi2_pri}
    \end{align}
\end{lemma}
\proof Since the server  knows $\bm{\Delta}, \bm{Q}$ and can test all possible realizations of $\bm{\theta}, \bm{\mathcal{S}}, \bm{\Lambda}$ for decodability. If there exists $(\theta, \mathcal{S}, \Lambda)$ such that $\bm{W}_{\theta}$ cannot be decoded, then that $(\theta, \mathcal{S}, \Lambda)$ can be ruled out by the server. This contradicts the joint $(\bm{\theta}, \bm{\mathcal{S}, \bm{\Lambda}})$ privacy constraint.$\hfill\square$

As a direct result of \eqref{eq:pcsi2_pri}, for any PIR-PCSI-II scheme that preserves joint $(\bm{\theta}, \bm{\mathcal{S}}, \bm{\Lambda})$ privacy, 
\begin{align}
    H(\bm{W}_{\mathcal{S}} \mid \bm{\Delta}, \bm{Y}^{[\mathcal{S},\Lambda]}, \bm{Q}=Q) = 0,\notag\\
    \forall (\mathcal{S}, \Lambda, Q) \in \mathfrak{S}\times\mathfrak{C}\times\mathcal{Q}.\label{eq:pcsi2_pri_dec}
\end{align}
Note that \eqref{eq:pcsi2_pri_dec} is a \emph{stronger} version of \eqref{eq:pcsi2_inf_dec} which is sufficient to bound $C_{\mbox{\tiny PCSI-II}}(q=2)$. Thus, we have $C_{\mbox{\tiny PCSI-II}}^{\mbox{\tiny pri}}(q) \leq C_{\mbox{\tiny PCSI-II}}(q=2) = C_{\mbox{\tiny PCSI-II}}^{\inf}$.

\subsection{Achievability}
The \emph{Generic Linear Combination Based Scheme} in Section \ref{sec:PCSI2_inf_ach} where $M-1$ linear combinations of each messages (represented in the extended field $\mathbb{F}_{q^l}$ where $L = Ml$) are downloaded, also works under $(\bm{\theta}, \bm{\mathcal{S}}, \bm{\Lambda})$ privacy, but with a slight modification. The only difference between the modified scheme and the infimum capacity achieving scheme of PIR-PCSI-II in Section \ref{sec:PCSI2_inf_ach} is that, instead of the matrix in \eqref{eq:F2_inv}, the following matrix 
\begin{align}
    \mathbf{G}_{\mathcal{S}}^{(\gamma_1, \gamma_2, \cdots,\gamma_M)} = 
    \begin{bmatrix}
        \gamma_{1}\mathbf{I}_{M} & \cdots & \gamma_{M}\mathbf{I}_{M}\\
        & \mathbf{e}_{1}\otimes\mathbf{L}_{j_1}^{(1)} &\\
        & \cdots &\\
        & \mathbf{e}_{1}\otimes\mathbf{L}_{j_1}^{(M-1)} &\\
        & \cdots &\\
        & \mathbf{e}_{M}\otimes\mathbf{L}_{j_M}^{(1)} &\\
        & \cdots &\\
        & \mathbf{e}_{M}\otimes\mathbf{L}_{j_M}^{(M-1)} &
    \end{bmatrix},\label{eq:Fq_inv_arb}
\end{align}
must have full rank for every $\mathcal{S} = \{j_1, \cdots, j_{M}\} \in \mathfrak{S}$ and every realization of $(\gamma_1, \gamma_2, \cdots,\gamma_M) \in \mathfrak{C}$. Let us prove that the scheme is correct, jointly private and such  $\mathbf{L}_{\cdot}^{(\cdot)}$ vectors  exist when $l$ is large enough that,
\begin{align}
    q^l > (q-1)^{M}\tbinom{K}{M}M(M-1).
\end{align}

\proof For a particular realization of $(\gamma_1, \gamma_2, \cdots, \gamma_M)$, e.g., $(\gamma_1, \gamma_2, \cdots, \gamma_M) = (1, 1, \cdots, 1)$, \eqref{eq:Fq_inv_arb} yields a set of $\tbinom{K}{M}$ matrices 
\begin{align}
    \mathcal{G}^{(1,1,\cdots,1)} = \{\mathbf{G}_{\mathcal{S}_{1}}^{(1,1,\cdots,1)}, \mathbf{G}_{\mathcal{S}_{2}}^{(1,1,\cdots,1)}, \cdots, \mathbf{G}_{\mathcal{S}_{\tbinom{K}{M}}}^{(1,1,\cdots,1)}\}\notag
\end{align}
corresponding to all possible $\{j_1, j_2, \cdots, j_M\} \in \mathfrak{S}$. If all the $\tbinom{K}{M}$ matrices in $\mathcal{G}^{(1,1,\cdots,1)}$ are invertible, this scheme preserves the joint privacy of $(\bm{\theta}, \bm{\mathcal{S}})$ and enables the user to decode all the $M$ messages in the support set, when all the coefficients in CSI are $1$, according to Appendix \ref{app:existence}. 

Going over all the possible realizations of $(\gamma_1, \cdots, \gamma_M) \in \mathfrak{C}$ and $\{j_1, j_2, \cdots, j_M\} \in \mathfrak{S}$, \eqref{eq:Fq_inv_arb} yields $(q-1)^{M}$ sets of matrices 
\begin{align}
    \mathcal{G}^{(1,\cdots,1)}, \mathcal{G}^{(1,\cdots,1,2)}, \cdots, \mathcal{G}^{(q-1,\cdots,q-1)},
\end{align}
each of which contains $\tbinom{K}{M}$ matrices, i.e., there are in total $(q-1)^{M}\tbinom{K}{M}$ matrices. If all the $(q-1)^{M}\tbinom{K}{M}$ matrices are invertible, then for arbitrary realization of $(\gamma_1, \gamma_2, \cdots, \gamma_M)$, i.e., arbitrary $M$ coefficients in the CSI, this scheme enables the user to decode all the $M$ messages in the support set and preserves the joint $(\bm{\theta}, \bm{\mathcal{S}})$ privacy. Since this scheme works for arbitrary coefficients, from the server's perspective, all the realizations of $M$ coefficients are equally likely. Thus, the joint privacy of coefficients $\bm{\Lambda}$, index $\bm{\theta}$, and support set $\bm{\mathcal{S}}$, is preserved.

To prove the existence of such linear combinations, note that the determinant of each one of the $(q-1)^{M}\tbinom{K}{M}$ matrices yields a degree $M(M-1)$ multi-variate polynomial as proved in Appendix \ref{app:existence}. Thus, the product of the determinants of all the matrices $F$ is a multi-variate polynomial of degree $(q-1)^{M}\tbinom{K}{M}M(M-1)$. Again, as in Appendix \ref{app:existence}, according to the Schwartz-Zippel Lemma, when $q^l > (q-1)^{M}\tbinom{K}{M}$, there exists elements in $\mathbb{F}_{q^l}$ such that the polynomial $F$ does not evaluate to $0$, i.e., all the $(q-1)^{M}\tbinom{K}{M}M(M-1)$ matrices are invertible. 

Let us give an example.
\begin{example}
Consider $M=2, K=4, L=2l, q = 3$. The $4$ messages are $\bm{A},\bm{B},\bm{C},\bm{D}$ each of which has $L = 2l$ symbols in $\mathbb{F}_{3}$. Let $l\geq 2$.

$\bm{A}$ can be represented as a $2\times 1$ vector $V_{\bm{A}} = [V_{\bm{A}}(1) \quad V_{\bm{A}}(2)]^{\mathrm{T}}$ where $V_{\bm{A}}(1), V_{\bm{A}}(2) \in \mathbb{F}_{3^l}$. Similarly, $\bm{B}, \bm{C}, \bm{D}$ can be represented as $V_{\bm{B}}$, $V_{\bm{C}}$, $V_{\bm{D}}$, respectively. Choose $\alpha_1, \cdots, \alpha_4$ as elements of $\mathbb{F}_{3^l}$ 
such that $\alpha_1, \alpha_2, \alpha_3, \alpha_4,0,1,2$ are all distinct elements of $\mathbb{F}_{3^l}$. Note that $0,1,2,$ are the elements of $\mathbb{F}_3$, which are also elements of $\mathbb{F}_{3^l}$ because $\mathbb{F}_3$ is a sub-field of $\mathbb{F}_{3^l}$. Furthermore, since $\mathbb{F}_{3^l}$ has $3^l\geq 9$ distinct elements, such $\alpha_i$ are guaranteed to exist. For all possible realizations of $(\bm{\mathcal{S}}, \bm{\theta})$, the download $\bm{\Delta}$ remains the same as follows, 
\begin{align}
\bm{\Delta} = 
\begin{bmatrix}
    \bm{\Delta}_{A}\\
    \bm{\Delta}_{B}\\
    \bm{\Delta}_{C}\\
    \bm{\Delta}_{D}\\
\end{bmatrix}
=
\begin{bmatrix}
    V_{\bm{A}}(1)+\alpha_1 V_{\bm{A}}(2)\\
    V_{\bm{B}}(1)+\alpha_2 V_{\bm{B}}(2)\\
    V_{\bm{C}}(1)+\alpha_3 V_{\bm{C}}(2)\\
    V_{\bm{D}}(1)+\alpha_4 V_{\bm{D}}(2)
\end{bmatrix},\label{eq:delta_3ary}
\end{align}
The query is a constant as the $\bm{\Delta}$ remains unchanged for any realizations of $\bm{\mathcal{S}}, \bm{\theta}, \bm{\Lambda}$. Thus, the privacy is guaranteed. We then prove the correctness, i.e., this specific download enables the user to decode the desired message under all realizations of $\bm{\mathcal{S}}, \bm{\theta}, \bm{\Lambda}$. 

Let us consider the case where the support set is $\{\bm{A}, \bm{B}\}$ and the side information is $\bm{A} + 2\bm{B}$ (i.e., $\Lambda = [1~~2]$), and the desired message is $\bm{A}$. The side information can be represented as $V_{\bm{Y}} \in \mathbb{F}_{3^{l}}^{2\times 1}$ where
\begin{align}
    V_{\bm{Y}} = 
    \begin{bmatrix}
        1&0&2&0\\
        0&1&0&2
    \end{bmatrix}
    \begin{bmatrix}
    V_{\bm{A}}(1)\\
    V_{\bm{A}}(2)\\
    V_{\bm{B}}(1)\\
    V_{\bm{B}}(2)
    \end{bmatrix} =
    \begin{bmatrix}
        V_{\bm{A}}(1)+2V_{\bm{B}}(1)\\
        V_{\bm{A}}(2)+2V_{\bm{B}}(2)
    \end{bmatrix}.
\end{align}

$V_{\bm{Y}}$, together with $\bm{\Delta}_{A}, \bm{\Delta}_{B}$, can be written as follows
\begin{align}
\begin{bmatrix}
    V_{\bm{Y}}\\
    \bm{\Delta}_{A}\\
    \bm{\Delta}_{B}
\end{bmatrix}
=
\begin{bmatrix}
    1&0&2&0\\
    0&1&0&2\\
    1&\alpha_1&0&0\\
    0&0&1&\alpha_2
\end{bmatrix}
\begin{bmatrix}
    V_{\bm{A}}(1)\\
    V_{\bm{A}}(2)\\
    V_{\bm{B}}(1)\\
    V_{\bm{B}}(2)
\end{bmatrix},
\end{align}
where the matrix is invertible because $\alpha_1,\cdots,\alpha_4,0,1,2$ are distinct by design. The user is thus able to recover both $\bm{A}, \bm{B}$ by inverting the matrix.

Similarly, suppose the side information is instead $\bm{A} + \bm{B}$ (i.e., $\bm{\Lambda} = [1~~1]$), the vector representation of the side information is $V_{\bm{Y}^{\prime}} \in \mathbb{F}_{3^l}^{2 \times 1}$. 

$V_{\bm{Y}}^{\prime}$, together with $\bm{\Delta}_{A}, \bm{\Delta}_{B}$, can be written as follows
\begin{align}
\begin{bmatrix}
    V_{\bm{Y}^{\prime}}\\
    \bm{\Delta}_{A}\\
    \bm{\Delta}_{B}
\end{bmatrix}
=
\begin{bmatrix}
    1&0&1&0\\
    0&1&0&1\\
    1&\alpha_1&0&0\\
    0&0&1&\alpha_2
\end{bmatrix}
\begin{bmatrix}
    V_{\bm{A}}(1)\\
    V_{\bm{A}}(2)\\
    V_{\bm{B}}(1)\\
    V_{\bm{B}}(2)
\end{bmatrix}.
\end{align}
Since the matrix is invertible by design, the user is  able to recover both $\bm{A}, \bm{B}$.

Note that the recoverability of both $\bm{A}, \bm{B}$ when the support set is $\{\bm{A}, \bm{B}\}$ and $\bm{\Lambda} = [2 ~~ 1]$ or $\bm{\Lambda} = [2 ~~ 2]$ is automatically proved as $[2 ~~ 1] = 2[1 ~~2]$ and $[2 ~~ 2] = 2[1 ~~ 1]$ in $\mathbb{F}_{3}$. 

Thus when the support set is $\{\bm{A}, \bm{B}\}$, this scheme is correct for arbitrary realizations of $\bm{\theta}, \bm{\Lambda}$.

Similarly, the scheme is also correct for arbitrary realizations of $\bm{\theta}, \bm{\Lambda}$ when $\bm{\mathcal{S}} = \{\bm{A}, \bm{C}\}$, $\{\bm{A}, \bm{D}\}$, $\{\bm{B}, \bm{C}\}$, $\{\bm{B}, \bm{D}\}$, $\{\bm{C}, \bm{D}\}$. Thus, this scheme is correct for arbitrary realizations of $\bm{\mathcal{S}}, \bm{\theta}, \bm{\Lambda}$.

\end{example}

\section{Proof of Theorem \ref{thm:redundancy1}}\label{proof:redundancy1}
Here we bound the redundancy $\rho_{\mbox{\tiny PCSI-I}} $ from above (equivalently, lower-bound $\alpha^{*}$) for $1 \leq M \leq K-1$.

Recall that the supremum capacity for PIR-PCSI-I is $(K-M)^{-1}$, i.e., the optimal average download cost is $D/L = K-M$. Consider an achievable scheme such that $\alpha$-CSI is sufficient and the average download cost $D/L \leq K-M+\epsilon$ for some $L$. Since $D/L \leq K-M+\epsilon$, we have $L(K-M) + \epsilon L \geq D \geq H(\bm{\Delta} \mid \bm{Q})$. Thus, there exists a feasible $Q$ such that 
\begin{align}
    H(\bm{\Delta} \mid \bm{Q}=Q) \leq (K-M)L + \epsilon L.\label{eq:red1_deltabound}
\end{align}
For all $i \in [M]$, let $\mathcal{S}_{i} = [M+1] \setminus \{i\}$. ALso, for all $i \in [M+1:K]$, let $\mathcal{S}_{i} = [M]$. For all $i \in [K]$, let $\Lambda_i \in \mathfrak{C}$ satisfy  
\begin{align}
    H(\bm{W}_{i} \mid \bm{\Delta}, \overline{\bm{Y}}^{[\mathcal{S}_i, \Lambda_i]}, \bm{Q}=Q) = 0.\label{eq:redundancy1_1}
\end{align}
Such $\Lambda_i$'s must exist according to \eqref{eq:lemma1pcsi1} in Lemma \ref{lem:privacy}.

Writing $\overline{\bm{Y}}^{[\mathcal{S}_i, \Lambda_i]}$ as $\overline{\bm{Y}}_{i}$ for compact notation, we have 
\begin{align}
    &H(\bm{W}_{[K]} \mid \bm{\Delta}, \overline{\bm{Y}}_{[M]}, \bm{Q}=Q)\\
    &= H(\bm{W}_{[K]} \mid \bm{\Delta}, \overline{\bm{Y}}_{[M]}, \bm{W}_{[M]}, \bm{Q}=Q)\label{eq:redundancy1_2}\\
    &= H(\bm{W}_{[M+1:K]} \mid \bm{\Delta}, \overline{\bm{Y}}_{[K]}, \bm{W}_{[M]}, \bm{Q}=Q)\label{eq:redundancy1_3}\\
    &= 0\label{eq:redundancy1_4},
\end{align}
where \eqref{eq:redundancy1_2} follows from \eqref{eq:redundancy1_1}. \eqref{eq:redundancy1_3} is correct since $\overline{\bm{Y}}_{[M+1:K]}$ are functions of $\bm{W}_{[M]}$. \eqref{eq:redundancy1_4} follows from \eqref{eq:redundancy1_1}. Since we are considering the case where the supremum capacity is achieved, we have 
\begin{align}
    &(K-M)L + \epsilon L + M\alpha L\notag\\
    &\geq H(\bm{\Delta} \mid \bm{Q}=Q) + H(\overline{\bm{Y}}_{[M]} \mid \bm{Q}=Q)\label{eq:redundancy1_5}\\
    &\geq H(\bm{\Delta}, \overline{\bm{Y}}_{[M]} \mid \bm{Q}=Q)\\
    &\geq I(\bm{\Delta}, \overline{\bm{Y}}_{[M]}; \bm{W}_{[K]} \mid \bm{Q}=Q)\notag\\
    &= H(\bm{W}_{[K]} \mid \bm{Q}=Q) = KL.\label{eq:redundancy1_6}
\end{align}
\eqref{eq:redundancy1_5} follows from \eqref{eq:red1_deltabound} and \eqref{eq:invaYR}. Step \eqref{eq:redundancy1_6} follows from \eqref{eq:redundancy1_4} and the fact that the query and the messages are mutually independent according to \eqref{eq:indQ}. Thus we have $\alpha \geq 1 - \frac{\epsilon}{M}$. In order to approach capacity, we must have $\epsilon\rightarrow 0$, so we need $\alpha\geq 1$, and since this is true for any $\alpha$ such that $\alpha$-CSI is sufficient, it is also true for $\alpha^*$. Thus, the redundancy is bounded as $\rho_{\mbox{\tiny PCSI-I}}\leq 0$.

\section{Proof of Theorem \ref{thm:cap_PCSI1_inf}}\label{sec:cap_PCSI1_inf}
\subsection{Converse for $C_{\mbox{\tiny PCSI-I}}(q=2)$}
Again, \eqref{eq:lemma1pcsi1} is true for arbitrary $\mathbb{F}_{q}$. The only thing different in $\mathbb{F}_{2}$ is that $\bm{\Lambda}$ must be the vector of all ones. As a direct result of \eqref{eq:lemma1pcsi1}, for PIR-PCSI-I in $\mathbb{F}_{2}$,
\begin{align}
    H(\bm{W}_{[K]\setminus\mathcal{S}} \mid \bm{\Delta}, {\bm{Y}}^{[\mathcal{S}, 1_{M}]}, \bm{Q} = Q) = 0, \forall (Q,\mathcal{S}) \in \mathcal{Q} \times \mathfrak{S}\label{eq:dec_inf_PCSI1_1}
\end{align}
and thus 
\begin{align}
    &H(\bm{W}_{[K]\setminus\mathcal{S}} \mid \bm{\Delta}, \bm{Q} = Q) \\
    &\overset{\eqref{eq:dec_inf_PCSI1_1}}{=}I(\bm{W}_{[K]\setminus\mathcal{S}}; {\bm{Y}}^{[\mathcal{S}, 1_{M}]}\mid \bm{\Delta}, \bm{Q} = Q)\\
    &\leq H({\bm{Y}}^{[\mathcal{S}, 1_{M}]}\mid \bm{\Delta}, \bm{Q} = Q)\\
    &\overset{\eqref{eq:sideinfo_CSI}}{\leq} L, ~~\forall (Q,\mathcal{S}) \in \mathcal{Q} \times \mathfrak{S}.
\end{align}
Averaging over $\bm{Q}$ gives 
\begin{align}
    H(\bm{W}_{[K]\setminus\mathcal{S}} \mid \bm{\Delta}, \bm{Q}) \leq L, \forall \mathcal{S} \in\mathfrak{S}.\label{eq:dec_inf_PCSI1_2}
\end{align}
Also, for all $\mathcal{S} \in \mathfrak{S}$ and $Q \in \mathcal{Q}$, 
\begin{align}
    &H(\bm{W}_{[K]} \mid \bm{\Delta}, \bm{Q}=Q)\\ 
    &= H(\bm{W}_{\mathcal{S}}\mid \bm{\Delta}, \bm{Q}=Q)\notag\\
    &~~ + H(\bm{W}_{[K]\setminus\mathcal{S}}\mid \bm{\Delta}, \bm{W}_{\mathcal{S}}, \bm{Q}=Q)\\
    &= H(\bm{W}_{\mathcal{S}}\mid \bm{\Delta}, \bm{Q}=Q)\notag\\ 
    &~~ + H(\bm{W}_{[K]\setminus\mathcal{S}}\mid \bm{\Delta}, \bm{W}_{\mathcal{S}}, {\bm{Y}}^{[\mathcal{S}, 1_{M}]}, \bm{Q}=Q)\label{eq:pcsi1_inf_Ysum}\\
    &= H(\bm{W}_{\mathcal{S}}\mid \bm{\Delta}, \bm{Q}=Q),
\end{align}
where \eqref{eq:pcsi1_inf_Ysum} results from the fact that $\overline{\bm{Y}}^{[\mathcal{S}, 1_{M}]} = \sum_{s \in \mathcal{S}}\bm{W}_{s}$, and the last step follows from \eqref{eq:dec_inf_PCSI1_1}. Averaging over $\bm{Q}$, it follows that 
\begin{align}
    H(\bm{W}_{[K]} \mid \bm{\Delta}, \bm{Q}) = H(\bm{W}_{\mathcal{S}}\mid \bm{\Delta}, \bm{Q}), &&\forall \mathcal{S} \in \mathfrak{S}.\label{eq:pcsi1_inf_equ}
\end{align}

Let us first prove $C_{\mbox{\tiny PCSI-I}}(q=2) \leq (K-1)^{-1}$ in the regime where $1 \leq M \leq \frac{K}{2}$. 
\begin{align}
    &H(\bm{W}_{[K]} \mid \bm{\Delta}, \bm{Q}) \notag\\
    &=H(\bm{W}_{[M]} \mid \bm{\Delta}, \bm{Q})\label{eq:dec_inf_PCSI1_r1_1}\\
    &\leq H(\bm{W}_{[K-M]} \mid \bm{\Delta}, \bm{Q})\label{eq:MtoK-M}\\
    & \leq L \label{eq:K-MtoL},
\end{align}
where \eqref{eq:dec_inf_PCSI1_r1_1} is true according to \eqref{eq:pcsi1_inf_equ}, \eqref{eq:MtoK-M} follows from $(K-M \geq M)$ and \eqref{eq:dec_inf_PCSI1_1}, and \eqref{eq:K-MtoL} follows from \eqref{eq:dec_inf_PCSI1_2}. Thus
\begin{align}
    H(\bm{\Delta} \mid \bm{Q}) &\geq I(\bm{\Delta}; \bm{W}_{[K]} \mid \bm{Q})\\
    &= H(\bm{W}_{[K]} \mid \bm{Q}) - H(\bm{W}_{[K]} \mid \bm{\Delta}, \bm{Q})\\
    &\geq KL - L.
\end{align} 
Thus $D \geq H(\bm{\Delta} \mid \bm{Q}) \geq KL-L$ and since the rate $L/D\leq (K-1)^{-1}$ for every achievable scheme, we have shown that $C_{\mbox{\tiny PCSI-I}}(q=2) \leq (K-1)^{-1}$ when $K-M\geq M\geq 1$, i.e., $1\leq M\leq K/2$.

Next let us prove that $C_{\mbox{\tiny PCSI-I}}(q=2) \leq \big(K - \frac{M}{K-M}\big)^{-1}$ for the regime $\frac{K}{2} < M \leq K-1$. It suffices to prove $H(\bm{\Delta} \mid \bm{Q}) \geq KL - \frac{ML}{K-M}$. Define,
\begin{align}
    H_{m}^{K} = \frac{1}{\tbinom{K}{m}}\sum_{\mathcal{M}:\mathcal{M}\subset[K], |\mathcal{M}|=m}\frac{H(\bm{W}_{\mathcal{M}} \mid \bm{\Delta}, \bm{Q})}{m},
\end{align} 
we have
\begin{align}
    H_{K-M}^{K}&\geq H_{M}^{K}\label{eq:dec_inf_PCSI1_r2_1}\\
    &= \frac{H(\bm{W}_{[K]}\mid \bm{\Delta}, \bm{Q})}{M},\label{eq:dec_inf_PCSI1_r2_2}
\end{align}
where \eqref{eq:dec_inf_PCSI1_r2_1} follows from Han's inequality \cite{Cover_Thomas}, and \eqref{eq:dec_inf_PCSI1_r2_2} follows from \eqref{eq:pcsi1_inf_equ}. Note that according to \eqref{eq:dec_inf_PCSI1_2},
\begin{align}
    \frac{L}{K-M} \geq H_{K-M}^{K},
\end{align}
and therefore,
\begin{align}
    H(\bm{W}_{[K]}\mid \bm{\Delta}, \bm{Q}) \leq \frac{ML}{K-M}.
\end{align}
Thus, $H(\bm{\Delta} \mid \bm{Q}) \geq KL - \frac{ML}{K-M}$, which completes the converse proof for Theorem \ref{thm:cap_PCSI1_inf}. We next prove  achievability.

\subsection{Two PIR-PCSI-I Schemes  for Arbitrary $q$}\label{sec:PCSI1_inf_ach}
\subsubsection{Achieving rate $\frac{1}{K-1}$ when $1 \leq M \leq \frac{K}{2}$}\label{sec:PCSI1_inf_ach1}
The goal here is to download $K-1$ generic linear combinations so that along with the one linear combination already available as side-information, the user has enough information to retrieve all $K$ messages. Let $L$ be large enough that  $q^L > \tbinom{K}{M}(K-1)$. For all $k \in [K]$, message $\bm{W}_k \in \mathbb{F}_{q}^{L\times 1}$ can be represented as a scalar $\bm{w}_k \in \mathbb{F}_{q^L}$. Let 
\begin{align}
    \bm{w}_{[K]} = 
    \begin{bmatrix}
        \bm{w}_1 & \bm{w}_2 & \cdots & \bm{w}_K
    \end{bmatrix}^{\mathrm{T}} \in \mathbb{F}_{q^L}^{K\times 1},
\end{align}
be the length $K$ column vector whose entries are the messages represented in $\mathbb{F}_{q^{L}}$. Let $\Psi \in\mathbb{F}_{q^L}^{K\times (K-1)}$ be a  $K\times (K-1)$ matrix whose elements are the variables $\psi_{ij}$. The user downloads 
\begin{align}
    \bm{\Delta} =\Psi^T \bm{w}_{[K]} \in\mathbb{F}_{q^L}^{(K-1)\times 1}.
\end{align}
Suppose the realization of the coefficient vector is $\bm{\Lambda}=\Lambda$. The linear combination available to the user can be expressed as $\bm{Y}^{[{\Lambda},\bm{\mathcal{S}}]}=U_{{{\Lambda}},\bm{\mathcal{S}}}^T\bm{w}_{[K]}$ for some $K\times 1$ vector $U_{{\Lambda},\bm{\mathcal{S}}}$ that depends on $({\Lambda},\bm{\mathcal{S}})$. Combined with the download, the user has 
\begin{align}
[U_{{\Lambda},\bm{\mathcal{S}}}, \Psi]^T\bm{w}_{[K]},
\end{align}
so if the $K\times K$ matrix $G_{{\Lambda},\bm{\mathcal{S}}}=[U_{{\Lambda},\bm{\mathcal{S}}}, \Psi]$ is invertible (full rank) then the user can decode all $K$ messages. For all $\mathcal{S}\in\mathfrak{S}$, let $f_{\Lambda,\mathcal{S}}(\cdot)$ be the multi-variate polynomial of degree $K-1$ in variables $\psi_{ij}$, representing the determinant of $G_{{\Lambda},{\mathcal{S}}}$. This is not the zero polynomial because the $K-1$ columns of $\Psi$  can always be chosen to be linearly independent of the vector $U_{{\Lambda},{\mathcal{S}}}$ in a $K$ dimensional vector space. The product of all such polynomials, $f_\Lambda=\prod_{\mathcal{S}\in\mathfrak{S}}f_{\Lambda,\mathcal{S}}$ is itself a multi-variate non-zero polynomial of degree $(K-1)\binom{K}{M}$ in the variables $\psi_{ij}$. By Schwartz-Zippel Lemma, if the $\psi_{ij}$ are chosen randomly from $\mathbb{F}_{q^L}$ then the probability that the corresponding evaluation of $f_\Lambda$ is zero, is no more than $(K-1)\binom{K}{M}/q^L<1$, so there exists a choice of $\psi_{ij}$ for which all $f_{\Lambda,\mathcal{S}}$ evaluate to non-zero values, i.e.,  $G_{\Lambda, \mathcal{S}}$ is invertible for every $\mathcal{S}\in\mathfrak{S}$. Thus, with this choice of $\Psi$, we have a scheme with rate $1/(K-1)$ that is correct and private and allows the user to retrieve all $K$ messages. To verify privacy, note that the user constructs the query based on the realization of $\bm\Lambda$ alone, and does not need to know $(\bm{\mathcal{S}},\bm{\theta})$ before it sends the query, so the query is independent of $(\bm{\mathcal{S}},\bm{\theta})$. 

\begin{remark}\label{rmk:pcsi1_inf_pcsi_inf}
Since the scheme allows the user to decode all messages, the scheme also works if $\bm{\theta}$ is uniformly drawn from $[K]$, i.e., in the PIR-PCSI setting.
\end{remark}

\subsubsection{Achieving rate $(K-\frac{M}{K-M})^{-1}$ when $K/2 < M \leq K-1$}
Now let us present a scheme with rate $(K-\frac{M}{K-M})^{-1}$ which is optimal for the regime $\frac{K}{2} < M \leq K-1$. The scheme is comprised of two steps.

\emph{Step 1}: The user converts the $(M,K)$ PIR-PCSI-I problem to $(K-M,K)$ PIR-PCSI-II problem as follows.

The user first downloads
\begin{align}
    \bm{\Delta}_{1} = \sum_{k \in [K]}\bm{a}_{k}\bm{W}_{k},
\end{align}
where $\bm{a}_{\bm{i}_m} = \bm{\lambda}_m$ for $\bm{i}_{m} \in \bm{\mathcal{S}}$ while for $k \notin \bm{\mathcal{S}}$, $\bm{a}_k$'s are independently and uniformly drawn from $\mathbb{F}_{q}^{\times}$.
The user then computes
\begin{align}
    \bm{Y}^{\prime} = \bm{\Delta}_1 - \bm{Y}^{[\bm{\mathcal{S}}, \bm{\Lambda}]} = \sum_{k \in [K]\setminus \bm{\mathcal{S}}}\bm{a}_{k}\bm{W}_{k}.
\end{align} 
In this step, from the server's perspective, $\bm{a}_1, \cdots, \bm{a}_K$ are i.i.d. uniform over $\mathbb{F}_{q}^{\times}$, thus there is no loss of privacy. The download cost of this step is $H(\bm{\Delta}_1) = L$.

\emph{Step 2}: The user has $\bm{Y}^{\prime}$ as  coded side information and applies the fully private PIR-PCSI-II scheme described in Section \ref{proof:pcsi2_pub_pri} that  protects the privacy of all the coefficients. 

The reason to apply the PIR-PCSI-II scheme that maintains the privacy of coefficients is that in \emph{Step 1}, server knows $\bm{a}_1, \cdots, \bm{a}_K$. If in the second step, the Query is not independent of $\bm{a}_i, i \in [K]\setminus \bm{\mathcal{S}}$, then the server may be able to rule out some  realizations of $\bm{\mathcal{S}}$. The download cost of this step is $\frac{K(K-M-1)L}{K-M}$. Thus, the total download cost of this scheme is $KL - \frac{ML}{K-M}$ and the rate is $\big(K - \frac{M}{K-M}\big)^{-1}$.

\section{Proof of Theorem \ref{thm:pcsi1_pub_pri}}\label{proof:pcsi1_pub_pri}
\subsection{Proof of $C_{\mbox{\tiny PCSI-I}}^{\mbox{\tiny pri}, \sup}=C_{\mbox{\tiny PCSI-I}}^{\inf}$}
First let us prove the converse. As a direct result of \eqref{eq:pcsi1_pri} in Lemma \ref{lem:fullypri}, for any PIR-PCSI-I scheme that preserves joint $(\bm{\theta}, \bm{\mathcal{S}}, \bm{\Lambda})$ privacy, 
\begin{align}
    H(\bm{W}_{[K]\setminus\mathcal{S}} \mid \bm{\Delta}, \bm{Y}^{[\mathcal{S},\Lambda]}, \bm{Q}=Q) = 0, \notag\\
    \forall (\mathcal{S}, \Lambda, Q) \in \mathfrak{S}\times\mathfrak{C}\times\mathcal{Q}.\label{eq:pcsi1_pri_dec}
\end{align}
Note that \eqref{eq:pcsi1_pri_dec} is a stronger version of \eqref{eq:dec_inf_PCSI1_1} which is sufficient to bound $C_{\mbox{\tiny PCSI-I}}(q=2)$. Thus, we have $C_{\mbox{\tiny PCSI-I}}^{\mbox{\tiny pri}}(q) \leq C_{\mbox{\tiny PCSI-I}}(q=2) = C_{\mbox{\tiny PCSI-I}}^{\inf}$, which completes the proof of converse.

For achievability, let us note that $C_{\mbox{\tiny PCSI-I}}^{\mbox{\tiny pri}, \sup}\geq C_{\mbox{\tiny PCSI-I}}^{\mbox{\tiny pri}}(q=2)=C_{\mbox{\tiny PCSI-I}}(q=2)=C_{\mbox{\tiny PCSI-I}}^{\inf}$, because over $\mathbb{F}_2$, the $\bm{\Lambda}$ vector is constant (all ones) and therefore trivially private.

\subsection{Proof of the bound: $C_{\mbox{\tiny PCSI-I}}^{\mbox{\tiny pri}, \inf} \leq \min\bigg(C_{\mbox{\tiny PCSI-I}}^{\inf}, \frac{1}{K-2}\bigg)$}
Since privacy of $\bm\Lambda$ only further constrains PIR-PCSI, it is trivial that $C_{\mbox{\tiny PCSI-I}}^{\mbox{\tiny pri}, \inf} \leq C_{\mbox{\tiny PCSI-I}}^{\inf}$. For the remaining bound,  $C_{\mbox{\tiny PCSI-I}}^{\mbox{\tiny pri}, \inf}\leq \frac{1}{K-2}$, it suffices to show that $C_{\mbox{\tiny PCSI-I}}^{\mbox{\tiny pri}}(q\geq M) \leq \frac{1}{K-2}$, because $C_{\mbox{\tiny PCSI-I}}^{\mbox{\tiny pri}, \inf}\leq C_{\mbox{\tiny PCSI-I}}^{\mbox{\tiny pri}}(q\geq M)$. Note that by $C_{\mbox{\tiny PCSI-I}}^{\mbox{\tiny pri}}(q\geq M)$ we mean $C_{\mbox{\tiny PCSI-I}}^{\mbox{\tiny pri}}(q)$ for all $q\geq M$.

Let  
\begin{align}
    \bm{Y}_1 &= \bm{W}_2 + \alpha_3 \bm{W}_3 + \cdots \alpha_{M+1} \bm{W}_{M+1},\\
    \bm{Y}_2 &= \bm{W}_1 + \bm{W}_3 + \bm{W}_4 + \cdots \bm{W}_{M+1},
\end{align}
where $\alpha_3, \alpha_4, \cdots, \alpha_{M+1}$ are $M-1$ distinct elements in $\mathbb{F}_{q}^\times$.

Let $\beta_3, \beta_4, \dots \beta_{M+1}$ be $M-1$ distinct  elements in $\mathbb{F}_{q}^\times$ such that $\forall{m \in [3:M+1]}, \beta_{m}\alpha_{m} + 1 = 0$ in $\mathbb{F}_{q}$.

Note that such $\alpha$'s and $\beta$'s exist since $q \geq M$.

Then let 
\begin{align}
    \bm{Y}_{m} &= \beta_m \bm{Y}_1 + \bm{Y}_2 \notag\\
    &= \bm{W}_1 + \beta_m \bm{W}_2 + (\beta_m \alpha_3 + 1)\bm{W}_3 + \cdots \notag\\
    &\quad +(\beta_m \alpha_i + 1) \bm{W}_i + \cdots + (\beta_m \alpha_{M+1} + 1)\bm{W}_{M+1}, \notag\\
    &\forall m \in [3:M+1],
\end{align}
be $M-1$ linear combinations of the first $M+1$ messages $\bm{W}_{[M+1]}$. Note that for any $m \in [3:M+1]$, the coefficient for $\bm{W}_m$ in $\bm{Y}_m$ (i.e., $\beta_{m}\alpha_{m} + 1$) is $0$ while the coefficient for any $\bm{W}_i, i\in[M+1], i\neq m$ (i.e., $\beta_{m}\alpha_{i} + 1$) is non-zero\footnote{Since $\beta_{m}\alpha_{m}+1=0$, $\beta_{m}\alpha_{i}+1\neq 0$ for $i \neq m$.}. For example, 
\begin{align}
    \bm{Y}_3 &= \bm{W}_1 + \beta_3 \bm{W}_2 + 0\bm{W}_3 + (\beta_3\alpha_4 + 1)\bm{W}_4\notag\\
    &\quad + \cdots + (\beta_3\alpha_{M+1} + 1)\bm{W}_{M+1}.
\end{align}
Thus, for any $m \in [M+1]$, $\bm{Y}_m$ is a linear combination of $M$ messages $\bm{W}_{[M+1]\setminus\{m\}}$ with non-zero coefficients. For $\mathcal{S}_m=[M+1]/\{m\}$ and $\Lambda_m$ as the vector of coefficients that appear in $\bm{Y}_m$, we $\bm{Y}^{[\mathcal{S}_m,\Lambda_m]}=\bm{Y}_m$.

According to \eqref{eq:pcsi1_pri_dec}, 
\begin{align}
    H(\bm{W}_m, \bm{W}_{[M+2:K]} \mid \bm{\Delta}, \bm{Y}_m, \bm{Q} = Q) = 0,\notag\\
     \forall m \in [M+1], Q \in \mathcal{Q}. \label{eq:PCSI1_pri_dec1}
\end{align}
Thus, for all $Q\in\mathcal{Q}$,
\begin{align}
    &H(\bm{W}_{[K]} \mid \bm{\Delta}, \bm{Q} = Q) \notag\\
    &\leq H(\bm{W}_{[K]}, \bm{Y}_{[M+1]} \mid \bm{\Delta}, \bm{Q}=Q) \\
    &= H(\bm{Y}_{[M+1]} \mid \bm{\Delta}, \bm{Q}=Q) + H(\bm{W}_{[K]} \mid \bm{\Delta}, \bm{Y}_{[M+1]}, \bm{Q}=Q)\\
    &= H(\bm{Y}_1, \bm{Y}_2 \mid \bm{\Delta}, \bm{Q}=Q)\label{eq:PCSI1_pri_dec2}\\
    &\leq 2L,
\end{align}
where \eqref{eq:PCSI1_pri_dec2} follows from \eqref{eq:PCSI1_pri_dec1} and the fact that $\bm{Y}_{[3:M+1]}$ are functions of $\bm{Y}_{1}, \bm{Y}_{2}$. Averaging over $\bm{Q}$ we have 
\begin{align}
    H(\bm{W}_{[K]} \mid \bm{\Delta}, \bm{Q}) \leq 2L.
\end{align}

\noindent Therefore, the average download cost is bounded as,
\begin{align}
  D&\geq  H(\bm{\Delta} \mid \bm{Q}) \geq H(\bm{W}_{[K]}\mid\bm{Q}) - H(\bm{W}_{[K]} \mid \bm{\Delta}, \bm{Q}) \\
  & \geq (K-2)L.
\end{align}
Thus, for $q\geq M$, we have $C_{\mbox{\tiny PCSI-I}}^{\mbox{\tiny pri}}(q) \leq \frac{1}{K-2}$.

\subsection{Proof of $C_{\mbox{\tiny PCSI-I}}^{\mbox{\tiny pri}, \inf} \geq \frac{1}{K-1}$}\label{sec:pcsi1_pri_ach}
We need to show that $C_{\mbox{\tiny PCSI-I}}^{\mbox{\tiny pri}}(q)\geq \frac{1}{K-1}$ for all $\mathbb{F}_q$. The scheme is identical to the scheme with rate $(K-1)^{-1}$ in Section \ref{sec:PCSI1_inf_ach1} with a slight modification. Instead of fixing a realization $\bm{\Lambda}=\Lambda$, we will consider all possible realizations $\Lambda\in\mathfrak{C}$, and consider the product polynomial $f=\prod_{\Lambda\in\mathfrak{C}}f_{\Lambda}$ which is a multi-variate polynomial of degree $(K-1)\binom{K}{M}(q-1)^M$ in variables $\psi_{ij}$. Following the same argument based on the Schwartz-Zippel Lemma, we find that there exists a $\Psi$ for which all $G_{\Lambda,\mathcal{S}}$ are invertible matrices, provided that $L$ is large enough that $q^L>(q-1)^M(K-1)\binom{K}{M}$. Thus, with this choice of $\Psi$ we have a scheme that is allows the user to retrieve all $K$ messages. The scheme is also $(\bm{\mathcal{S}},\bm{\theta},\bm{\Lambda})$ private because we note that the user does not need to know the realization of $(\bm{\mathcal{S}},\bm{\theta},\bm{\Lambda})$ before it sends the query, so the query is independent of $(\bm{\mathcal{S}},\bm{\theta},\bm{\Lambda})$. 

\begin{remark}\label{rmk:pcsi1_pri_pcsi_pri}
Since the scheme allows the user to decode all messages, and the query does not depend on $(\bm{\theta}, \bm{\mathcal{S}}, \bm{\Lambda})$, the scheme also works if $\bm{\theta}$ is uniformly drawn from $[K]$, i.e., in the PIR-PCSI setting.
\end{remark}

\section{Proof of Theorem \ref{thm:cap_PCSI_sup}}\label{sec:cap_PCSI_sup}
\subsection{Converse}
The converse is  divided into two regimes.

\textbf{Regime 1}: $2 \leq M \leq K$. The proof relies on \eqref{eq:lemma1pcsi} in Lemma \ref{lem:privacy}.
Consider any particular realization $Q \in \mathcal{Q}$ of $\bm{Q}$. For all $i \in [K]$, consider $\mathcal{S} = [M], \theta = i$, and let $\Lambda_i$ be a coefficient vector that satisfies \eqref{eq:lemma1pcsi} according to Lemma \ref{lem:privacy}, so that 
\begin{align}
    H(\bm{W}_i \mid \bm{\Delta}, \bm{Y}^{[[M],\Lambda_i]}, \bm{Q} = Q) = 0.\label{eq:con_PCSI_0}
\end{align}
Writing $\bm{Y}^{[[M],\Lambda_i]}$ as $\bm{Y}_{i}$ for compact notation, we have
\begin{align}
    &H(\bm{W}_{[K]} \mid \bm{\Delta}, \bm{Y}_{[M-1]}, \bm{Q} = Q)\notag\\
    &= H(\bm{W}_{[K]} \mid \bm{\Delta}, \bm{Y}_{[M-1]}, \bm{W}_{[M-1]}, \bm{Q} = Q)\label{eq:con_PCSI_1}\\
    &= H(\bm{W}_{[K]} \mid \bm{\Delta}, \bm{W}_{[M]}, \bm{Q} = Q)\label{eq:con_PCSI_2}\\
    &= H(\bm{W}_{[M+1:K]} \mid \bm{\Delta}, \bm{W}_{[M]}, \bm{Y}_{[M+1:K]}, \bm{Q} = Q)\label{eq:con_PCSI_3}\\
    &= 0,\label{eq:con_PCSI_3a}
\end{align}
where \eqref{eq:con_PCSI_1} holds according to \eqref{eq:con_PCSI_0}, and \eqref{eq:con_PCSI_2} follows from the fact that $\bm{W}_M$ is decodable by subtracting $\bm{W}_{[M-1]}$ terms from $\bm{Y}_1$. Then, \eqref{eq:con_PCSI_3} uses the fact that $\bm{Y}_{[M+1:K]}$ are functions of $\bm{W}_{[M]}$. Finally, \eqref{eq:con_PCSI_3a} follows from \eqref{eq:con_PCSI_0}. 

Averaging over $\bm{Q}$, 
\begin{align}
    H(\bm{W}_{[K]} \mid \bm{\Delta}, \bm{Y}_{[M-1]}, \bm{Q}) = 0.\label{eq:con_PCSI_4}
\end{align}
Then we have 
\begin{align}
    &H(\bm{W}_{[K]} \mid \bm{\Delta}, \bm{Q})\\
    &= H(\bm{W}_{[K]}, \bm{Y}_{[M-1]} \mid \bm{\Delta}, \bm{Q})\label{eq:con_PCSI_5}\\
    &= H(\bm{Y}_{[M-1]} \mid \bm{\Delta}, \bm{Q}) + H(\bm{W}_{[K]} \mid \bm{\Delta}, \bm{Q}, \bm{Y}_{[M-1]})\\
    &\leq H(\bm{Y}_{[M-1]})\label{eq:con_PCSI_6}\\
    &\leq (M-1)L,
\end{align}
where \eqref{eq:con_PCSI_5} follows from the fact that $\bm{Y}_{[M-1]}$ are linear combinations of $\bm{W}_{[M]}$. Step \eqref{eq:con_PCSI_6} holds because of  \eqref{eq:con_PCSI_4}, and because conditioning reduces entropy.

Thus $D \geq H(\bm{\Delta} \mid \bm{Q}) \geq H(\bm{W}_{[K]}) - H(\bm{W}_{[K]} \mid \bm{\Delta}, \bm{Q}) \geq (K-M+1)L$, which implies that $C_{\mbox{\tiny PCSI}}^{\sup}  \leq (K-M+1)^{-1}$ for $2 \leq M \leq K$.

\textbf{Regime 2}: $M=1$.

Consider any particular realization $Q \in \mathcal{Q}$ of $\bm{Q}$. Since $M=1$, $\bm\Lambda$ is irrelevant, e.g., we may assume $\bm{\Lambda}=\Lambda=1$ without loss of generality. For all $j \in [2:K]$, consider $\mathcal{S} = \{1\}, \theta = j$, and apply \eqref{eq:lemma1pcsi} according to Lemma \ref{lem:privacy} so that 
\begin{align}
    H(\bm{W}_j \mid \bm{\Delta}, \bm{Y}^{[\{1\},1]}, \bm{Q}=Q) = 0\\
\implies        H(\bm{W}_{[2:K]} \mid \bm{\Delta}, \bm{Y}^{[\{1\},1]}, \bm{Q}=Q) = 0\label{eq:dec_PCSI_corner}
\end{align}
\begin{align}
    &H(\bm{W}_{[K]} \mid \bm{\Delta}, \bm{Q}=Q)\\
    &{\leq H(\bm{W}_{1}, \bm{W}_{[2:K]}, \bm{Y}^{[\{1\}]} \mid \bm{\Delta}, \bm{Q}=Q)}\\
    &= H(\bm{W}_1, \bm{Y}^{[\{1\},1]} \mid \bm{\Delta}, \bm{Q} = Q)\label{eq:corner_PCSI_1}\\
    &= H(\bm{W}_1 \mid \bm{\Delta}, \bm{Q}=Q)\label{eq:corner_PCSI_2}\\
    &\leq L,
\end{align}
where \eqref{eq:corner_PCSI_1} holds since \eqref{eq:dec_PCSI_corner} holds, and \eqref{eq:corner_PCSI_2} is true as $\bm{Y}^{[\{1\},1]}$ is simply $\bm{W}_1$. Averaging over $\bm{Q}$, $H(\bm{W}_{[K]} \mid \bm{\Delta}, \bm{Q}) \leq L$. Thus $D \geq H(\bm{\Delta} \mid \bm{Q}) \geq H(\bm{W}_{[K]}) - H(\bm{W}_{[K]} \mid \bm{\Delta}, \bm{Q}) \geq KL-L$, which implies that $C_{\mbox{\tiny PCSI}}(q) \leq (K-1)^{-1}$ for $M=1$.

\subsection{Achievability}
For $2 \leq M \leq K$, the achievable scheme will be a combination of \emph{Specialized GRS Codes} and \emph{Modified Specialized GRS Codes} which are schemes in \cite{PIR_PCSI} for PIR-PCSI-I and PIR-PCSI-II setting, respectively.

The rate $(K-M)^{-1}$ is achievable by \emph{Specialized GRS Codes} for PIR-PCSI-I setting and the rate $(K-M+1)^{-1}$ is achievable by \emph{Modified Specialized GRS Codes} for the PIR-PCSI-II setting. Both  schemes work for $L=1$, so let us say $L=1$ here. Intuitively, these two achievable schemes have the same structures as explained below. 

For the PIR-PCSI-I setting,  the desired message is not contained in the support set. The download will be $K-M$ linear equations of $K$ unknowns ($K$ messages). These $K-M$ linear equations are independent by design, so they allow the user to eliminate any $K-M-1$ unknowns and get an equation in the remaining $K-(K-M-1) = M+1$ unknowns (messages). Let these $M+1$ unknowns be the $M$ messages in the support set and the desired message. With careful design, the equation will be equal to $\bm{Y}^{[\bm{\mathcal{S}},\bm{\Lambda}]} + \bm{\lambda}^{\prime}\bm{W}_{\bm{\theta}}$ for some non-zero $\bm{\lambda}^\prime$. Thus by subtracting CSI from the equation the user is able to recover $\bm{W}_{\bm{\theta}}$.

For the PIR-PCSI-II setting the desired message is contained in the support set. The download will be $K-M+1$ linear equations in $K$ unknowns (messages). These $K-M+1$ linear equations are independent by design, so they allow the user to eliminate any $K-M$ unknowns and get an equation in the remaining $K-(K-M) = M$ unknowns (messages). Let these $M$ unknowns be the $M$ messages in the support set. With careful design, the equation will be equal to $\bm{Y}^{[\bm{\mathcal{S}},\bm{\Lambda}]} + \bm{\lambda}^{\prime}\bm{W}_{\bm{\theta}}$ for some $\bm{\lambda}^{\prime} \neq 0$. Thus by subtracting CSI from the equation the user is able to recover $\bm{W}_{\bm{\theta}}$.

Consider a scheme where the user applies \emph{Specialized GRS Codes} when $\bm{\theta} \notin \bm{\mathcal{S}}$ and applies \emph{Modified Specialized GRS Codes} when $\bm{\theta} \in \bm{\mathcal{S}}$. This scheme is obviously correct but not private because the server can tell if $\bm{\theta} \in \bm{\mathcal{S}}$ or not from the download cost since the download cost of the two schemes are different. However, if the user always downloads one more redundant equation when applying \emph{Specialized GRS Codes}, then there is no difference in the download cost. This is essentially the idea for the achievable scheme.

Let us first present the \emph{Specialized GRS Codes} in \cite{PIR_PCSI} here for ease of understanding. There are $K$ distinct evaluation points in $\mathbb{F}_{q}$, namely $\omega_{1}, \cdots, \omega_{K}$. A polynomial $\bm{p}(x)$ is constructed as 
\begin{align}
    \bm{p}(x) &\triangleq \prod_{k \in [K]\setminus(\bm{\mathcal{S}} \cup \{\bm{\theta}\})}(x - \omega_{k})\\
    & = \sum_{i=1}^{K-M}\bm{p}_i x^{i-1}.\label{eq:polyGRS}
\end{align}
The query $\bm{Q}$ is comprised of $K-M$ row vectors, each $1\times K$, namely $\bm{Q}_{1}, \cdots, \bm{Q}_{K-M}$ such that 
\begin{align}
    \bm{Q}_i = [\bm{v}_1\omega_{1}^{i-1}~~ \cdots~~ \bm{v}_K\omega_{K}^{i-1}], \forall i \in [K-M],
\end{align}
where for $\bm{i}_m \in \bm{\mathcal{S}}, m \in [M]$, $\bm{v}_{\bm{i}_m} = \frac{\bm{\lambda}_m}{p(\omega_{\bm{i}_m})}$ ($\bm{\lambda}_m$ is the $m^{th}$ coefficient in the CSI), while for $k \notin \bm{\mathcal{S}}$, $\bm{v}_{k}$ is randomly drawn from $\mathbb{F}_{q}^{\times}$. Upon receiving $\bm{Q}$, the server sends  
\begin{align}
    \bm{\Delta} = 
    \begin{bmatrix}
        \bm{\Delta}_1\\
        \vdots\\
        \bm{\Delta}_{K-M}
    \end{bmatrix}
    =
    \begin{bmatrix}
        \bm{Q}_1\\
        \vdots\\
        \bm{Q}_{K-M}
    \end{bmatrix}
    \begin{bmatrix}
        \bm{W}_1\\
        \bm{W}_2\\
        \vdots\\
        \bm{W}_K
    \end{bmatrix}
\end{align}
to the user. Let us call $[\bm{Q}_1^{\mathrm{T}} ~ \cdots ~ \bm{Q}_{K-M}^{\mathrm{T}}]^{\mathrm{T}}$ the \emph{Specialized GRS Matrix} and $[\bm{\Delta}_1 ~ \cdots ~ \bm{\Delta}_{K-M}]^{\mathrm{T}}$  \emph{Specialized GRS Codes} of $\bm{W}_{[K]}$ for ease of reference. Note that the \emph{Specialized GRS Matrix} is uniquely defined by $\bm{v}_{1}, \cdots, \bm{v}_{K}$ as $\omega$'s are constants.

The user gets $\bm{W}_{\bm{\theta}}$ by subtracting $\bm{Y}^{[\bm{\mathcal{S}}, \bm{\Lambda}]}$ from 
\begin{align}
    \sum_{i=1}^{K-M}\bm{p}_i\bm{\Delta}_{i} = \bm{Y}^{[\bm{\mathcal{S}}, \bm{\Lambda}]} + \bm{v}_{\bm{\theta}}\bm{p}(\omega_{\bm{\theta}})\bm{W}_{\bm{\theta}}.
\end{align}

Our PIR-PCSI scheme is as follows.
For any realization $(\theta, \mathcal{S})$ of $(\bm{\theta}, \bm{\mathcal{S}})$, 
\emph{1)} When $\theta \in [K]\setminus\mathcal{S}$, first apply the Specialized GRS Codes in \cite{PIR_PCSI}. Besides $Q_1, Q_2, \cdots, Q_{K-M}$ as specified in the \emph{Specialized GRS Codes} of \cite{PIR_PCSI}, the user also has 
\begin{align}
    Q_{K-M+1} = [v_1\omega_{1}^{K-M}, \cdots, v_K\omega_{K}^{K-M}]
\end{align}
as part of the query. And the answer $\bm{\Delta}_{K-M+1} = \sum_{j=1}^{K}v_j \omega_j^{K-M} \bm{W}_j$ will be generated for $Q_{K-M+1}$ and downloaded by the user as a redundant equation. Note that the matrix $[Q_1^{\mathrm{T}}, Q_2^{\mathrm{T}}, \cdots, Q_{K-M+1}^{\mathrm{T}}]^{\mathrm{T}}$ is the generator matrix of a $(K, K-M+1)$ GRS code \cite{Coding_Theory}.

\emph{2)} When $\theta \in \mathcal{S}$, the user will directly apply \emph{Modified Specialized GRS Codes} where the queries also form a generator matrix of a $(K, K-M+1)$ GRS code as specified in \cite{PIR_PCSI}.

Such a scheme is private since the queries in both cases form a generator matrix of a $(K,K-M+1)$ GRS code, and the $v_1, \cdots, v_{K}$ in both cases are identically uniform over $\mathbb{F}_{q}^{\times}$ for any realization of $\bm{\theta}, \bm{\mathcal{S}}$.

For the corner case $M=1$, it suffices to download $K-1$ generic linear combinations of all the $K$ messages such that from the $K-1$ downloaded linear combinations and the CSI, all the $K$ messages are decodable as noted in Remark \ref{rmk:pcsi1_inf_pcsi_inf}.

\section{Proof of Theorem \ref{thm:redundancy}}\label{proof:redundancy}
Here we bound the redundancy $\rho_{\mbox{\tiny PCSI}}$ from above (equivalently, lower-bound $\alpha^{*}$) for $1 \leq M \leq K$. For $\frac{K+2}{2} < M \leq K$, the proof for $\rho_{\mbox{\tiny PCSI}} = 0$ is the same as in Section \ref{proof:red} show that  so it will not be repeated.

Consider an achievable scheme such that $\alpha$-CSI is sufficient and the average download cost, $D \leq \frac{1}{C_{\mbox{\tiny PCSI}}^{\sup}}L+\epsilon L$ for some $L$. Note that $D\geq H(\bm{\Delta\mid \bm{Q}})$, therefore,
\begin{align}
H(\bm{\Delta\mid \bm{Q}})\leq \frac{1}{C_{\mbox{\tiny PCSI}}^{\sup}}L+\epsilon L\label{eq:deltabound}
\end{align}

It follows from \eqref{eq:deltabound} that there exists a feasible $Q \in \mathcal{Q}$ such that 
\begin{align}
    H(\bm{\Delta} \mid \bm{Q}=Q) \leq \frac{1}{C_{\mbox{\tiny PCSI}}^{\sup}}L+\epsilon L.\label{eq:deltabound_Q}
\end{align}
For all $i \in [K]$, let $\Lambda_{i} \in \mathfrak{C}$ satisfy 
\begin{align}
    H(\bm{W}_{i} \mid \bm{\Delta}, \overline{\bm{Y}}^{[[M], \Lambda_i]}, \bm{Q} = Q) = 0.\label{eq:red_M1_dec}
\end{align}
The argument that such $\Lambda_i$'s must exist is identical to the proof of Lemma \ref{lem:privacy}. 
Writing $\overline{\bm{Y}}^{[[M], \Lambda_i]}$ as $\overline{\bm{Y}}_{i}$ for compact notation,
\begin{align}
    &H(\bm{W}_{[K]} \mid \bm{\Delta}, \overline{\bm{Y}}_{[M]}, \bm{Q} = Q)\\
    &= H(\bm{W}_{[M]} \mid \bm{\Delta}, \overline{\bm{Y}}_{M}, \bm{Q} = Q)\notag\\
    &~~~+ H(\bm{W}_{[M+1:K]} \mid \bm{\Delta}, \overline{\bm{Y}}_{[M]}, \bm{W}_{[M]}, \bm{Q} = Q)\\
    &= 0 + H(\bm{W}_{[K]} \mid \bm{\Delta}, \bm{W}_{[M]}, \overline{\bm{Y}}_{[K]}, \bm{Q} = Q)\label{eq:red_M1_funcW}\\
    &=0.
\end{align}
where \eqref{eq:red_M1_funcW} follows from \eqref{eq:red_M1_dec} and the fact that $\overline{\bm{Y}}_{[K]}$ are functions of $\bm{W}_{[M]}$. The last step also follows from \eqref{eq:red_M1_dec}. Thus,
\begin{align}
    &\frac{1}{C_{\mbox{\tiny PCSI}}^{\sup}}L + \epsilon L + M\alpha L\notag\\
    &\geq H(\bm{\Delta} \mid \bm{Q}=Q) + H(\overline{\bm{Y}}_{[M]} \mid \bm{Q}=Q)\label{eq:red_M1_indY}\\
    &\geq H(\bm{\Delta}, \overline{\bm{Y}}_{[M]} \mid \bm{Q}=Q)\\
    &\geq I(\bm{\Delta}, \overline{\bm{Y}}_{[M]}; \bm{W}_{[K]} \mid \bm{Q}=Q)\\
    &=H(\bm{W}_{[K]} \mid \bm{Q}=Q) = KL.\label{eq:red_M1_indW}
\end{align}
\eqref{eq:red_M1_indY} is true because \eqref{eq:deltabound_Q}, \eqref{eq:invaYR} hold. Step \eqref{eq:red_M1_indW} follows from \eqref{eq:red_M1_dec} and the fact that the query and messages are mutually independent according to \eqref{eq:indQ}. Thus, $\alpha \geq (K-\frac{1}{C_{\mbox{\tiny PCSI}}^{\sup}})/M - \epsilon/M$. In order to achieve capacity, we must have $\epsilon \rightarrow 0$, so we must have $\alpha \geq (K-\frac{1}{C_{\mbox{\tiny PCSI}}^{\sup}})/M$, for all $1\leq M\leq K$.

Now note that for $M=1$, since $C_{\mbox{\tiny PCSI}}^{\sup} = (K-1)^{-1}$, we have shown that $\alpha \geq 1$, which implies $\rho_{\mbox{\tiny PCSI}} = 0$ in this case. 

For $2 \leq M \leq \frac{K+2}{2}$, since $C_{\mbox{\tiny PCSI}}^{\sup} = (K-M+1)^{-1}$, we have shown that $\alpha \geq \frac{M-1}{M}$, which implies $\rho_{\mbox{\tiny PCSI}} \leq \frac{1}{M}$ in this case.

It only remains to show that for $M=2$,  $\rho_{\mbox{\tiny PCSI}} = \frac{1}{2}$ is achievable, or equivalently, $\alpha^{*} = \frac{1}{2}$. For this case, let us present a PIR-PCSI scheme that achieves the rate $(K-M/2)^{-1}$ for arbitrary $1 \leq M \leq K$. Note that $K-M/2 = K-M+1$ when $M=2$, which is the only case where the supremum capacity is achieved by this scheme. The rate of this scheme is strictly smaller than $C_{\mbox{\tiny PCSI}}^{\sup}$ for other $M \neq 2$.

Let the size of the base field $q$ be an even power of a prime number such that $\sqrt{q}$ is a prime power and $\sqrt{q} \geq K$. For arbitrary realization $(\theta, \mathcal{S}) \in [K]\times\mathfrak{S}$ of $(\bm{\theta},\bm{\mathcal{S}})$, if $\theta \in \mathcal{S}$, the user can apply the \emph{Interference Alignment} based PIR-PCSI-II scheme where half of each message is downloaded. If $\theta \in [K]\setminus\mathcal{S}$, then user can apply the \emph{Specialized GRS Codes} based scheme for the halves of the messages corresponding to the CSI  dimension that is retained (while the other half of the CSI dimensions is discarded as redundant) and download the other half dimension of all the messages directly. Note that in both cases, a half-dimension of each of the $K$ messages is directly downloaded. The other halves are involved in the download corresponding to the \emph{Specialized GRS Codes} which is not needed for decodability/correctness if $\theta \in \mathcal{S}$, but is still included for privacy, i.e., to hide whether or not $\bm\theta\in\bm{\mathcal{S}}$. The download cost required is $K\left(\frac{L}{2}\right)$ for the direct downloads of half of every message, plus  $(K-M)\frac{L}{2}$ for the \emph{Specialized GRS Codes} based scheme that usually requires $K-M$ downloads per message symbol, but is applied here to only half the symbols from each message, for a total download cost of $(K-M/2)L$ which achieves the supremum capacity of PIR-PCSI for $M=2$. The details of the scheme are presented next.

For all $k \in [K]$, let $V_{\bm{W}_k} \in \mathbb{F}_{\sqrt{q}}^{2\times 1}$ be the length $2$ vector representation of $\bm{W}_k \in \mathbb{F}_{q}$. For all $m \in [M]$, let $M_{\bm{\lambda}_m} \in \mathbb{F}_{\sqrt{q}}^{2\times 2}$ be the matrix representation of $\bm{\lambda}_m \in \mathbb{F}_{q}^{\times}$ where $\bm{\lambda}_m$ is the $m^{th}$ entry of the coefficient vector $\bm{\Lambda}$. Let 
\begin{align}
    \overline{\bm{Y}}^{[\bm{\mathcal{S}}, \bm{\Lambda}]} = M_{\bm{\lambda}_1}(1,:)V_{\bm{W}_{\bm{i}_1}} +\cdots+ M_{\bm{\lambda}_M}(1,:)V_{\bm{W}_{\bm{i}_M}},
\end{align}
where $\bm{\mathcal{S}} = \{\bm{i}_1, \bm{i}_2, \cdots, \bm{i}_{M}\}$ is the support index set, be the processed CSI where $H(\overline{\bm{Y}}^{[\bm{\mathcal{S}}, \bm{\Lambda}]}) = \frac{1}{2}H(\bm{W}_k)$. Note that $\forall m \in [M], M_{\bm{\lambda}_m}(1,:)$ is uniform over $\mathbb{F}_{\sqrt{q}}^{1\times 2} \setminus \{[0~~0]\}$ according to Lemma \ref{lem:uniform12}.

The query $\bm{Q} = \{\bm{Q}_1, \bm{Q}_2, \bm{Q}_3\}$,
\begin{align}
    \bm{Q}_1 &= \{\mathbf{L}_1, \mathbf{L}_2, \cdots, \mathbf{L}_{K}\},\\
    \bm{Q}_2 &= \{\mathbf{L}_1^{\prime}, \mathbf{L}_2^{\prime}, \cdots, \mathbf{L}_{K}^{\prime}\},\\
    \bm{Q}_3 &= \{\bm{v}_1, \bm{v}_2, \cdots, \bm{v}_{K}\}.
\end{align}
where $\mathbf{L}_{k}, \mathbf{L}_{k}^{\prime} \in \mathbb{F}_{\sqrt{q}}^{1\times 2} \setminus \{[0~~0]\}$. $\mathbf{L}_{k}, \mathbf{L}_{k}^{\prime}$ serve as two linearly independent projections that ask the server to split $\bm{W}_k$ into two halves 
\begin{align}
    \bm{w}_{k}(1) = \mathbf{L}_{k}V_{\bm{W}_k} \in \mathbb{F}_{\sqrt{q}},\\
    \bm{w}_{k}(2) = \mathbf{L}_{k}^{\prime}V_{\bm{W}_k} \in \mathbb{F}_{\sqrt{q}}.
\end{align}
$\bm{Q}_3$ uniquely defines a \emph{Specialized GRS Matrix} whose elements are in $\mathbb{F}_{\sqrt{q}}$.

The user will download the first halves of all the $K$ messages after projection, i.e., $\bm{w}_{[K]}(1)$ and apply the \emph{Specialized GRS Matrix} to download a \emph{Specialized GRS Codes} of the second halves of all the $K$ messages after projection, i.e., $\bm{w}_{[K]}(2)$.  

Let us specify $\mathbf{L}_{k}, \mathbf{L}_{k}^{\prime}, \bm{v}_{k}$. Consider any realization $(\theta, \mathcal{S}) \in [K]\times\mathfrak{S}$ of $(\bm{\theta},\bm{\mathcal{S}})$. Let us say $\mathcal{S} = \{i_1, i_2, \cdots, i_M\}$. For the messages not involved in the CSI, they are randomly projected to two linearly independent directions, i.e., for any $k \in [K] \setminus \mathcal{S}$, $\mathbf{L}_{k}, \mathbf{L}_{k}^{\prime}$ are linearly independent and are randomly drawn from $\mathbb{F}_{\sqrt{q}}^{1 \times 2} \setminus \{[0~~0]\}$. Also, for any $k \in [K] \setminus \mathcal{S}$, $\bm{v}_{k}$ is uniformly distributed in $\mathbb{F}_{\sqrt{q}}^{\times}$. 

For messages involved in the CSI, the construction of projections and $\bm{v}$'s depends on whether $\theta$ is in $\mathcal{S}$ or not.
\begin{enumerate}
    \item When $\theta \in \mathcal{S}$, for any $m \in [M]$,
    \begin{align}
        \mathbf{L}_{i_m} = 
        \begin{cases}
            M_{\bm{\lambda}_m}(2,:), i_m = \theta,\\
            M_{\bm{\lambda}_m}(1,:), i_m \neq \theta.
        \end{cases}
    \end{align}
    $\mathbf{L}_{i_m}^{\prime}$ is then chosen randomly from $\mathbb{F}_{\sqrt{q}}^{1 \times 2} \setminus \{[0~~0]\}$ such that it is linearly independent with $\mathbf{L}_{i_m}$. Meanwhile, $\bm{v}_{i_m}$ is randomly drawn from $\mathbb{F}_{\sqrt{q}}^{\times}$. Under this case, the user has 
    \begin{align}
        \overline{\bm{Y}}^{[\mathcal{S}, \bm{\Lambda}]} = \sum_{i_m \in \mathcal{S}\setminus\{\theta\}}\bm{w}_{i_m}(1) + \bm{w}_{\theta}(2)
    \end{align} 
    according to the construction of $\mathbf{L}_{i_m}$. $\bm{w}_{\theta}(1)$ is directly downloaded and $\bm{w}_{\theta}(2)$ can be recovered by subtracting $\{\bm{w}_{i_m}(1)\}_{i_m \neq \theta}$ from $\overline{\bm{Y}}^{[\mathcal{S}, \bm{\Lambda}]}$. The user is then able to recover $\bm{W}_{\theta}$ as the two projections are linearly independent. $\bm{Q}_3$ uniquely defines a \emph{Specialized GRS Matrix} and applying $\bm{Q}_3$ to download a \emph{Specialized GRS Codes} of $\bm{w}_{[K]}(2)$ is just for privacy.

    \item When $\theta \in [K]\setminus\mathcal{S}$, for any $m \in [M]$, 
    \begin{align}
        \mathbf{L}_{i_m}^{\prime} = \frac{1}{\bm{a}_m}M_{\bm{\lambda}_m}(1,:),
    \end{align}
    where $\bm{a}_m$ is randomly drawn from $\mathbb{F}_{\sqrt{q}}^{\times}$. $\mathbf{L}_{i_m}$ is then chosen randomly from $\mathbb{F}_{\sqrt{q}}^{1 \times 2} \setminus \{[0~~0]\}$ such that they are linearly independent with $\mathbf{L}_{i_m}^{\prime}$. Under this case, the user has 
    \begin{align}
        \sum_{m\in[M]}\bm{a}_{m}\bm{w}_{i_m}(2) = \overline{\bm{Y}}^{[\mathcal{S}, \bm{\Lambda}]},
    \end{align}
    and sets 
    \begin{align}
        \bm{v}_{i_m} = \frac{\bm{a}_m}{p(\omega_{i_m})}, \forall m \in [M],
    \end{align}
    where $p(\omega_{i_m})$ is the evaluation of the polynomial specified in \eqref{eq:polyGRS} (when $(\bm{\theta},\bm{\mathcal{S}}) = (\theta$, $\mathcal{S})$) at $\omega_{i_m}$, which is a non-zero constant given $(\theta, \mathcal{S})$. Thus, given $(\theta, \mathcal{S})$, $\bm{v}_{i_m}$ is still uniform over $\mathbb{F}_{\sqrt{q}}^{\times}$. $\bm{Q}_{3}$ uniquely defines a \emph{Specialized GRS Matrix}. Applying $\bm{Q}_3$ to download a \emph{Specialized GRS Codes} of $\bm{w}_{[K]}(2)$, together with $\sum_{m\in[M]}\bm{a}_{m}\bm{w}_{i_m}(2)$ as the side information, enable the user to recover $\bm{w}_{\theta}(2)$. Since the first halves of all the projected messages are also downloaded, the user also has $\bm{w}_{\theta}(1)$, thus, is able to decode $\bm{W}_{\theta}$.
\end{enumerate}

Note that for arbitrary realization $(\theta, \mathcal{S})$ of $(\bm{\theta}, \bm{\mathcal{S}})$, no matter $\theta \in \mathcal{S}$ or not, $\mathbf{L}_1, \cdots, \mathbf{L}_{K}$, $\mathbf{L}_1^{\prime}, \cdots, \mathbf{L}_{K}^{\prime}$, $\bm{v}_1, \cdots, \bm{v}_{K}$ are independent, and for any $k \in [K]$, the matrix whose first row is $\mathbf{L}_{k}$ and second row is $\mathbf{L}_{k}^{\prime}$ is uniform over the set that contains all the full-rank matrix in $\mathbb{F}_{\sqrt{q}}^{2\times 2}$, $\bm{v}_{k}$ is uniform over $\mathbb{F}_{\sqrt{q}}^{\times}$. Thus, the scheme is private.

\section{Proof of Theorem \ref{thm:cap_PCSI_inf}}\label{sec:cap_PCSI_inf}
The rate $\frac{1}{K-1}$ PIR-PCSI-I scheme in Section \ref{sec:PCSI1_inf_ach} is also the infimum capacity achieving PIR-PCSI scheme as noted in Remark \ref{rmk:pcsi1_inf_pcsi_inf}, so we just prove the converse here.

As a result of \eqref{eq:lemma1pcsi} and the fact that in $\mathbb{F}_{2}$, we can only have $\bm{\Lambda} = 1_M$, i.e., the length-$M$ vector all of whose elements are equal to $1$, we have
\begin{align}
    H(\bm{W}_{[K]} \mid \bm{\Delta}, \bm{Y}^{[\mathcal{S},1_M]}, \bm{Q}=Q) = 0, \notag\\
    \forall (Q,\mathcal{S}) \in \mathcal{Q}\times\mathfrak{S}. \label{eq:pcsi_inf_dec}
\end{align}
Writing $\bm{Y}^{[[M],1_M]}$ as $\bm{Y}$ for compact notation, for any $Q \in \mathcal{Q}$, we have
\begin{align}
    &H(\bm{W}_{[K]} \mid \bm{\Delta}, \bm{Q}=Q)\notag\\
    &= H(\bm{W}_{[K]}, \bm{Y} \mid \bm{\Delta}, \bm{Q}=Q) \label{eq:dec_K_1}\\
    &= H(\bm{Y} \mid \bm{\Delta}, \bm{Q}=Q) + H(\bm{W}_{[K]} \mid \bm{\Delta}, \bm{Y}, \bm{Q} = Q)\label{eq:dec_K_2}\\
    &\leq H(\bm{Y}) = L.
\end{align}
\eqref{eq:dec_K_1} is true since $\bm{Y}$ is a summation of the first $M$ messages, and \eqref{eq:dec_K_2} follows from \eqref{eq:pcsi_inf_dec}. Averaging over $\bm{Q}$ we have,
\begin{align}
    H(\bm{W}_{[K]} \mid \bm{\Delta}, \bm{Q}) \leq L.
\end{align}
Thus, $D \geq H(\bm{\Delta} \mid \bm{Q}) \geq I(\bm{\Delta}; \bm{W}_{[K]} \mid \bm{Q}) = H(\bm{W}_{[K]}) - H(\bm{W}_{[K]} \mid \bm{\Delta}, \bm{Q}) \geq KL-L$ which implies that $C_{\mbox{\tiny PCSI}}^{\inf}(q = 2) \leq (K-1)^{-1}$.

\section{Proof of Theorem \ref{thm:pcsi_pub_pri}}\label{proof:pcsi_pub_pri}
The rate $\frac{1}{K-1}$ PIR-PCSI-I scheme which preserves $(\bm{\theta}, \bm{\mathcal{S}}, \bm{\Lambda})$ in Section \ref{sec:pcsi1_pri_ach} is also the  capacity achieving PIR-PCSI scheme with private coefficients as noted in Remark \ref{rmk:pcsi1_pri_pcsi_pri}, so we just prove the converse here. Specifically, we prove that $C_{\mbox{\tiny PCSI}}^{\mbox{\tiny pri}}(q)\leq C_{\mbox{\tiny PCSI}}(q=2) = C_{\mbox{\tiny PCSI}}^{\inf}$.

According to \eqref{eq:pcsi_pri} in Lemma \ref{lem:fullypri}, for a fully private PIR-PCSI scheme,
\begin{align}
    H(\bm{W}_{[K]} \mid \bm{\Delta}, \bm{Y}^{[\mathcal{S}, \Lambda]}, \bm{Q} = Q) = 0, \notag\\
    \forall (Q,\mathcal{S},\Lambda) \in \mathcal{Q} \times \mathfrak{S} \times \mathfrak{C}.\label{eq:pcsi_pri_dec}
\end{align}
Note that \eqref{eq:pcsi_pri_dec} is a \emph{stronger} version of \eqref{eq:pcsi_inf_dec} which is sufficient to bound $C_{\mbox{\tiny PCSI}}(q=2) = C_{\mbox{\tiny PCSI}}^{\inf}$. Thus, $C_{\mbox{\tiny PCSI}}^{\mbox{\tiny pri}}(q) \leq C_{\mbox{\tiny PCSI}}^{\inf}$.

\section{Conclusion} \label{sec:con}
Side-information is a highly valuable resource for PIR in general, and for single-server PIR in particular. Building on the foundation laid by Heidarzadeh et al. in \cite{PIR_PCSI}, this work  presents a more complete picture, as encapsulated in Table \ref{tab:capacity}, revealing new insights that are described in the introduction. The redundancy of side-information is particularly noteworthy, because it allows the user to save storage cost, which may be used to store additional non-redundant side-information, e.g., multiple linear combinations instead of just one, as assumed in this work and in \cite{PIR_PCSI}. An interesting direction for future work is to understand the trade-off between the size of side information and the efficiency of single-server PIR, e.g., by characterizing the $\alpha$-CSI constrained capacity of PIR-PCSI-I, PIR-PCSI-II, PIR-PCSI. Other questions that remain open include  issues that are field-specific. For example, is the supremum capacity of PIR-PCSI-II for $M>2$ achievable for all fields except $\mathbb{F}_2$?  Are there other fields besides $\mathbb{F}_2$ over which the capacity is equal to the infimum capacity? Can the capacity over certain fields take values other than the supremum and infimum capacities? Progress on these issues may require field-dependent constructions of interference alignment schemes for achievability, and combinatorial arguments for converse bounds, both of which may be of broader interest.

\appendix

\subsection{Proof of Lemma \ref{lem:existence}}\label{app:existence}
For all $k \in [K], m \in [M-1]$, let us say 
\begin{align}
    \mathbf{L}_{k}^{(m)} = 
    \begin{bmatrix}
        x_{k,1}^{(m)} & x_{k,2}^{(m)} & \cdots & x_{k,M}^{(m)}
    \end{bmatrix},
\end{align}
where $x_{\cdot,\cdot}^{(\cdot)} \in \mathbb{F}_{q^l}$.
Let 
\begin{align}
    \mathbf{H}_{k} = 
    \begin{bmatrix}
        {\mathbf{L}_{k}^{(1)}}^{\mathrm{T}} & {\mathbf{L}_{k}^{(2)}}^{\mathrm{T}} & \cdots & {\mathbf{L}_{k}^{(M-1)}}^{\mathrm{T}}
    \end{bmatrix}^{\mathrm{T}}.
\end{align}

Let us denote by $\mathcal{S}_1, \mathcal{S}_2, \cdots \mathcal{S}_{\tbinom{K}{M}}$, the $\tbinom{K}{M}$ distinct elements of $\mathfrak{S}$. Let $\mathcal{S}_1 = [M]$.
Then $\mathbf{G}_{\mathcal{S}_1}$ can be written as
\begin{align}
    \mathbf{G}_{\mathcal{S}_1} = 
    \begin{bmatrix}
        \lambda_{1}\mathbf{I}_{M} & \lambda_{2}\mathbf{I}_{M} & \cdots & \lambda_{M}\mathbf{I}_{M}\\
        \mathbf{H}_{1} & \mathbf{0}_{(M-1)\times M} & \cdots & \mathbf{0}_{(M-1)\times M}\\
        \mathbf{0}_{(M-1)\times M} & \mathbf{H}_{2} & \cdots & \mathbf{0}_{(M-1)\times M}\\
        \vdots &  & \ddots & \\
        \mathbf{0}_{(M-1)\times M} & \mathbf{0}_{(M-1)\times M} & \cdots & \mathbf{H}_{M}
    \end{bmatrix}\label{eq:G_stru}
\end{align}
which is an $M^2 \times M^2$ matrix. Note that 
\begin{align}
    \det(\mathbf{G}_{\mathcal{S}_1}) = f_{1}(x_{1,1}^{(1)}, \cdots, x_{M,M}^{(M-1)}),
\end{align}
where $f_{1}(\cdot)$ is an $M^{2}(M-1)$-variate polynomial with degree $\deg(f_1) = M(M-1)$. To verify that $f_{1}(\cdot)$ is not the zero polynomial, note that if each $\mathbf{H}_m, m\in[M]$ is chosen as the $(M-1)\times M$ matrix obtained by inserting the all-zero column  into the $(M-1)\times (M-1)$ identity matrix after its first $m-1$ columns, then $\det(\mathbf{G}_{\mathcal{S}_1})=\lambda_1\lambda_2\cdots\lambda_M\neq 0$.

Similarly, $\forall j \in [2:\tbinom{K}{M}]$,
\begin{align}
    \det(\mathbf{G}_{\mathcal{S}_j}) = f_{j}\Big(\big(x_{k,1}^{(m)}, \cdots x_{k,M}^{(m)}\big)_{k \in \mathcal{S}_j, m \in [M-1]}\Big),
\end{align}
where $f_{j}(\cdot)$ is an $M^{2}(M-1)$-variate polynomial with degree $\deg(f_j) = M(M-1)$.

Now, to satisfy the correctness and privacy constraints, we must choose all $ \mathbf{L}_{k}^{(m)}$ to simultaneously have all the polynomials $f_j(\cdot)$ evaluate to non-zero values. Equivalently, the polynomial $f$ that is the product of all  $f_j(\cdot)$ should evaluate to a non-zero value.
\begin{align}
    \prod_{j \in [\tbinom{K}{M}]}\det(\mathbf{G}_{\mathcal{S}_j}) = \prod_{j \in [\tbinom{K}{M}]}f_{j} = f \neq 0,
\end{align}
where $f$ is a $KM(M-1)$-variate polynomial with degree 
\begin{align}
    \deg(f) = \prod_{j \in [\tbinom{K}{M}]} \deg(f_j) = \tbinom{K}{M}M(M-1).
\end{align}
Now, since it is a product of non-zero polynomials, $f$ is also a non-zero polynomial. Therefore, by Schwartz-Zippel Lemma, if the values of the $KM(M-1)$ variables are randomly chosen from $\mathbb{F}_{q^l}$, then the probability of the polynomial $f$ evaluating to $0$ is bounded as,
\begin{align}
    \text{Pr}(f = 0) \leq \frac{\deg(f)}{q^l} = \frac{\tbinom{K}{M}M(M-1)}{q^l}.
\end{align} 
Therefore, if $q^l > \tbinom{K}{M}M(M-1)$, then $\text{Pr}(f = 0) < 1$, which implies that there exists a choice of the $KM(M-1)$ variables such that $f \neq 0$. That choice satisfies the condition of Lemma \ref{lem:existence}, thus completing the proof of  Lemma \ref{lem:existence}.

For ease of understanding, consider the following example.
\begin{example}
    Consider $M=3$, $K=4$ messages: $\bm{A},\bm{B},\bm{C},\bm{D}$, each of which consists $Ml = 3l$ symbols in $\mathbb{F}_3$. Message $\bm{A}$ can be represented as a length $M=3$ column vector with all the $3$ entries in $\mathbb{F}_{3^l}$, i.e., $V_{\bm{A}} \in \mathbb{F}_{3^l}^{3 \times 1}$. $V_{\bm{B}},V_{\bm{C}},V_{\bm{D}}$ are similarly defined.

    Let us say $\bm{A},\bm{B},\bm{C}$ are in the support set and $\bm{Y} = 2\bm{A}+\bm{B}+\bm{C}$. $\bm{Y}$ can also be represented by $V_{\bm{Y}} \in \mathbb{F}_{3^l}^{3 \times 1}$ where 
    \begin{align}
        V_{\bm{Y}} = 2\mathbf{I}_{3}V_{\bm{A}} + \mathbf{I}_{3}V_{\bm{B}} + \mathbf{I}_{3}V_{\bm{C}}.
    \end{align}
    For each one of  $V_{\bm{A}},V_{\bm{B}},V_{\bm{C}},V_{\bm{D}}$, the user will download $M-1 = 2$ linear combinations. For example, the download corresponding to $V_{\bm{A}}$ is,
    \begin{align}
        \bm{\Delta}_{\bm{A}} = 
        \begin{bmatrix}
            \mathbf{L}_{1}^{(1)} \\
            \mathbf{L}_{1}^{(2)}
        \end{bmatrix}V_{\bm{A}} = \mathbf{H_{1}}V_{\bm{A}},
    \end{align}
    where $\mathbf{L}_{1}^{(1)}, \mathbf{L}_{1}^{(2)} \in \mathbb{F}_{3^l}^{1\times 3}$, and $\bm{\Delta}_{\bm{A}} \in \mathbb{F}_{3^l}^{2\times 1}$.
    Similarly, the user downloads 
    \begin{align}
        \bm{\Delta}_{\bm{B}} = \mathbf{H_{2}}V_{\bm{B}}, \bm{\Delta}_{\bm{C}} = \mathbf{H_{3}}V_{\bm{C}}, \bm{\Delta}_{\bm{D}} = \mathbf{H_{4}}V_{\bm{D}}
    \end{align}
\end{example}

Regarding messages $\bm{A},\bm{B},\bm{C}$, the user has 
\begin{align}
    \begin{bmatrix}
        V_{\bm{Y}}\\
        \bm{\Delta}_{\bm{A}}\\
        \bm{\Delta}_{\bm{B}}\\
        \bm{\Delta}_{\bm{C}}
    \end{bmatrix}=
    \underbrace{
    \begin{bmatrix}
        2\mathbf{I}_3 & \mathbf{I}_3 & \mathbf{I}_3\\
        \mathbf{H}_1 & \mathbf{0}_{2\times 3} & \mathbf{0}_{2\times 3}\\
        \mathbf{0}_{2\times 3} & \mathbf{H}_{2} & \mathbf{0}_{2\times 3}\\
        \mathbf{0}_{2\times 3} & \mathbf{0}_{2\times 3} & \mathbf{H}_{3}
     \end{bmatrix}}_{\mathbf{G}_{\{1,2,3\}}}
     \begin{bmatrix}
        V_{\bm{A}}\\
        V_{\bm{B}}\\
        V_{\bm{C}}
     \end{bmatrix}
\end{align}

To recover $V_{\bm{A}},V_{\bm{B}},V_{\bm{C}}$ and thus recover $\bm{A},\bm{B},\bm{C}$, $\mathbf{G}_{\{1,2,3\}} \in \mathbb{F}_{3^l}^{9 \times 9}$ must have full rank. Let us explicitly write down $\mathbf{G}_{\{1,2,3\}}$ as 
\begin{align}
    \begin{bmatrix}
        {\color{blue}2} & 0 & 0 & 1 & 0 & 0 & 1 & 0 & 0\\
        0 & 2 & 0 & 0 & {\color{blue}1} & 0 & 0 & 1 & 0\\
        0 & 0 & 2 & 0 & 0 & 1 & 0 & 0 & {\color{blue}1}\\
        x_{1,1}^{(1)} & \mbox{\color{red}$\mathtt{x}_{1,2}^{(1)}$} & x_{1,3}^{(1)} & 0 & 0 & 0 & 0 & 0 & 0\\
        x_{1,1}^{(2)} & x_{1,2}^{(2)} & \mbox{\color{red}$\mathtt{x}_{1,3}^{(2)}$} & 0 & 0 & 0 & 0 & 0 & 0\\
        0 & 0 & 0 & \mbox{\color{red}$\mathtt{x}_{2,1}^{(1)}$}& x_{2,2}^{(1)} & \mbox{\color{black}$\mathtt{x}_{2,3}^{(1)}$} & 0 & 0 & 0\\
        0 & 0 & 0 & \mbox{\color{black}$\mathtt{x}_{2,1}^{(2)}$} & x_{2,2}^{(2)} & \mbox{\color{red}$\mathtt{x}_{2,3}^{(2)}$} & 0 & 0 & 0\\
        0 & 0 & 0 & 0 & 0 & 0 & \mbox{\color{red}$\mathtt{x}_{3,1}^{(1)}$} & x_{3,2}^{(1)} & x_{3,3}^{(1)}\\
        0 & 0 & 0 & 0 & 0 & 0 & x_{3,1}^{(2)} & \mbox{\color{red}$\mathtt{x}_{3,2}^{(2)}$} & x_{3,3}^{(2)}
    \end{bmatrix}.\notag
\end{align}
Now note that $\det(\mathbf{G}_{\{1,2,3\}}) = f_1$ is an $M^{2}(M-1) = 18$-variate non-zero polynomial of degree $6$. The polynomial is non-zero because
e.g., setting the variables shown in red color as $1$ and the rest of the variables to $0$, yields the evaluation $f_1=\lambda_1\lambda_2\lambda_3=2$.

To ensure the joint privacy of $(\bm{\theta}, \bm{\mathcal{S}})$, the matrix
\begin{align}
    \mathbf{G}_{\{1,2,4\}} = 
    \begin{bmatrix}
        2\mathbf{I}_3 & \mathbf{I}_3 & \mathbf{I}_3\\
        \mathbf{H}_1 & \mathbf{0}_{2\times 3} & \mathbf{0}_{2\times 3}\\
        \mathbf{0}_{2\times 3} & \mathbf{H}_{2} & \mathbf{0}_{2\times 3}\\
        \mathbf{0}_{2\times 3} & \mathbf{0}_{2\times 3} & \mathbf{H}_{4}
     \end{bmatrix}
\end{align}
should also be invertible, which enables the user to recover $\bm{A},\bm{B},\bm{D}$ if the CSI is $2\bm{A}+\bm{B}+\bm{D}$. Similarly, $\mathbf{G}_{\{1,3,4\}}, \mathbf{G}_{\{2,3,4\}}$ should also be invertible. Let 
\begin{align}
    f_2 = \det(\mathbf{G}_{\{1,2,4\}}), f_3 = \det(\mathbf{G}_{\{1,3,4\}}), f_4 = \det(\mathbf{G}_{\{2,3,4\}}).\notag
\end{align}
Similarly, $f_2,f_3,f_4$ are $18$-variate polynomials of degree $6$. Thus $f = f_1f_2f_3f_4$ is a $KM(M-1) = 24$-variate non-zero polynomial of degree $\tbinom{K}{M}M(M-1) = 24$. According to Schwartz-Zippel Lemma, if the $24$ variables are randomly chosen from $\mathbb{F}_{3^l}$, 
\begin{align}
    \Pr(f=0) \leq \frac{\deg(f)}{3^l} = \frac{24}{3^l}.
\end{align} 
When $l \geq 3$ we have $\Pr(f=0) < 1$ which implies that there exists a choice of the $24$ variable such that $f\neq 0$ and $\mathbf{G}_{\{1,2,3\}}, \mathbf{G}_{\{1,2,4\}},\mathbf{G}_{\{1,3,4\}}, \mathbf{G}_{\{2,3,4\}}$ are all invertible.

\bibliographystyle{IEEEtran}
\bibliography{Thesis}

\end{document}